\documentclass[fleqn,usenatbib,useAMS]{mnras}

\usepackage{newtxtext,newtxmath}
\usepackage[T1]{fontenc}
\usepackage{ae,aecompl}

\usepackage{graphicx}	% Including figure files
\usepackage{amsmath}	% Advanced maths commands
\usepackage{amssymb}	% Extra maths symbols
\usepackage{xcolor,xspace}

                \usepackage{threeparttable} %table with notes
                 \usepackage{threeparttablex} %longtable with notes
\newcommand{\lya}{\relax \ifmmode {\mbox Ly}\alpha\else Ly$\alpha$\fi}
\newcommand{\ha}{\relax \ifmmode {\mbox H}\alpha\else H$\alpha$\fi}
\newcommand{\hg}{\relax \ifmmode {\mbox H}\gamma\else H$\gamma$\fi}
\newcommand{\hd}{\relax \ifmmode {\mbox H}\delta\else H$\delta$\fi}
\newcommand{\hb}{\relax \ifmmode {\mbox H}\beta\else H$\beta$\fi}
\newcommand{\sii}{\relax \ifmmode {\mbox S\,{\scshape ii}}\else S\,{\scshape ii}\fi}
\newcommand{\nii}{\relax \ifmmode {\mbox N\,{\scshape ii}}\else N\,{\scshape ii}\fi}
\newcommand{\neiii}{\relax \ifmmode {\mbox Ne\,{\scshape iii}}\else Ne\,{\scshape iii}\fi}
\newcommand{\oii}{\relax \ifmmode {\mbox O\,{\scshape ii}}\else O\,{\scshape ii}\fi}
\newcommand{\oi}{\relax \ifmmode {\mbox O\,{\scshape i}}\else O\,{\scshape i}\fi}
\newcommand{\oiii}{\relax \ifmmode {\mbox O\,{\scshape iii}}\else O\,{\scshape iii}\fi}
\newcommand{\hii}{\relax \ifmmode {\mbox H\,{\scshape ii}}\else H\,{\scshape ii}\fi}
\newcommand{\hi}{\relax \ifmmode {\mbox H\,{\scshape ii}}\else H\,{\scshape i}\fi}

\newcommand{\kms}{\mbox{km\,s$^{-1}$}}
\newcommand{\cgs}{\mbox{erg\,s$^{-1}$}}
\newcommand{\ourobject}{\mbox{$J$\,1429}}
\title[Chemodynamics of Green Peas]{Chemodynamics of Green Pea galaxies - I.  Outflows and turbulence driving the escape of ionising photons and chemical enrichment\thanks{Based on observations made with the William Herschel Telescope operated on the island of La Palma by the Isaac Newton Group of Telescopes in the Spanish Observatorio del Roque de los Muchachos of the Instituto de Astrof\'isica de Canarias.} }

\author[L. Hogarth et al.]{
L. Hogarth$^{1}$\thanks{E-mail: lucy.hogarth.18@ucl.ac.uk},
R. Amor\'in$^{2,3}$\thanks{E-mail: ricardo.amorin@userena.cl},
J.~M. V\'ilchez$^{4}$,
G.~F. H\"agele$^{5}$,
M. Cardaci$^{5}$, \and
E. P\'erez-Montero$^{4}$,
V. Firpo$^{6}$,
A. Jaskot$^{7}$,
R. Ch\'avez$^{8}$
\\
$^{1}$University College London (UCL), London WC1E 6BT, UK;\\
$^{2}$Instituto de Investigaci\'on Multidisciplinar en Ciencia y Tecnolog\'ia, Universidad de La Serena, Ra\'ul Bitr\'an 1305, La Serena, Chile;\\
$^{3}$Departamento de Astronom\'ia, Universidad de La Serena, Av. Juan Cisternas 1200 Norte, La Serena, Chile;\\
$^{4}$Instituto de Astrof\'isica de Andaluc\'ia - CSIC, Glorieta de la Astronom\'ia s.n., E-18008 Granada, Spain;\\
$^{5}$Instituto de Astrof\'isica de La Plata (UNLP - CONICET), La Plata, Argentina;\\
$^{6}$NSF's National Optical-Infrared Astronomy Research Laboratory/Gemini Observatory\\
$^{7}$Astronomy Department, Williams College, Williamstown, MA 01267, USA;\\
$^{8}$CONACYT-Instituto de Radioastronom\'ia y Astrof\'isica, UNAM, Campus Morelia, C.P. 58089, Morelia, M\'exico.
}

\date{Accepted XXX. Received YYY; in original form ZZZ}

\pubyear{2020}

\begin{document}
\label{firstpage}
\pagerange{\pageref{firstpage}--\pageref{lastpage}}
\maketitle

\begin{abstract}
  We investigate the ionised gas kinematics, physical properties and chemical abundances of \mbox{SDSS\,$J$142947}, a Green Pea galaxy at redshift  z$\sim$\,0.17 with strong, double-peak \lya\ emission and indirect evidence of Lyman continuum (LyC) leakage.
  Using high-dispersion spectroscopy, we perform a multi-component analysis of emission-line profiles.
  Our model consistently fits all lines as a narrow component with intrinsic velocity dispersion $\sigma \sim$\,40\,\kms, and two broader blue-shifted components with $\sigma \sim$\,90\,\kms\ and $\sim$\,250\,\kms.
  We find electron densities and temperatures, ionisation conditions, and direct O/H and N/O abundances for each component.
  A highly ionised, metal-poor, young and compact starburst dominates narrow emission, showing evidence of hard radiation fields and elevated N/O.
  The blue-shifted broader components are consistent with highly turbulent, possibly clumpy ionised gas at the base of a strong photoionised outflow, which accounts for $\gtrsim$\,50\% of the integrated emission-line fluxes.
  The outflow is dense and metal-enriched compared to the \hii\ regions, with expansion velocities larger than those obtained from UV interstellar absorption lines under standard assumptions.
  Some of these metals may be able to escape, with outflows loading factors comparable to those found in high-$z$ galaxies of similar SFR/Area.
  Our findings depict a two-stage starburst picture; hard radiation fields from young star clusters illuminate a turbulent and clumpy ISM that has been eroded by SNe feedback.
  While UV data suggest an extended \lya\ halo with high average \hi\ column density, LyC photons could only escape from \mbox{SDSS\,$J$142947} through low \hi\ density channels or filaments in the ISM approaching density-bounded conditions, traced by outflowing gas.
\end{abstract}


%%%%%%%%%%%%%%%%%%%%%%%%%%%%%%%%%%%%%%%%%%%%%%%%%%

%%%%%%%%%%%%%%%%% BODY OF PAPER %%%%%%%%%%%%%%%%%%

\section{Introduction}
\label{intro}

Young low-mass galaxies are key in the cosmological context, playing a role in
early growth of galaxies and cosmic reionisation \citep{Bouwens2015,Castellano2016}.
While normal galaxies at high redshift become more compact
\citep[e.g.][]{Shibuya2015,Paulino-Afonso2018}
and rapidly star-forming \citep[e.g.][]{Atek2014,Tasca2015,Santini2017}, increasing
observational evidence points toward ubiquitous extreme emission-line properties
amongst UV-bright galaxies at $z\gtrsim$\,6
\citep[e.g.][]{Smit2014,Smit2015,Huang2016,Khostovan2016,
Roberts-Borsani2016,Bowler2017,Castellano2017}.
This suggests more extreme interstellar medium (ISM) conditions compared to
lower redshift counterparts
--such as lower metallicity and harder radiation fields
\citep[e.g.][]{Stark2016,Stark2017,Sobral2018,
Dors2018}.
In order to evaluate the role of these younger star-forming systems in
the cosmic reionisation, it is crucial to be able to probe the production and
escape of ionising radiation, processes that are still poorly understood.
However, a comprehensive characterisation of the ISM properties of the first
galaxies is still not possible.

Analogues of early galaxies are exceedingly rare in the local universe, but
they may provide invaluable insight into the above open questions since they
can be studied in much greater detail. A class of low-metallicity galaxies, dubbed
Green Pea (GP) galaxies \citep{Cardamone2009}, probably has the closest resemblance
to high-z galaxies
\citep[see also e.g.][]{Nakajima2014,Amorin2017}.
These objects are undergoing galaxy-wide starburst activity and possess
both extreme emission-line properties and faint continuum luminosities.
Being on average more luminous (massive) than nearby \hii\ galaxies and
extragalactic \hii\ regions \citep[e.g.][]{Terlevich1981,Kunth2000,
Micheva2017,Yang2017c}, GPs were identified in the Sloan Digital Sky Survey
(SDSS) at 0.15$\lesssim z \lesssim$\,0.35 due to their extreme compactness and
distinctive green appearance, owing to unusually strong
[\oiii]$\lambda$\,4959,5007 (equivalent widths EW$\sim$\,200-2000 \AA)
redshifted to the SDSS $r$-band \citep{Cardamone2009}.

Consequently,  GP luminosity output is dominated by strong nebular and stellar
emission from very young compact starbursts with high specific star formation
rate \mbox{sSFR$\gtrsim$10$^{-9}$yr$^{-1}$} \citep{Cardamone2009} and low
($\sim$\,20\% solar in average) gas-phase metallicity \citep{Amorin2010,Amorin2012a}.
Compared to main sequence galaxies of similar stellar mass,
\mbox{10$^{8}$M$_{\odot}\lesssim$M$_{*}\lesssim$\,10$^{10}$M$_{\odot}$ \citep{Izotov2011}},
GPs are characterised by a lower O/H and higher N/O abundances, as studied
from the mass-metallicity and N/O-O/H relations \citep[][]{Amorin2010,Amorin2012a}.
This suggests a more rapid chemical evolution phase where gas inflows and outflows
may be highly relevant due to the increased SF activity
\citep[e.g.][and references therein]{Amorin2012a,SanchezAlmeida2015}.

The warm gas \ha\ kinematics of GPs has been shown to be highly complex
\citep[][hereinafter A12b]{Amorin2012b}.
Using deep high-dispersion spectroscopy, A12b demonstrated that the bright
emission lines from a sample of six GPs displayed largely non-Gaussian profiles.
These are well modelled as an ensamble of multiple narrow and broad kinematic
components superposed on kiloparsec scales.
The narrower \ha\ components show high intrinsic velocity dispersions suggesting
a strongly turbulent ISM. A12b interpret these as kinematically resolved ionised
regions, which can be spatially associated to resolved UV-bright clumps in
\textit{Hubble Space Telescope}  (HST) imaging.
The kinematics of GPs appear, therefore, consistent with the high velocity
dispersion of extreme emission-line galaxies (EELGs) at $z\sim$\,1-2
\citep{Maseda2014,Masters2014,Terlevich2015}, as well as with the internal
kinematics found in other dispersion-dominated compact star-forming galaxies
studied with integral field spectroscopy
\citep[e.g.][]{Law2009, Kassin2012, Mason2017,Turner2017}.
From the theoretical perspective, the turbulent nature of the warm ionised gas
in galaxies is typically associated to gravitational instabilities and/or
stellar feedback from intense star formation
\citep[e.g.][and references therein]{Krumholz2016}.

Moreover, A12b showed that the high velocity wings in the bright emission lines
of GPs require the presence of a broad component ($\sigma_{\rm broad}>$\,100\kms),
which comprises a high fraction of the total line luminosity
(up to $\sim$\,40\% in H$\alpha$).
This is consistent with the imprint of strong stellar feedback
--i.e. the collective effect of young massive star cluster winds and previous
SNe (A12b). Similarly, the presence and nature of broad components in other
young compact starburst galaxies at lower
\citep[e.g.][]{Castaneda1990,Roy1992,Martin1998,Izotov2007b,James2009,Firpo2010,
Firpo2011,Hagele2012,Terlevich2014,Arribas2014, Lagos2014, Olmo-Garcia2017,Rodriguez2019}
and higher redshift  \citep[e.g.][]{Genzel2011,Newman2012,Freeman2019,Davies2019}
has typically been ascribed to star-formation driven gas outflows.
However, this may not always be the cause of broad component emission and other
effects are not simple to disentangle. Alternative interpretations, such as the
presence of faint AGN activity \citep[e.g.][]{IzotovThuan2008} and complementary
interpretations, such as the effects of turbulent mixing layers (TML) on the
surface of nebular clumps embedded within the stellar outflow
\citep{Westmoquette2007,Binette2009}, might produce similar features.
However, these hypotheses are difficult to confirm or rule out based solely
on the analysis of a single (or a few) integrated emission lines.

In terms of ionisation properties, GPs are again extreme when compared to normal
star-forming galaxies and appear to have greater similarity to %Lyman-$\alpha$
\lya\ emitters at $z>$\,2 \mbox{\citep{Nakajima2013}}.
They show strong excitation at low metallicity, i.e. high [\oiii]/\hb\ and
low [\nii]/\ha\ in BPT diagnostics \citep{Baldwin1981}, and large [\oiii]/[\oii] ratios.
In the most extreme cases, this would imply ISM conditions approaching those of
optically thin, density-bounded HII regions \citep{Jaskot2013,Nakajima2014}
and/or the presence of a hard radiation field \citep{Stasinska2015}.
These extreme conditions may favour a higher ionising
radiation efficiency and higher escape fraction ($f_{\rm esc}$) of ionizing
photons \citep[e.g.][]{Jaskot2014}.
Similar to strong [\oiii]-emitters at $z\sim$\ 2-3
\citep{Erb2016,Nakajima2018,Vanzella2017,Amorin2017}, rest-frame UV spectroscopy
with HST-COS have confirmed strong Ly$\alpha$ emission in a large fraction of
GPs, showing moderate to high $f_{\rm esc}$
and a preference for double-peak Ly$\alpha$ profiles
\citep{Jaskot2014,Henry2015,Yang2017a, Verhamme2017}.
About a dozen such \lya\ emitting GPs have been probed for direct Lyman continuum
(LyC) detection at $\lambda<$\,912\AA\ using deep HST-COS spectra and all of them
have been detected with signal-to-noise (S/N) ratios $>3$
\citep{Izotov2016a,Izotov2016b,Izotov2018a,Izotov2018b}.
The leaking GPs show LyC escape fractions ($f_{\rm esc}$) ranging a few per cent
to more than 50\% \citep{Schaerer2016}, comparable to the $f_{\rm esc}$ values
found in the very few confirmed LyC leakers at $z\sim$\,3-4
\citep{deBarros2016,Shapley2016,Vanzella2016,
Vanzella2018,Bian2017,Ji2019}.
Most of these LyC leakers share a high \lya\ EW (>70\AA) with double-peak profiles
and small peak separation, suggesting low \hi\ column densities and hence giving
credence to theoretical predictions for LyC leakage \citep{Verhamme2015,Verhamme2017}.

The GPs and other similar compact UV-bright galaxies at low redshift with large
Ly$\alpha$ halos are typically dispersion-dominated in H$\alpha$
\citep[e.g.][]{Herenz2016}, suggesting a correlation between turbulence and $f_{\rm esc}$.
Moreover, strong outflows are expected to favour the escape of ionising photons
by clearing HI gas away from HII regions
\citep[e.g.][]{Heckman2011,Borthakur2014,Alexandroff2015,Erb2015}.
However, the ionised gas kinematics of leaking and non-leaking GPs traced by
UV absorption lines suggest that  extreme outflows may not be a sufficient
condition for LyC leakage \citep{Chisholm2017}.
Furthermore, the most extreme GPs show the smallest outflow velocities, suggesting
a suppression of strong superwinds in the younger and more optically thin compact
starbursts \citep{Jaskot2017}.

While it seems clear that several global properties such as compactness, starburst
ages, hardening and ionisation conditions, HI density and covering fraction and
gas kinematics may play a role in LyC leakage, divining which ISM properties
are most correlated with $f_{\rm esc}$(LyC) in galaxies remains elusive
\citep[][and references therein]{Izotov2018b}.
Confirming both the turbulent nature of the ISM and the ubiquitous presence of
strong, large-scale outflows in leaking and non-leaking GPs is, therefore, crucial
in understanding how these effects can shape the ISM. Such activity would promote
the chemical enrichment of the galaxy and contribute to open channels from which
Ly$\alpha$ and LyC photons may eventually escape into the galaxy's halos.

In this paper, we further the work of A12b by performing a \textit{chemodynamical}
study of a GP galaxy classified as a LyC leaking candidate by \citet{Alexandroff2015}.
The chemodynamical technique \citep{EstebanVilchez1992} relies on the combination
of chemical abundance and kinematic analysis, using deep high-dispersion spectroscopy
over the entire optical range, which includes faint key lines such as [\oiii]\,4363.
The study is further supplemented with high resolution imaging.
To date, such chemodynamical studies have been limited exclusively to nearby (bright)
star-forming regions \citep{EstebanVilchez1992,James2009,James2013a,James2013b,Hagele2012}.
Previous attempts for similar analysis using spatially-resolved spectroscopy of
GPs have been restricted by spatial and spectral resolution as well as the depth
in the visible and NIR \citep{Goncalves2010,Herenz2016,Lofthouse2017}.
While this paper is based on deep long-slit spectroscopy, a detailed 3D kinematic
and chemical abundance analysis using deep integral field spectroscopy are presented
in separate papers \citep[][H\"agele et al., in prep.]{Bosch2019}.

The paper is organised as follows: in Section 2 we present the general properties
for the galaxy under study and describe the complete dataset and data reduction.
In Section~3 we describe the methodology applied for the emission-line fitting,
while in Sect.~4 we present its main results. In Section~5 a discussion about
our chemodynamical analysis and its interpretation in the context of star formation
feedback and escape of ionising photons is presented.
Throughout this paper we adopt the standard $\Lambda$-CDM cosmology, $h=$\,0.7,
$\Omega_{\rm m}=$\,0.3, and $\Omega_{\Lambda}=$\,0.7 \citep{Spergel2007} and a
solar metallicity value of $12+\log$(O/H)$=$8.69 \citep{Asplund2009}.
Magnitudes are given in the AB system.

\section{SAMPLE AND DATA} % First session devoted to present data and basic analysis
\label{sec:data} % used for referring to this section from elsewhere

Our target galaxy is \mbox{SDSS\,$J$\,142947.00+064334.9}, hereinafter \ourobject, which is found at $z\sim$\,0.17 and belongs to a larger sample of GPs selected for a high-resolution spectroscopy survey (Amor\'in et al., in prep), whose first results were presented in A12b. \ourobject\ is representative of the most extreme GPs studied in the literature both in terms of sSFR and emission-line EWs \citep[cf. e.g.][]{Cardamone2009}.
\ourobject\ has been also included in samples presented in previous studies focusing on chemical abundances and physical properties of the ISM, both in the optical using {\it Sloan Digital Sky Survey} (SDSS) spectroscopy \citep{Shirazi2012,Shim2013} and in the UV using the {\it Hubble Space Telescope} (HST), the latter including Far-UV (FUV) spectroscopy with the {\it Cosmic Origins Spectrograph} (COS) \citep{Heckman2015,Alexandroff2015,Chisholm2015}.

\begin{figure}
\centering
\includegraphics[width=8cm, angle=0]{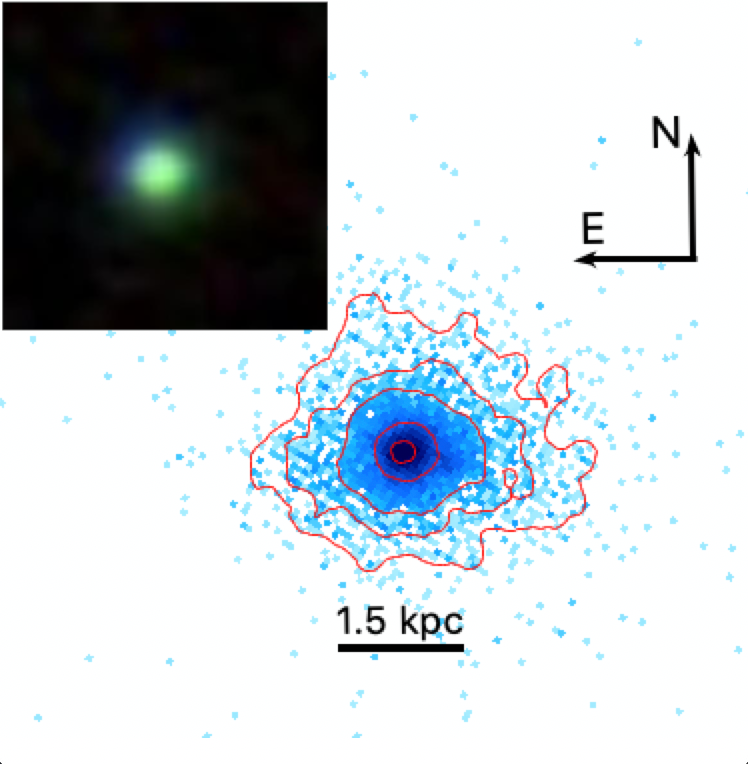}
\caption{HST-COS acquisition image ($T_{\rm exp}=90s)$ of \ourobject in the NUV ($\lambda_{\rm c} \sim$\,2000\AA). The figure has 3$"$ on a side --nearly the diameter of the COS aperture. The inner two contours indicate regions with radius of $\sim$\,150\,pc and $\sim$\,300\,pc, which nearly corresponds to the half-light radius of the galaxy (Table~1). The image is in log flux scale and the three outer isophotes indicate fluxes $\geq$\,1$\sigma_{\rm rms}$, 2$\sigma_{\rm rms}$, and 3$\sigma_{\rm rms}$, respectively.
These limits highlight the absence of both significant substructure in the inner kpc and tidal features in the outskirts. A \textit{gri} color composite SDSS image of $J1429$ is also presented on the upper left inset.}
\label{fig1}
\end{figure}

Table~\ref{tab1} summarises basic information on the target compiled from the literature. In Figure~\ref{fig1} we also present a SDSS {\it gri} colour image and a HST/COS acquisition image in the UV of \ourobject.
The galaxy is a low-mass (M$_{\*}\sim$\,4$\times$10$^{9}$\,M$_{\odot}$) low-metallicity ($\sim$20\% solar) starburst galaxy, showing very large EW in optical emission lines such as H$\beta$, which puts a strong constraint on the age of the ongoing star formation to only a few Myr \citep{Dottori1981,Leitherer1999}.
It is extremely compact in the UV, with a star formation rate per unit area of about 50 M$_{\odot}$\,yr$^{-1}$\,kpc$^{-2}$, which qualifies \ourobject\ as a Lyman-Break Analogue \citep[LBA,][]{Heckman2011,Overzier2009,Alexandroff2015}.
According to \citet{Alexandroff2015}, \ourobject\ appears to host a single massive UV clump of M$_{*} \sim$\,10$^{9}$M$_{\odot}$ resolved by HST (Fig.~\ref{fig1}), with no obvious signs of recent interactions in the stellar halo. The galaxy is a Ly$\alpha$ emitter (EW(Ly$\alpha$)$\sim$\,32\AA) showing a double-peak \lya\ profile with a small velocity shift between peaks, which makes \ourobject\ a good LyC candidate (see Section~\ref{sec:discussion}).

\begin{table}
	\centering
	\caption{Main Properties of GP\,$J1429$ from literature}
	\label{tab1}
	\begin{tabular}{ll|ll} % four columns, alignment for each
		\hline
$R.A.^a$   & 14:29:47.03 & $Dec.^a$ & +06:43:34.9 \\[2pt]
$z_{\rm SDSS}^a$&  0.17350  &  $M_{u^\prime}^b$  & -20.81 \\[2pt]
$\log$(M$_{\star}$/M$_{\odot})^c$ & 9.4 & $r_{50,\rm FUV}^e$ & 0.29\,kpc \\[2pt]
SFR$_{\rm \ha}^d$& 36.0\,M$_{\odot}$\,yr$^{-1}$ & SFR$_{\rm FUV}^d$ & 26.8\,M$_{\odot}$yr$^{-1}$\\[2pt]
$EW_{\rm [OIII]5007}^f$[\AA]		& 1038$\pm$7  &  $EW_{\rm H\beta}^f$[\AA]   & 178$\pm$3 \\[2pt]
$12+\log$(O/H)$_{\rm SDSS}^f$		&  8.20$\pm$0.07  &  $\log$(N/O)$_{\rm SDSS}^f$   & $-$1.17$\pm$0.06 \\[2pt]
 %        &    &     & \\
        \hline
	\end{tabular}
    \begin{tablenotes}
\footnotesize
\item $^a${\it J2000} coordinates and redshift from SDSS-DR7.
\item $^b$Obtained as $M_{u^\prime}= m_{u^\prime} - 5 \log(D_L/10 \rm pc) + 2.5\log(1+z_{\rm sdss})$, where $D_L$ is the luminosity distance and $m_{u^\prime}$ is the extinction corrected $u^{\prime}$ apparent magnitude from SDSS-DR14. Galactic extinction was taken from NED: \url{http://skyserver.sdss.org}.
\item $^c$Stellar mass was taken from the MPA-JHA SDSS catalog.
\item $^d$Attenuation-corrected star formation rates based on H$\alpha$ and FUV luminosities from \citet{Alexandroff2015}. These values assume a \citet{Kroupa2003} IMF and the \citet{Kennicutt2012} calibrations.
\item $^e$Petrosian half-light radius computed from the HST-COS NUV image (see Fig.~\ref{fig1}), from \citet{Heckman2015}.
\item $^f$Equivalent widths and oxygen and nitrogen-to-oxygen abundance derived from SDSS-DR7 fiber spectra following the direct method in Section~\ref{sec:results}.
		\end{tablenotes}
\end{table}
Observations of $J1429$ were conducted with the Intermediate Dispersion Spectrograph and Imaging System (ISIS)\footnote{http://www.ing.iac.es/astronomy/instruments/isis/} on the 4.2 m William Herschel Telescope (WHT) of the Isaac Newton Group (ING) at the Roque de los Muchachos Observatory (La Palma, Spain) in the night of June 3rd, 2012 (P25, PI: R. Amor\'in).
We used the EEV12 and REDPLUS CCD detectors attached to the blue and red arms of the spectrograph, respectively.  An instrumental configuration including a dichroic (D6100) allows us to split the incoming light into a blue and red arms of the spectrograph using the R1200B and R1200R gratings, respectively. For the blue arm, three setups using different central wavelengths were used to obtain spectra in a large fraction of the optical spectral range, from $\sim$4000\AA\ to $\sim$6000\AA, including lines from [\oii]\,3727 to [\oiii]\,5007. In the red arm, instead, we fixed the central wavelength at 7580\AA\ to obtain high S/N spectra within the wavelength range $\sim$7200\AA-8000\AA, thus including lines from [\oi]\,6300 to [\sii]\,6731.
The spatial scale of the data is 0.20 and 0.22 arcsec per pixel for the blue and red detectors, respectively. The average spectral dispersion and Full Width Half Maximuum (FWHM), as measured on bright sky lines and lamp lines, were 0.24 \AA\,pixel$^{-1}$ and 0.65\AA, respectively. Observations were taken using a long slit 0.9$"$ wide, oriented at the parallactic angle to reduce the effects of atmospheric dispersion. Observations were conducted under non-photometric conditions, with an average seeing of $\sim$1.1$"$. Three exposures of 1200s per setup were collected in both the blue and red arms. In the blue arm, each setup is across a different spectral range and, therefore, has a total exposure of $\sim$1h on source. In the red arm, the setups are over the same spectral range and have, consequently, a combined total exposure time of $\sim$3h on source.

\begin{figure*}
\centering
\includegraphics[scale=0.65, angle=0]{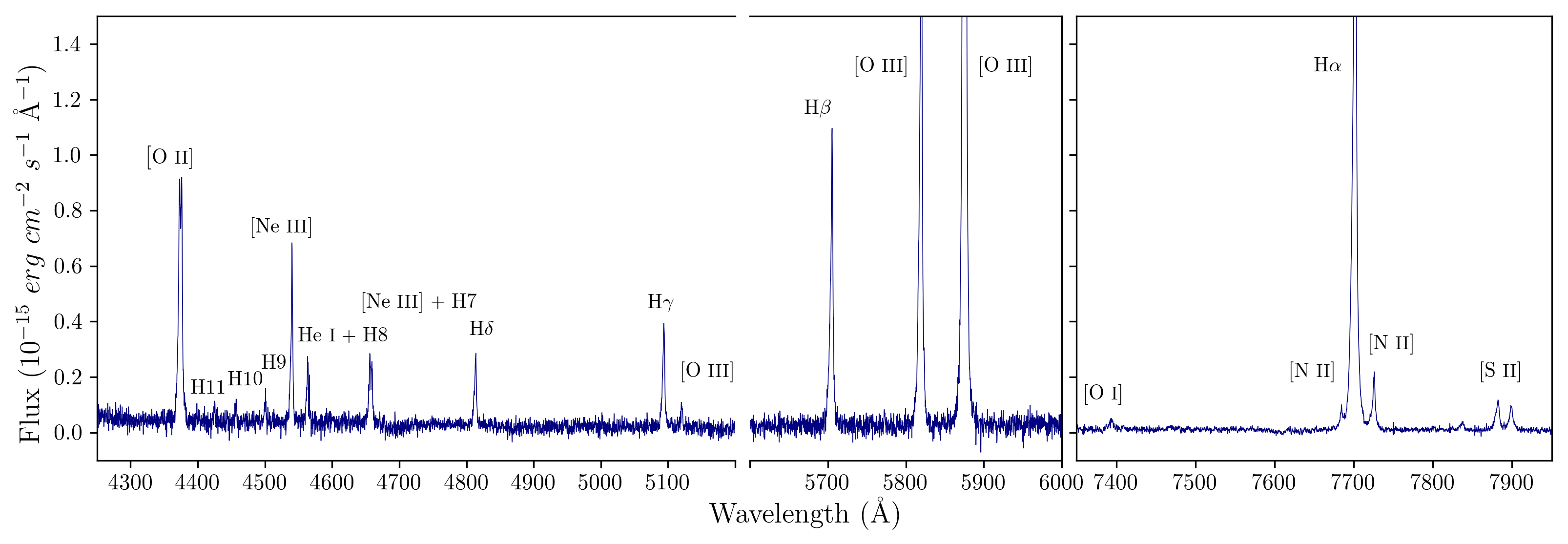}
\caption{Reduced 1D spectrum (from the blue and red arms) in the observed frame with significant spectral lines indicated.}
\label{fig2}
\end{figure*}
The raw ISIS long slit data were reduced and calibrated using standard IRAF subroutines by following the steps outlined by \citet{Hagele2007}. In brief, the 2D spectra were bias frame subtracted and corrected for both flat field inhomogeneities and background sky emission. Spectral wavelength calibration was performed using CuNe+CuAr lamp arcs with the R1200B and R1200R gratings. The uncertainty in wavelength calibration was about 0.3\AA. The spectra were corrected for atmospheric extinction
and flux calibrated using one spectrophotometric standard star (BD33) observed with the same ISIS setup immediately after the science data acquisition. One-dimensional spectra were extracted from the flux-calibrated frames using a spatial aperture of 20 pixels (the maximum extent of the emission lines in the spatial direction). In Figure~\ref{fig2}, we present the 1D spectrum of \ourobject\ after combining all the single exposures of the three spectral setups.

\section{Multi-component analysis of complex emission-line profiles}
\label{sec:method}

Having obtained two sets of reduced spectra from the blue (4000 - 6000 \AA) and red (7000 - 8500 \AA) configurations, the continuum was subtracted locally by taking emission-line free spectral windows of  $\approx 20$\AA \ either side of each line. The continuum level is then modelled in each window as a zeroth-order polynomial.
The continuum subtraction is, therefore, independent of the multi-component analysis for each emission line. We note that this method of continuum subtraction may be highly inaccurate when fitting the continuum below H recombination lines in normal galaxies, particularly \hb, due to underlying stellar absorption. The spectrum of $J1429$, however, does not show any signature of underlying absorption, as expected from its very high equivalent width (EW(\hb)$=$178\AA), which is indicative of a young stellar age of the dominant emitting regions \citep[e.g.][]{Dottori1981,Terlevich2004}. The error associated with continuum subtraction is, therefore, negligible compared to the errors in the line fitting procedure.

The continuum-subtracted data was then input into a custom-built, multi-component Gaussian fitting routine written in Python. By observation, it is clear the emission spectra are kinematically complex (see e.g. Figure~\ref{fig3}),  so it is assumed that a single Gaussian model will fail for all the observed lines. Our code uses curve-fitting routines from the scipy.optimize package \citep{Jones2001} and has been tailored to enable physical, and kinematic constraints to multiple Gaussian components, whilst also allowing simultaneous fitting of partially blended lines. The fitting parameters in the multi-component Gaussian model are the peak amplitude (flux), peak position (\AA) and velocity dispersion (\AA) for each respective component. These parameters can be left completely free, subject to an initial guess and constraints, or completely fixed, depending on the input criteria. Our code performs $\chi ^2$ minimisation, and the uncertainty in each emission-line component is assumed to sum in quadrature to the total uncertainty for the individual line.

\subsection{Methodology}

\subsubsection{Free and Fixed Parameters}

Following the methodology developed in \citet{Hagele2012}, the fitting process assumes that emission-line components will have the same kinematics and velocity as either H$\alpha$ or [\oiii] $\lambda$5007 (i.e. the brightest lines in our data), depending on the excitation structure and the energies involved. The first step is, therefore, to fit these two lines accurately.

H$\alpha$ is fitted simultaneously with [\nii] $\lambda \lambda $6548, 6584, which are located well within the wings of H$\alpha$. It is crucial to fit the H$\alpha$ and [\nii] lines concurrently to avoid over-estimating the flux in the wings for each respective line. During the simultaneous fitting, the peak velocity %($v_{peak}$)
and velocity dispersion of H$\alpha$ and [\nii] lines are fixed to each other. Furthermore, the amplitude of the [\nii] $\lambda$6548 components are fixed to the theoretical ratio of $1/3$ of those in [\nii] $\lambda$6584, which are free parameters within the fitting procedure (see Figure~\ref{fig3}).

The central velocity and velocity dispersion of components that minimises $\chi^2$ in this fit are copied to the other low-ionisation lines predicted to fall within the same ionisation region, leaving their amplitude as the only free parameter. This follows the assumption that the gas traced by [\nii], [\sii], [\oii] and [\oi] is the same as Hydrogen recombination lines (i.e. H$\alpha$, H$\beta$, H$\gamma$ and H$\delta$), whereas all high ionisation lines, such as [\oiii] $\lambda$4363 and [\neiii]$\lambda$3968 follow the same kinematics of [\oiii] $\lambda \lambda$4959, 5007. The [\oiii] $\lambda$5007 line is the only other line (alongside H$\alpha$) which has its amplitude, dispersion and
kinematics all as free parameters (see Figure~\ref{fig3}).
The [\oiii] $\lambda$5007 solution is copied in the same manner as H$\alpha$ to lines of the same ionisation region. In order to verify the validity of fixing these parameters, they are permitted to vary in a separate fit. This leads to a similar solution, therefore, to reduce the number to free parameters, the kinematics and dispersion are fixed for all faint emission lines. Furthermore, leaving all the parameters free to vary typically produces better fits in terms of $\chi^2$ minimisation, but it increases the possibility of degenerate solutions that may lead to unphysical results (i.e. non-consistent line ratios). By letting fewer parameters vary, the risk of nonsensical interpretations in this work is reduced.

\subsubsection{Two-Component Solution}

 We first fit the two bright emission lines, H$\alpha$ and [\oiii] $\lambda$5007, with two Gaussian components. The two-component solutions for H$\alpha$ and [\oiii]\,$\lambda$\,5007 both display a similar component structure comprised of a narrow component and blue-shifted broad component. The observed velocity dispersion of the two components are fairly consistent between the two lines; the narrow component having a dispersion $\sigma \sim$\,40\,\kms\ and the broad component having $\sigma \sim$\,130-140\,\kms.
 They also share a similar kinematic structure with a narrow component red-shifted by only $\Delta$v\,$\sim$\,5 - 10\,\kms$\ $and a broad component blue-shifted by $\Delta$v\,$\sim$\,50\,\kms, with respect to the peak wavelength of the composite profile.

\begin{figure*}
\centering
%\begin{subfigure}[b]{\columnwidth}
\includegraphics[scale=0.2]{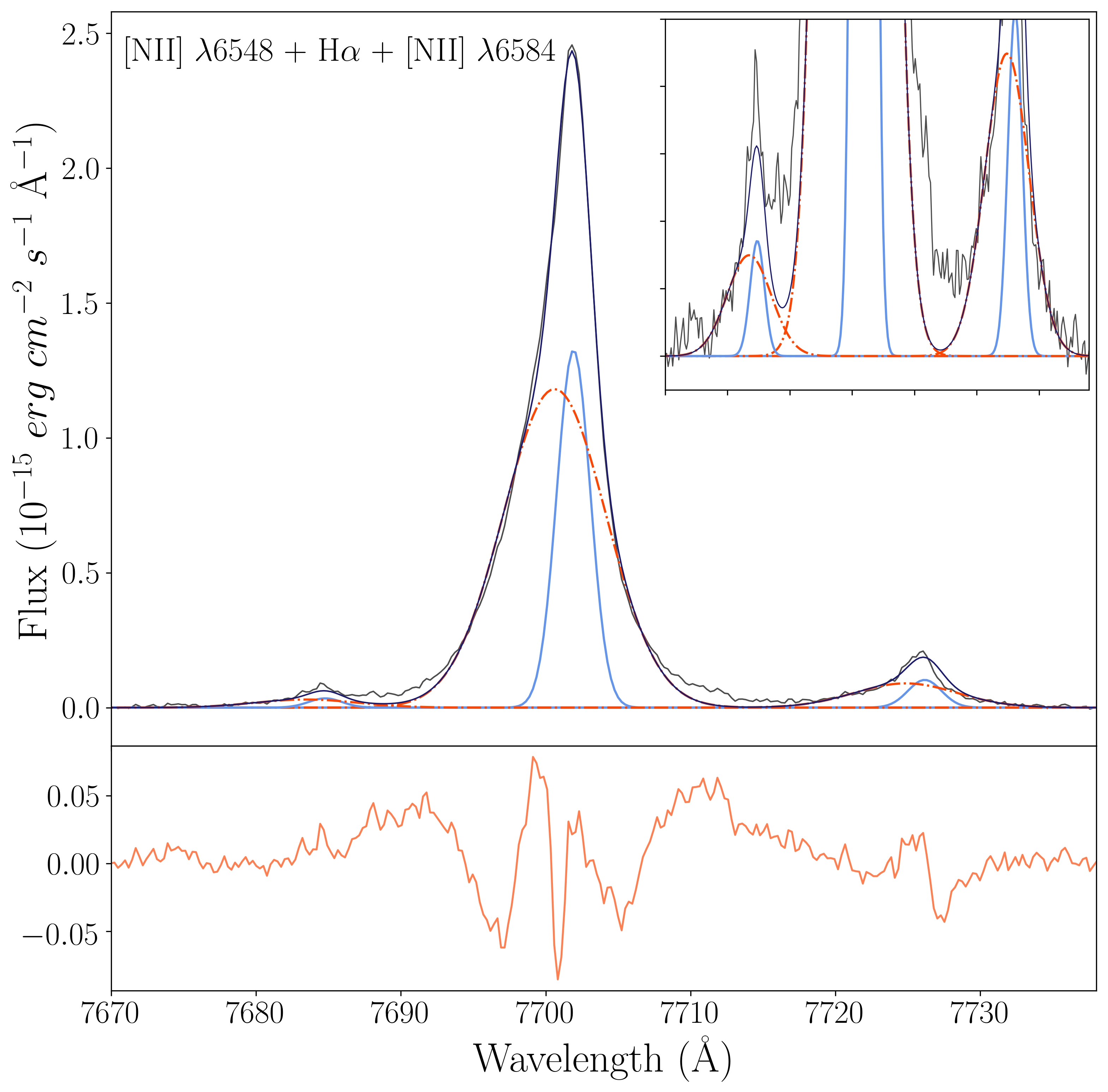}
%\end{subfigure}
%\centering
%\begin{subfigure}[b]{\columnwidth}
\includegraphics[scale=0.2]{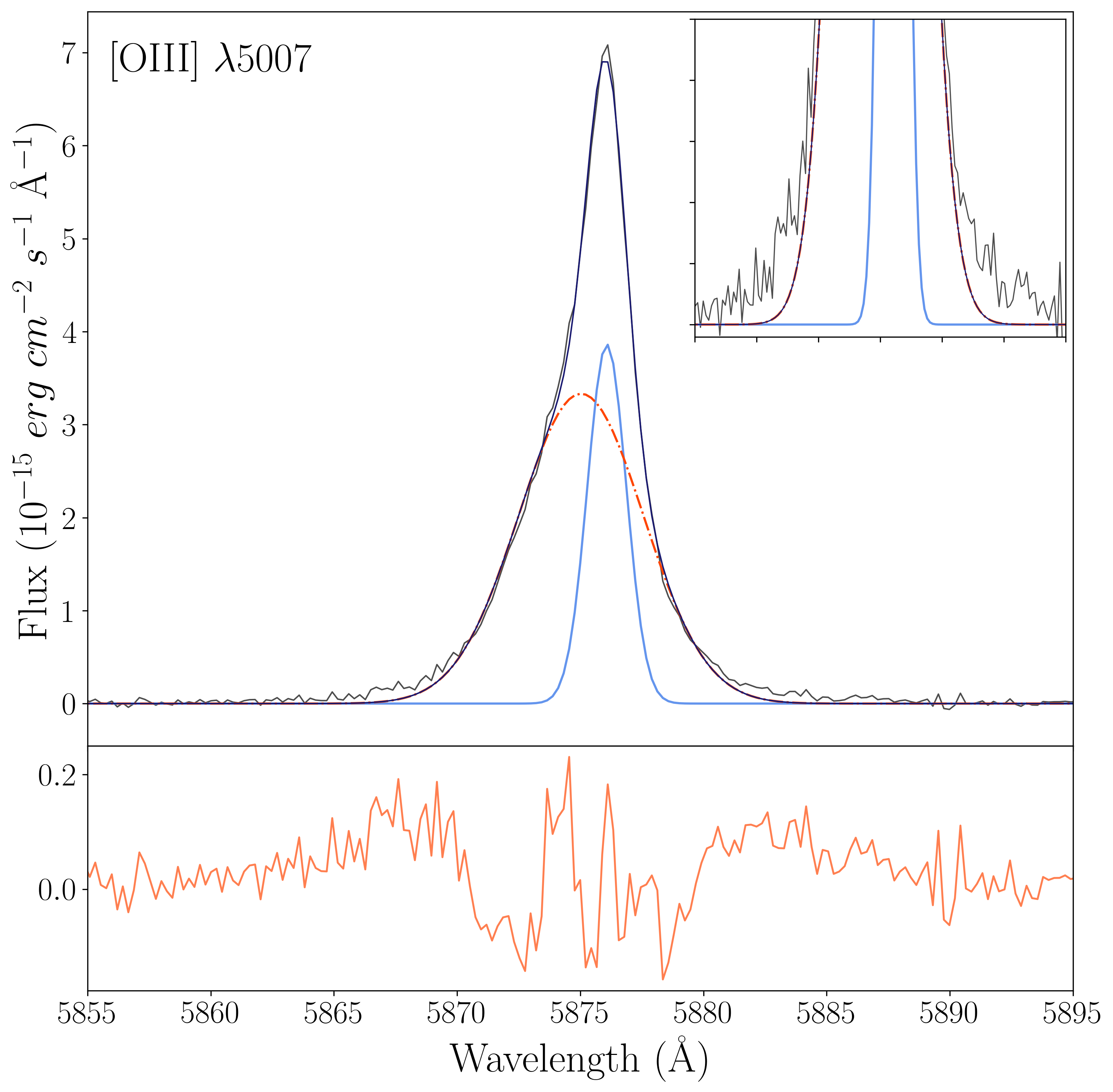}
%\end{subfigure}
\caption{Two-component fits of bright spectral lines  H$\alpha$ and [\nii]\,$\lambda \lambda$6548, 6584 (top) and [\oiii]\,$\lambda$\,5007 (bottom). Within these plots, the black solid line is used for the original data, the pale blue solid line for the narrow component and the red dashed line for the broader component. The composite two-component profiles are depicted by the dark blue solid line.}
\label{fig3}
\end{figure*}

Visually, the two component fit for H$\alpha$ and [\oiii]\,$\lambda$\,5007 plotted in Figure~\ref{fig3} is poor, as evidenced by the strong systematics in the residuals. Specifically, the broader component does not accurately describe the wings of either emission line and the residuals have a definite structure, implying the line has not been well reproduced. It can be inferred from the residuals in the bright lines, therefore, that the kinematic structure of the emission spectra is even more complex than a two-component model can describe, requiring the addition of at least one further Gaussian component.

\subsubsection{Three-Component Solution}

\begin{figure*}
\centering
\includegraphics[scale=.21, angle=0]{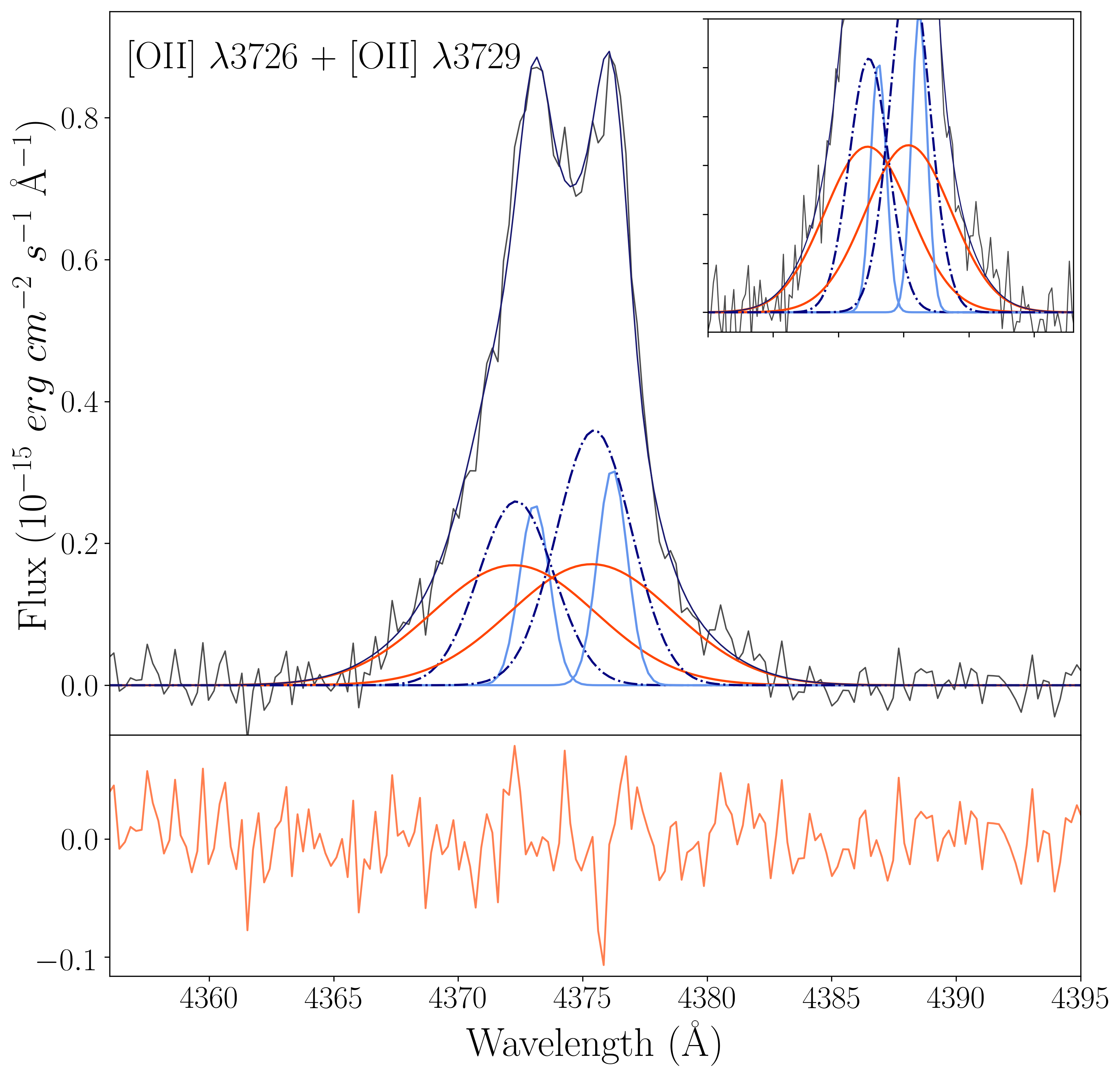}
\includegraphics[scale=.21, angle=0]{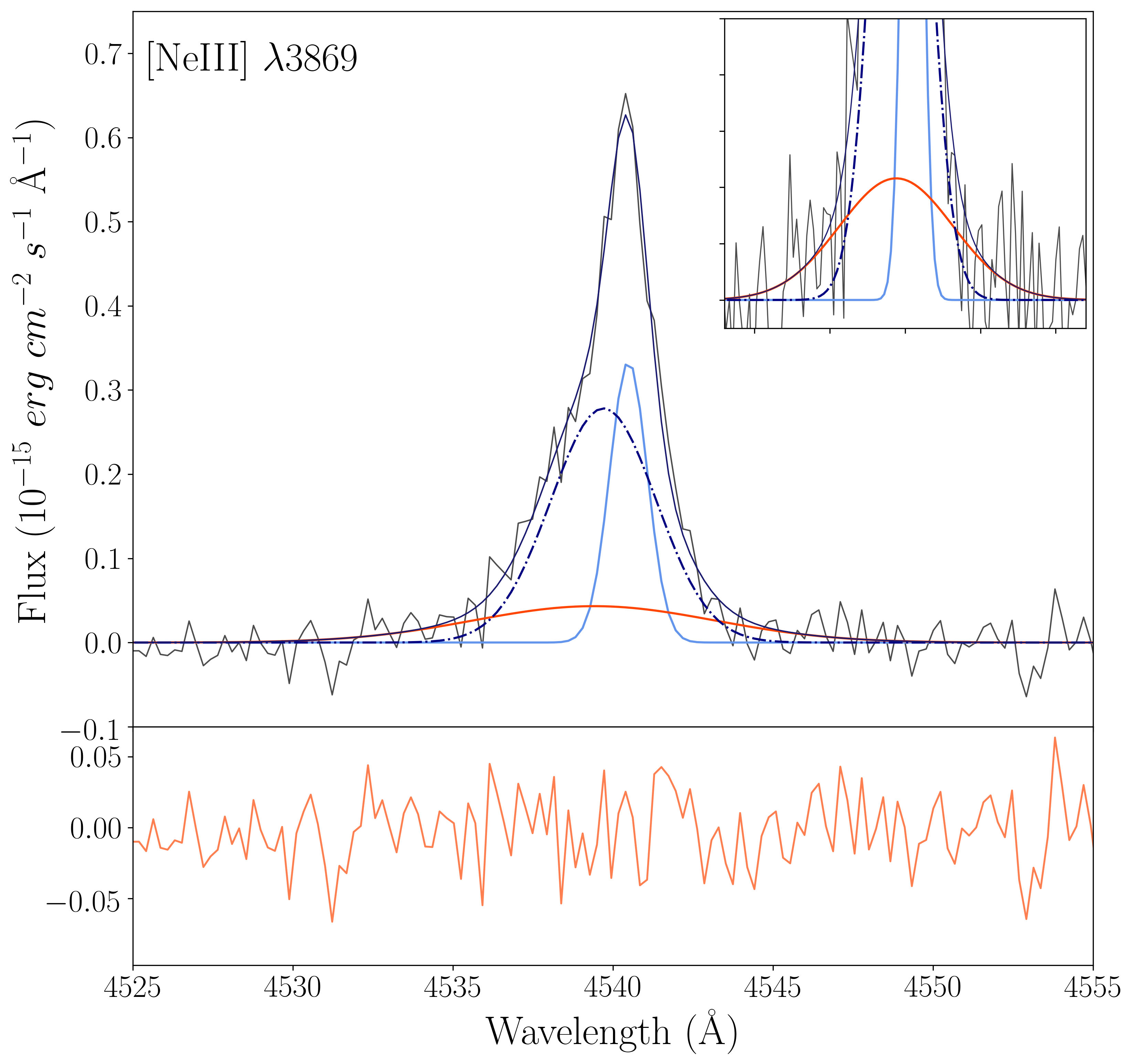}
\includegraphics[scale=.21, angle=0]{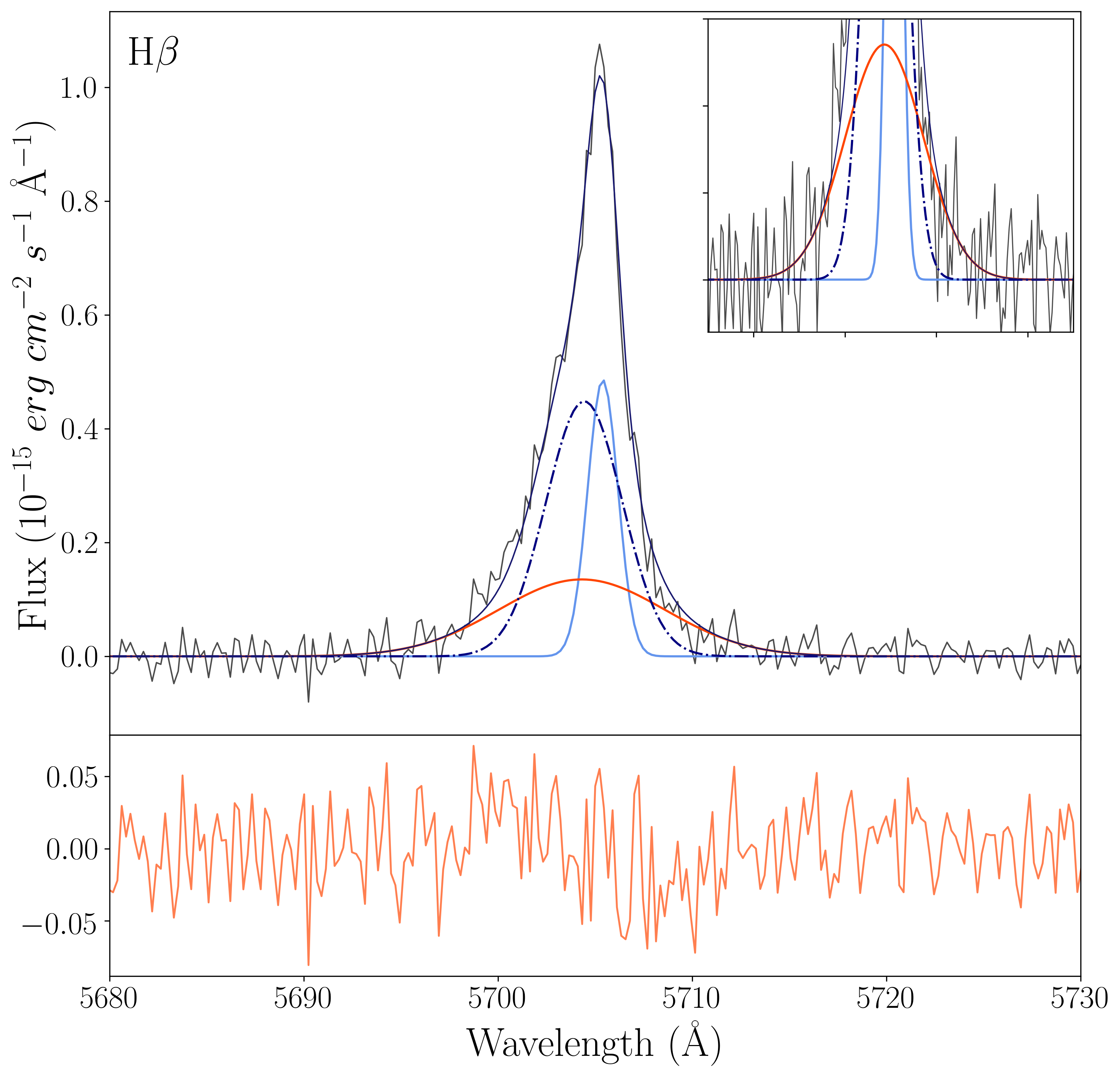}\\
\includegraphics[scale=.21, angle=0]{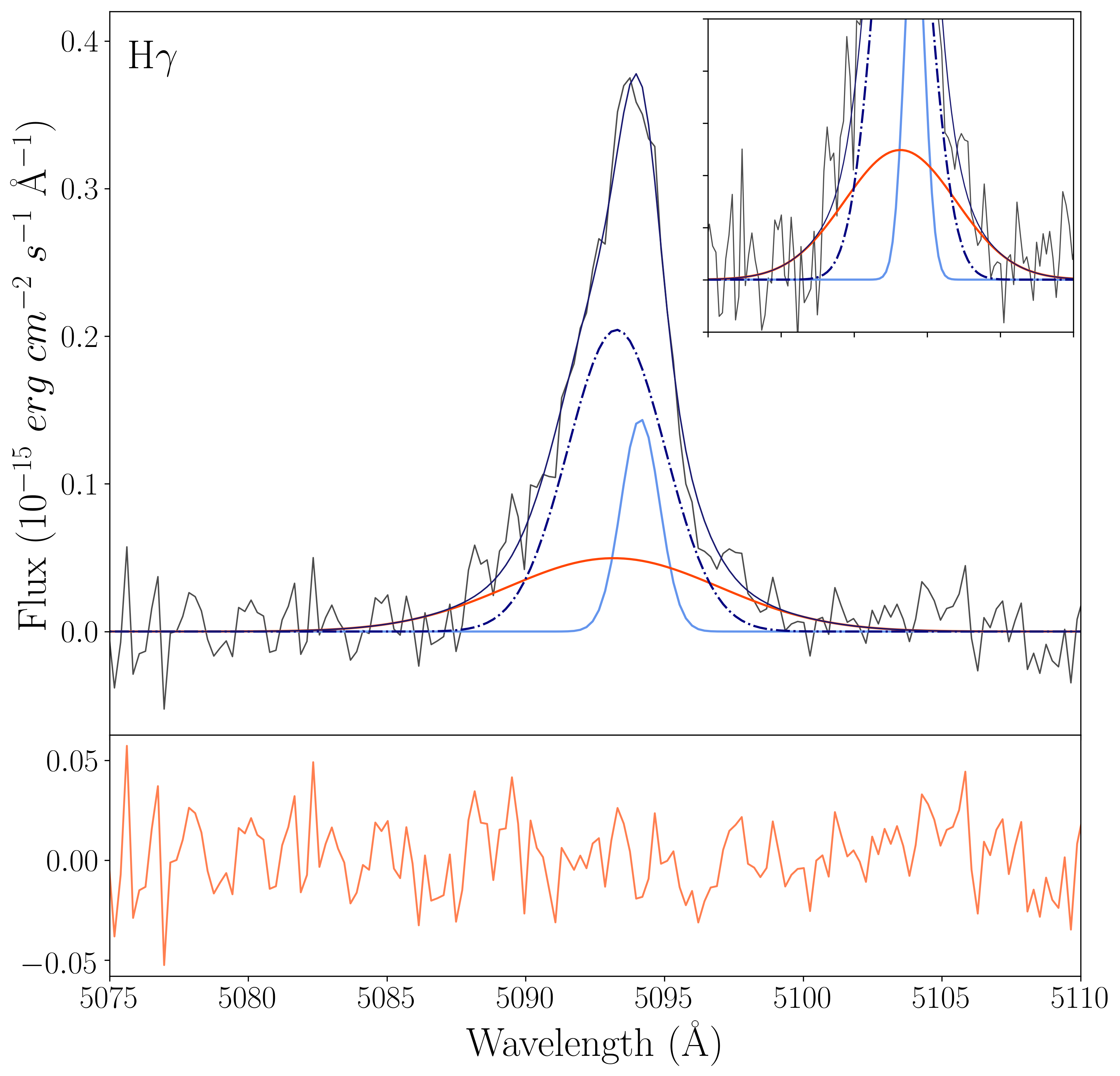}
\includegraphics[scale=.21, angle=0]{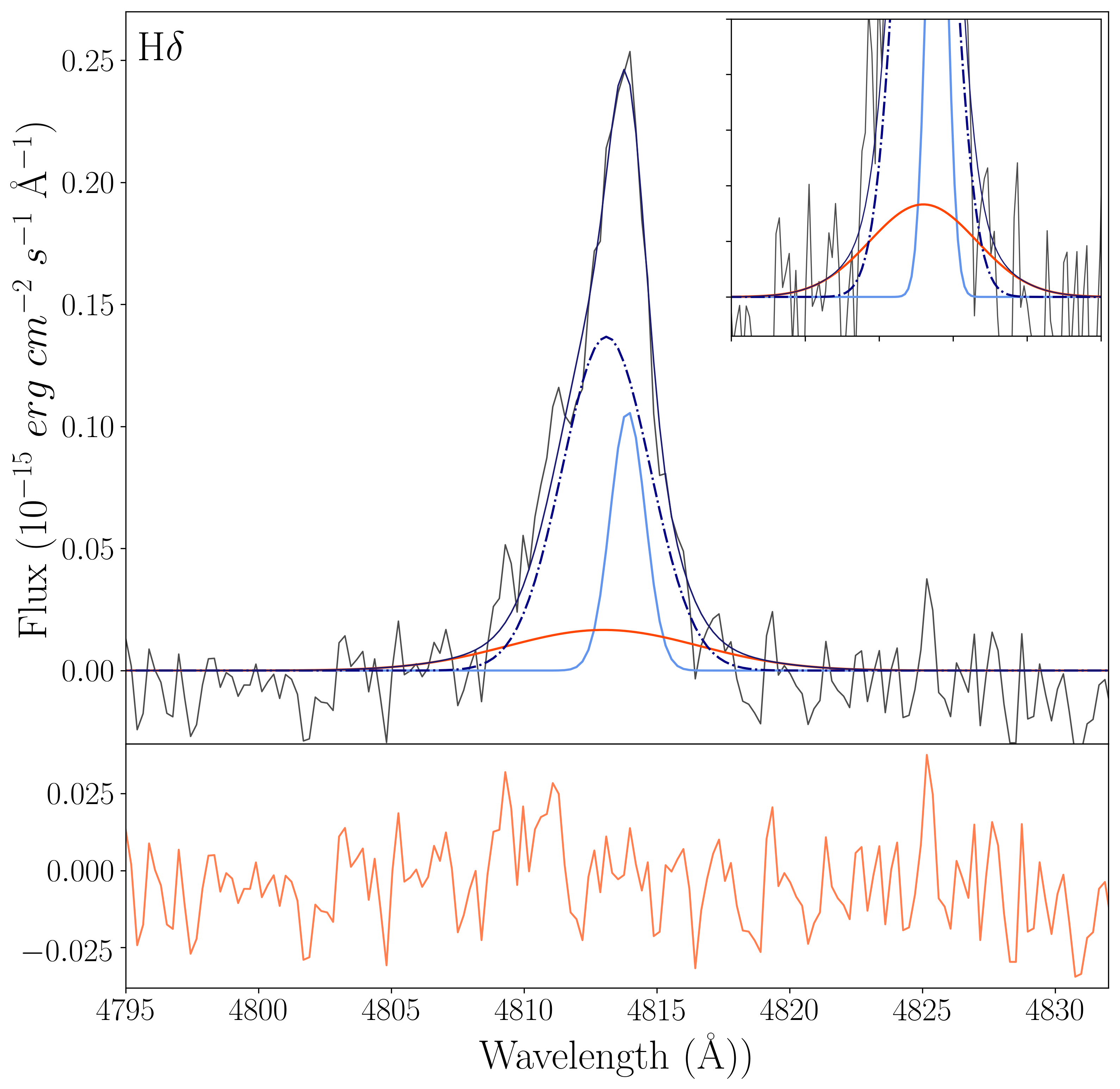}
\includegraphics[scale=.21, angle=0]{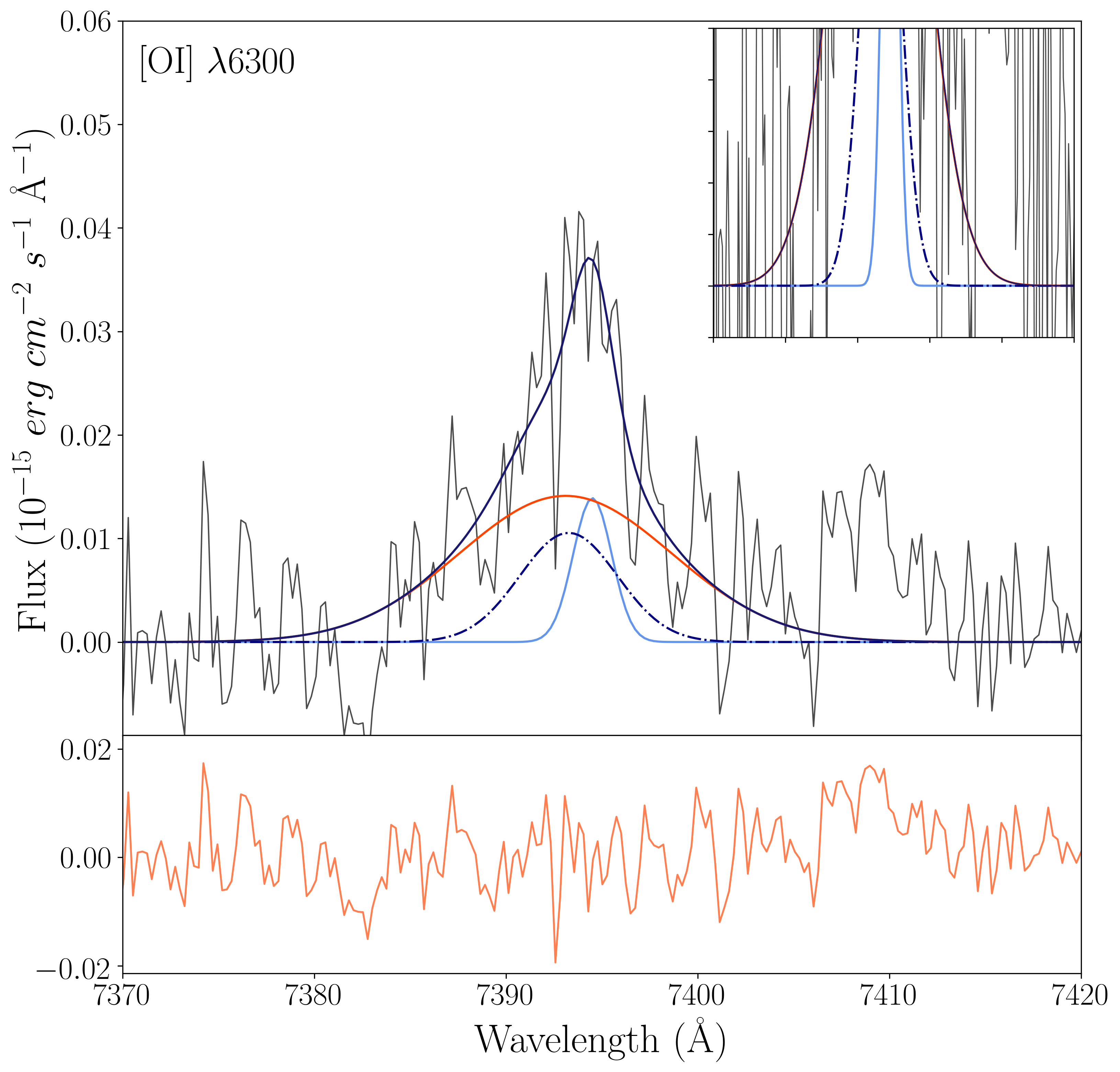}\\
\includegraphics[scale=.21, angle=0]{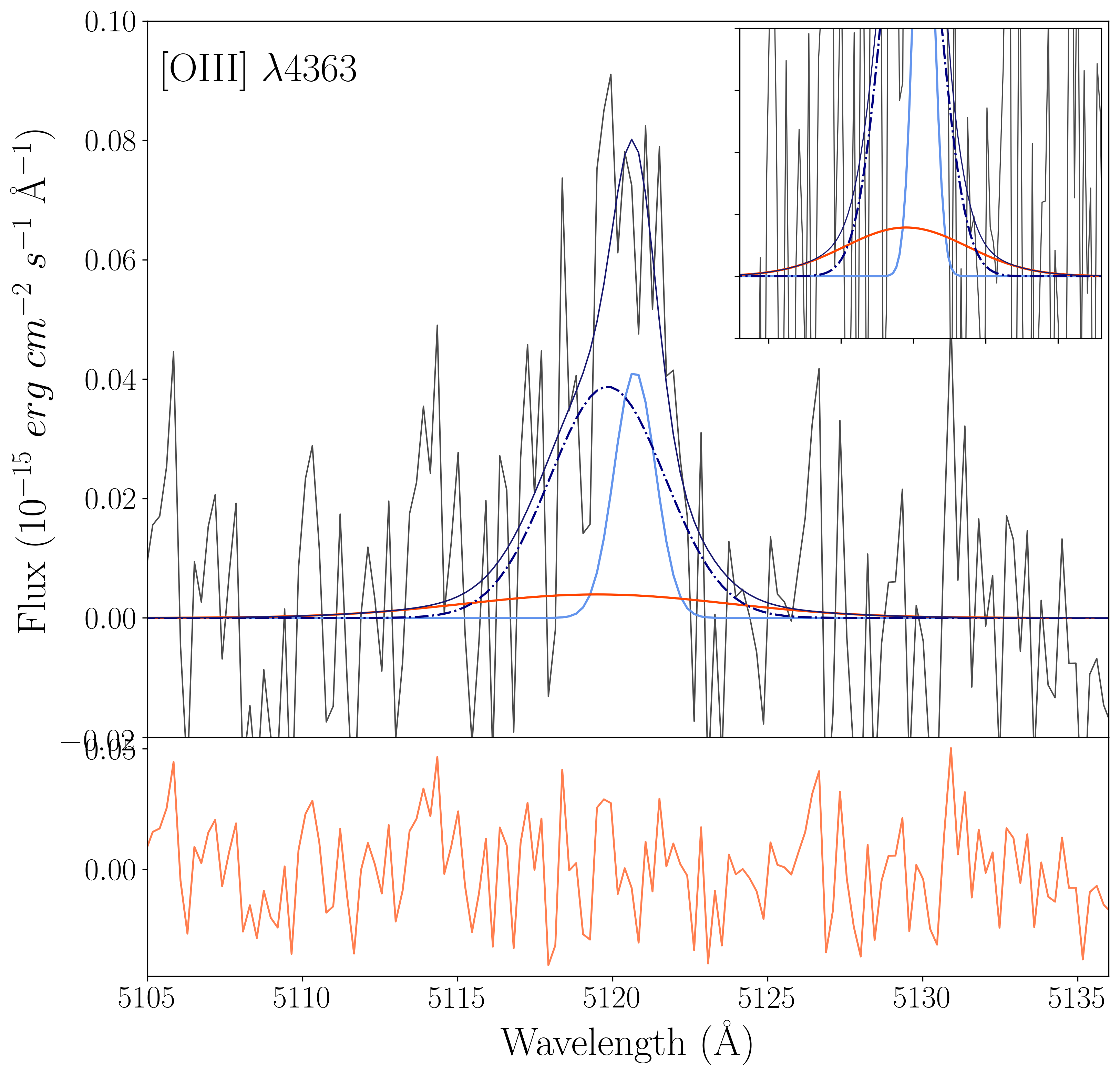}
\includegraphics[scale=.21, angle=0]{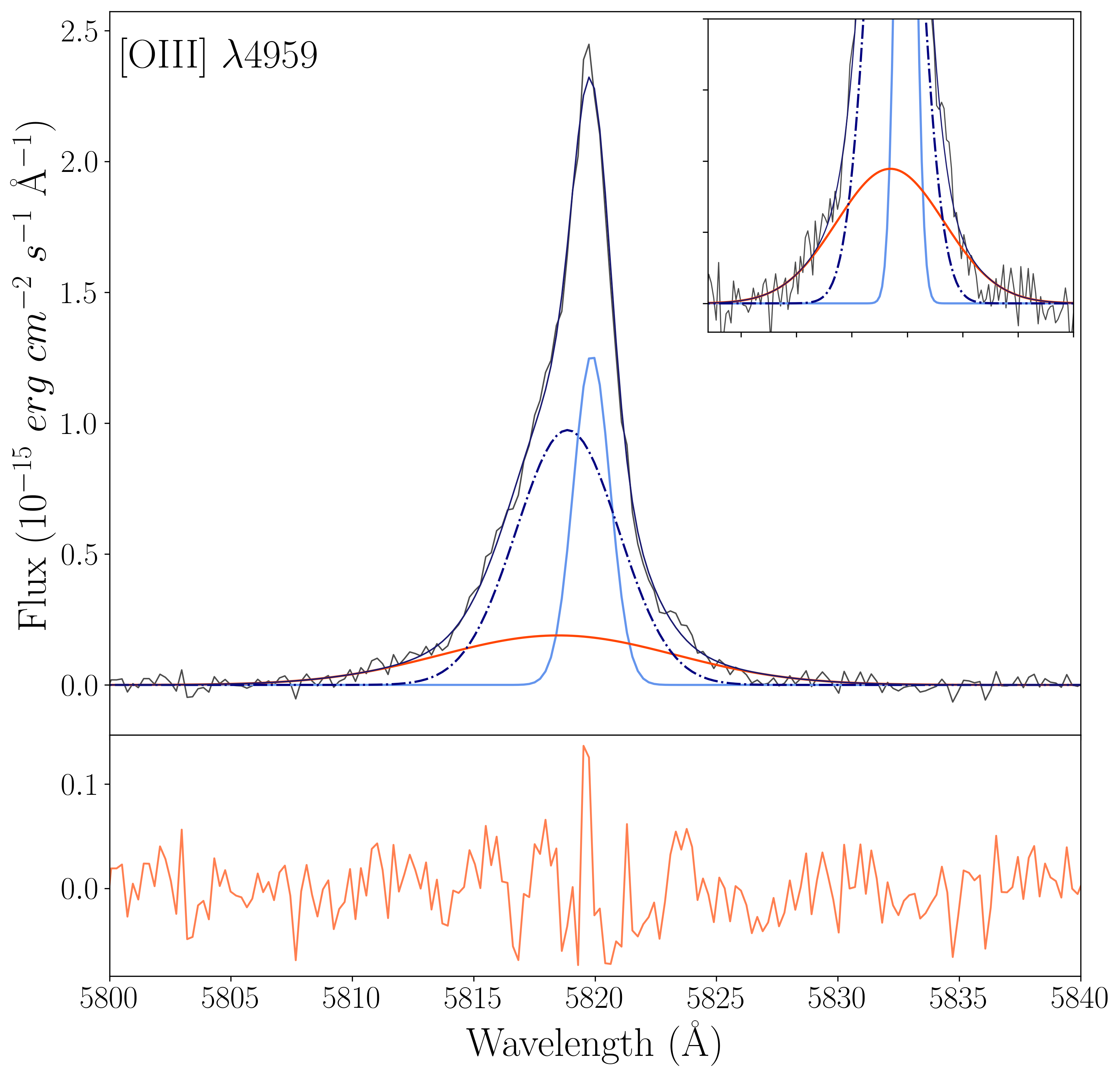}
\includegraphics[scale=.21, angle=0]{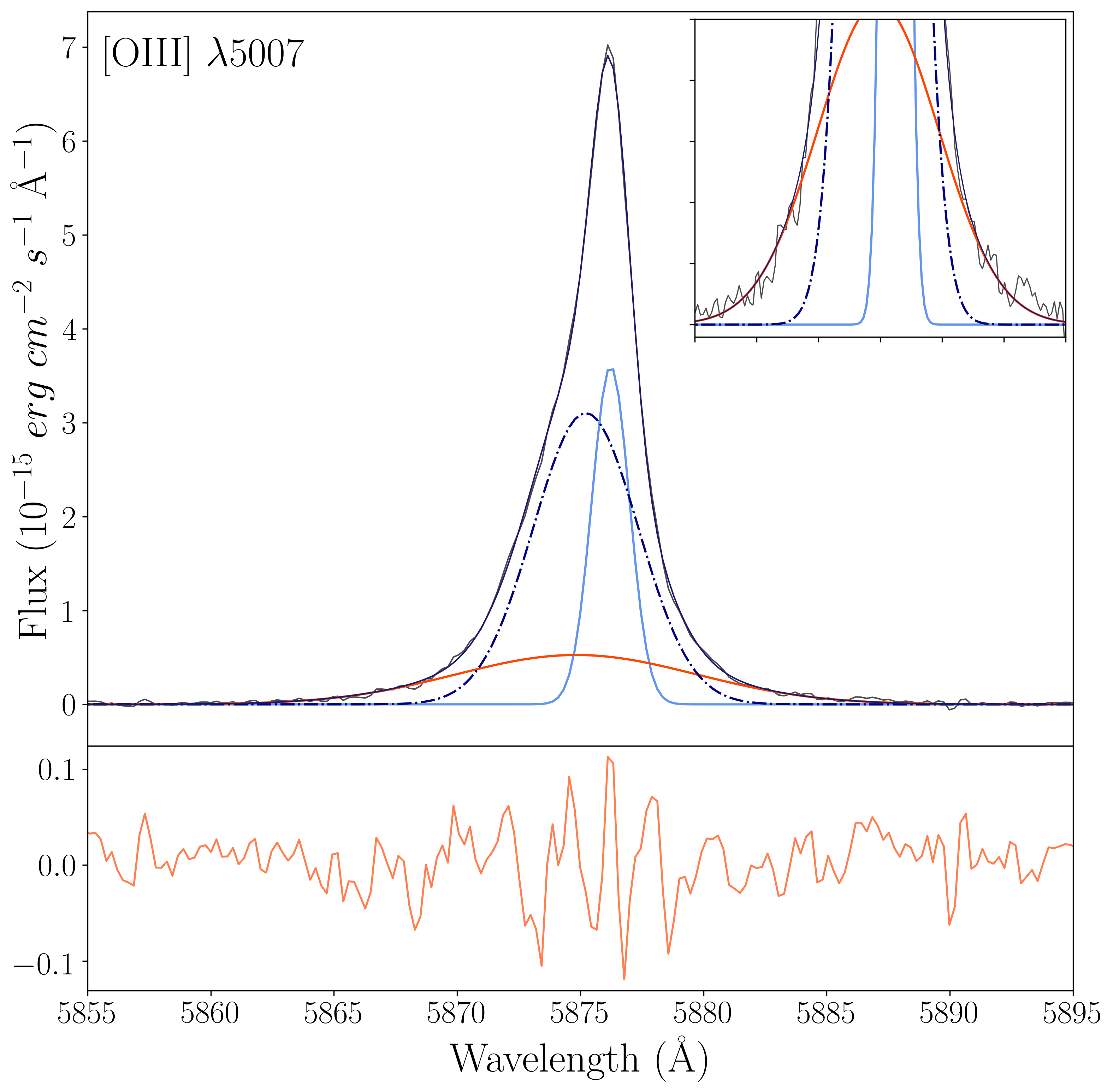}\\
\includegraphics[scale=.21, angle=0]{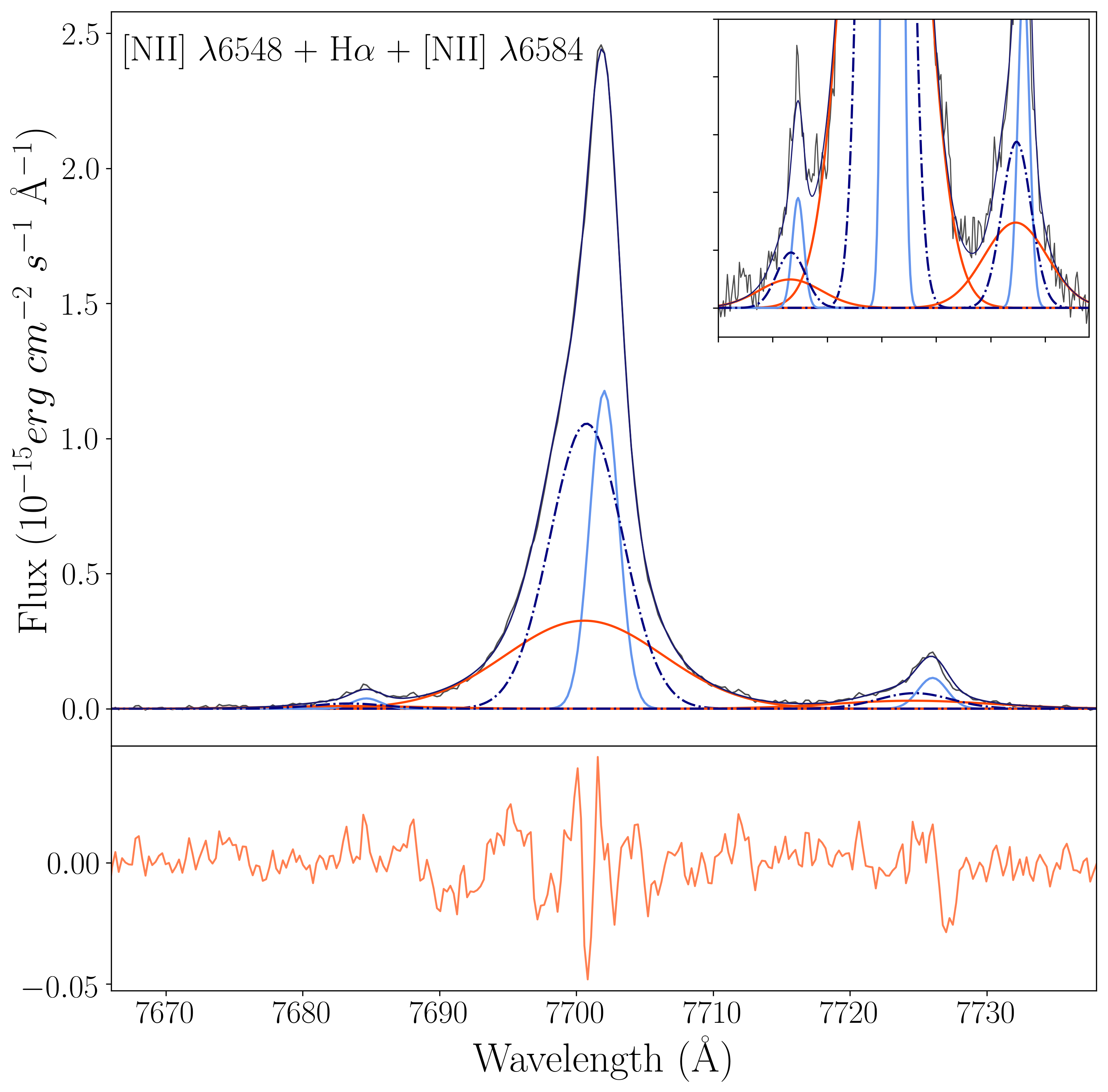}\hspace{1cm}
\includegraphics[scale=.21, angle=0]{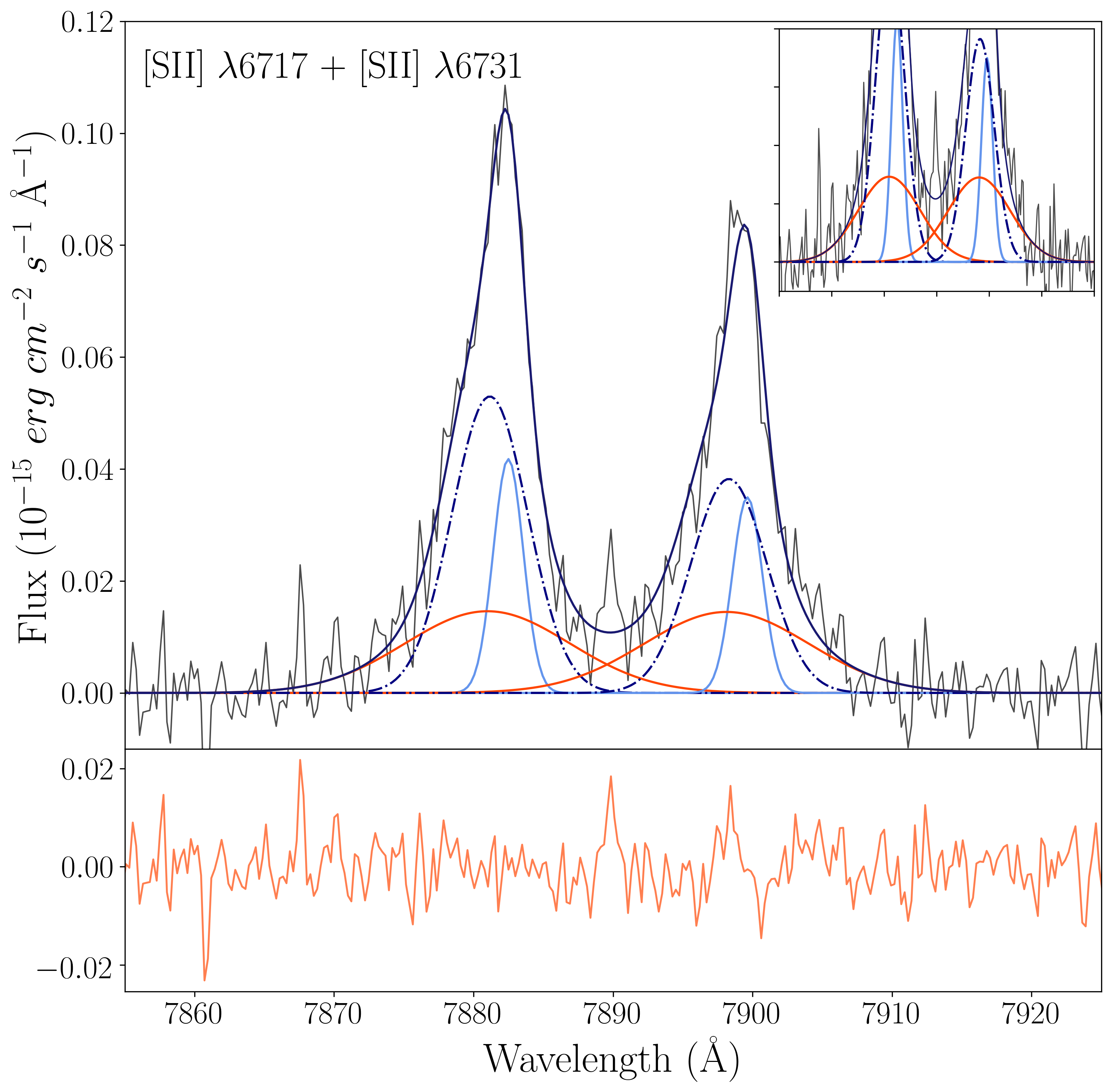}
\caption{Constrained three-component fits of the observed spectral lines across the spectral range derived from the free-parameter modelling of H$\alpha$ and [\oiii]\,$\lambda$\,5007. In these figures, the black solid line represents the original data, the light blue solid line the narrow component, the blue dashed line the mid-width component and the red solid line the broad component. The composite three-component profiles are depicted by the dark blue solid line.}
\label{fig4}
\end{figure*}

In Fig.~\ref{fig4}, we add a further Gaussian component to the emission spectra and present the resulting multi-component models across our spectral range. As in the instance of the two-component fitting procedure, the peak velocity and velocity dispersion of spectral lines are fixed to that of H$\alpha$ (which is again fitted simultaneously with [\nii]\,$\lambda \lambda$6548, 6584) or [\oiii]\,$\lambda$\,5007. For these subsequent lines, therefore, only the amplitude of the individual components is allowed to vary.

Once again, the kinematics and dispersion modelled in H$\alpha$ and [\oiii]\,$\lambda$\,5007 are broadly similar. The narrow component is largely unaffected by the addition of the third component in terms of both its width $\sigma \sim$\,40\,\kms\ and peak velocity, which is red-shifted by $\Delta$v$\sim$\,10 \kms$\ $with respect to the peak of the composite profile. However, the broader component is now split into two distinct blue-shifted components ($\Delta$ v\,$\sim$ 50 - 70 \kms).
Again this component structure is replicated in the unconstrained modelling of both H$\alpha$ and [\oiii]\,$\lambda$\,5007, the narrower of the components having a velocity dispersion of $\sigma \sim$\,100\,\kms\ and the broader having $\sigma \sim$\,220 - 250\,\kms\ in the rest frame. The separation of these two components improves the global fit by allowing both the wings and asymmetric bulk of the emission spectra to be more accurately modelled as distinct entities.

When applying this solution to the fainter emission lines, an additional constraint was imposed upon the simultaneous fitting of the [\oii]\,$\lambda \lambda$\,3726, 3729 doublet. Due to the significant blending between these lines, which imply a larger number of free parameters to be simultaneously fitted, a multi-component model with physical viability is difficult to achieve. The flux ratio between the [\oii] doublet components is, therefore, fixed to the inverted ratio between the more distinct [\sii]\,$\lambda \lambda$\,6717, 6731 doublet to facilitate deblending while assuring [\oii] electron densities consistently tied to the ones traced by [\sii], as we discuss in Section~4.2.
Within the [\oii]\,$\lambda \lambda$\,3726, 3729 fit, therefore, there are only three free parameters, which govern the absolute amplitude of the now coupled doublet components. The result of this fitting is shown in Figure~\ref{fig4}. Despite the heavy constraints applied during this fit, the model reproduces the data well, with no clear systematic signature visible in the residuals. Good results in terms of $\chi^2$ minimisation are also obtained when the flux ratio between the two [\oii] lines is left unconstrained. However, we note that the spectral resolution of our data is not sufficient to guarantee a physically meaningful unconstrained model for [\oii].

\begin{figure*}
\centering
\includegraphics[scale=1.45]{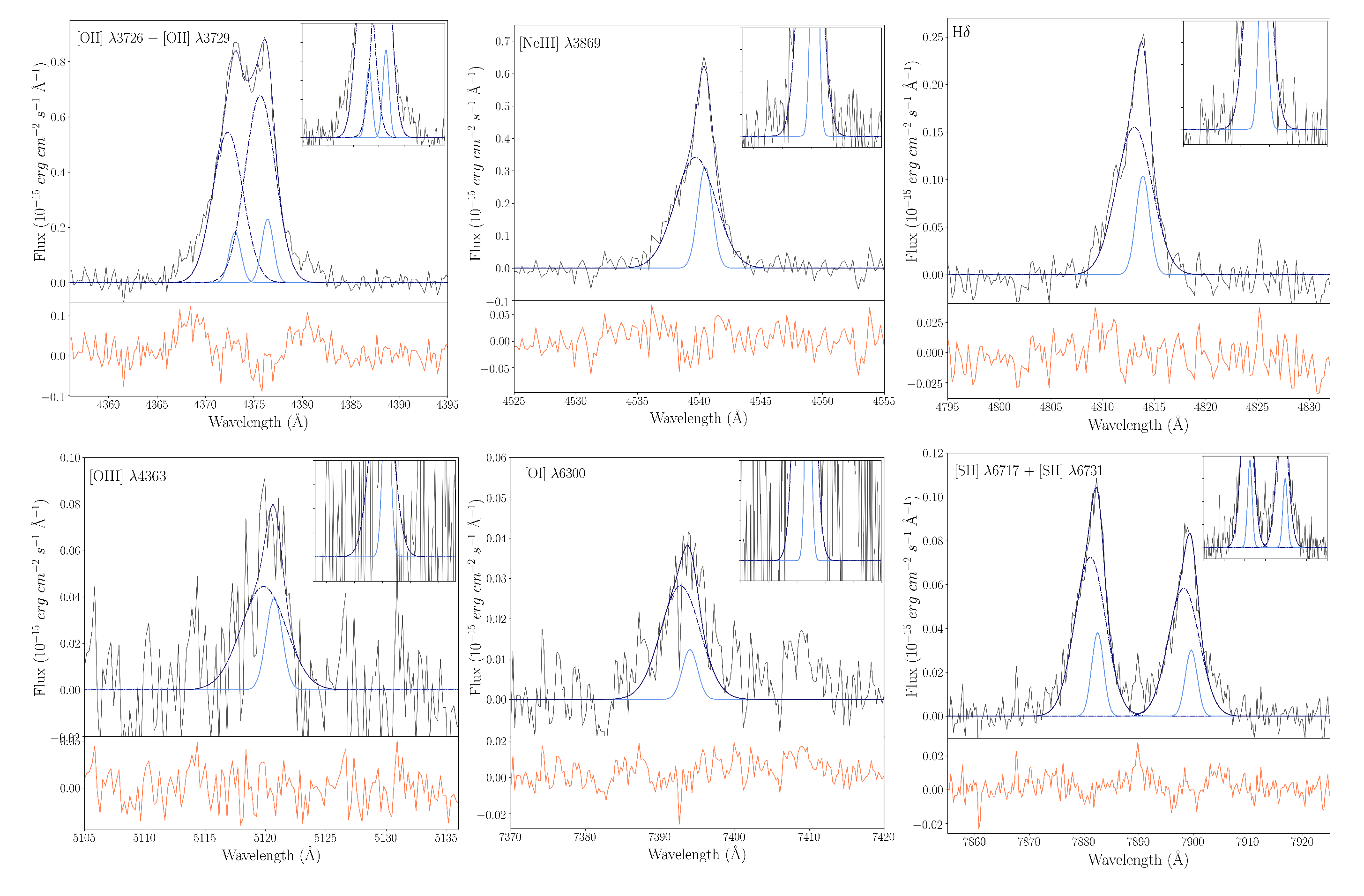}
\caption{Constrained two-component fit of the faint observed spectral lines. In these figures, the black solid lines represents that original data, the light blue solid line the narrow component and the blue dashed line the mid-width component. The composite three-component profiles are depicted by the dark blue solid line. These components are copied from the three-component solutions given in Figure 3, with only their respective amplitudes free to vary.}
\label{fig5}
\end{figure*}

To further validate the three-component model that is proposed to describe the emission-line kinematics, we investigate the necessity for three components in the fainter spectral lines.
Comparing the two- and three-component models, we find that the $\chi^2$ and residual structure of these fits are very similar in the faintest emission lines, which we ascribe to insufficient S/N ratio to detect the wings of the broad component. Consequently, in the two-component fit, most of the broad component detected in the brighter lines is assigned to the mid-width component observed in the three-component model.

To confirm this argument, the fainter emission lines shown in Fig.~\ref{fig5} were refit with only the narrow and mid-width components from the three-component fit of H$\alpha$ and [\oiii]$\lambda$\,5007. As shown in Fig.~\ref{fig5}, without the addition of a low amplitude broad component, a reasonable fit is produced in the majority of the faint observed spectral lines, with the flux mainly redistributed to the mid-width component. Possible exceptions to this are [\oii] and [\sii], which are relatively brighter than e.g. [\neiii] or [\oi], and where some systematic residual from the line wings is detected. We note that due to the difference in the total exposure time between the red and blue arm setups, the blue continuum has a S/N a factor of $\sim$\,1.5 lower than the red one (see Fig.~\ref{fig2}), strongly affecting the low-surface brightness wings of emission lines.
We, therefore, conclude that for faint emission lines, the data S/N cannot adequately distinguish between two and three kinematic components and physically motivated constraints must be taken into account.

\section{Results}
  \label{sec:results}

\begin{table*}
	\centering
	\caption{Multi-Component Fitting: Line intensities}
	\label{tab2}
	\begin{tabular}{l c c c c c c c} % four columns, alignment for each
		\hline
	Em. Line  &   $I_{\rm Tot}^a$ &  $I_{\rm N}^a$ &  $I_{\rm M}^a$  & $I_{\rm B}^a$ & EM$_{f,\rm N}^{b}$ & EM$_{f,\rm M}^{b}$ & EM$_{f,\rm B}^{b}$ \\[3pt]
 %   	     & (\AA) &  &  & & & \% &  \%   \\
 %    \multicolumn{3}{c}{10$^{-16}$ erg s$^{-1}$ cm$^{-2}$}\\
    		\hline
        %     \multicolumn{14}{|c|}{Two-Component fitting} \\
              \hline
[O\sc{ii}]\,$\lambda$3726  & 69.0$\pm$4.5 & 47.3$\pm$6.3 & 50.5$\pm$4.4 &112.1$\pm$9.9 &13.9&35.6&50.5  \\[1pt]
[O\sc{ii}]\,$\lambda$3729  & 80.7$\pm$5.1 & 56.6$\pm$6.5 & 70.1$\pm$5.7 &113.2$\pm$9.9 &14.2&42.2&43.6  \\[1pt]
[Ne\sc{iii}]\,$\lambda$3869& 50.8$\pm$3.3 & 60.0$\pm$6.8 & 58.6$\pm$4.7 & 32.5$\pm$3.6 &24.1&56.0&19.9  \\[1pt]
H$\delta$         		   & 21.5$\pm$1.6 & 20.6$\pm$3.2 & 28.3$\pm$2.6 & 11.6$\pm$1.8 &19.6&63.6&16.8  \\[1pt]
H$\gamma$         		   & 37.9$\pm$2.5 & 28.7$\pm$4.3 & 43.4$\pm$3.8 & 35.4$\pm$4.1 &15.5&55.3&29.2  \\[1pt]
[O\sc{iii}]\,$\lambda$4363 &  6.8$\pm$0.9 &  7.8$\pm$2.1 &  8.7$\pm$1.4 &  3.1$\pm$1.0 &23.6&62.0&14.4  \\[1pt]
H$\beta$        	       &100.0$\pm$4.6 &100.0$\pm$10.1&100.0$\pm$6.6 &100.0$\pm$8.2 &20.7&47.9&31.4  \\[1pt]
[O\sc{iii}]\,$\lambda$4959 &211.2$\pm$8.4 &248.0$\pm$21.2&231.3$\pm$13.1&156.6$\pm$11.8&24.3&52.4&23.3  \\[1pt]
[O\sc{iii}]\,$\lambda$5007 &644.1$\pm$23.0&716.3$\pm$55.6&746.5$\pm$37.8&441.2$\pm$29.0&22.9&55.5&21.6  \\[1pt]
[O\sc{i}]\,$\lambda$6300   &  5.8$\pm$0.5 &  3.2$\pm$0.6 &  2.7$\pm$0.4 & 11.9$\pm$1.2 &11.8&22.5&65.7  \\[1pt]
[N\sc{ii}]\,$\lambda$6548  &  7.0$\pm$0.5 &  9.1$\pm$1.2 &  5.1$\pm$0.6 &  8.5$\pm$1.0 &27.2&34.4&38.4  \\[1pt]
H$\alpha$                  &281.9$\pm$14.9&282.1$\pm$23.8&281.7$\pm$17.8&282.0$\pm$20.5&20.9&47.3&31.8  \\[1pt]
[N\sc{ii}]\,$\lambda$6584  & 21.2$\pm$1.3 & 27.4$\pm$2.8 & 15.4$\pm$1.3 & 25.6$\pm$2.2 &27.2&34.4&38.4  \\[1pt]
[S\sc{ii}]\,$\lambda$6716  & 13.0$\pm$0.9 & 10.2$\pm$1.3 & 14.3$\pm$1.2 & 12.8$\pm$1.3 &16.4&52.3&31.3  \\[1pt]
[S\sc{ii}]\,$\lambda$6731  & 10.7$\pm$0.8 &  8.5$\pm$1.1 & 10.4$\pm$1.0 & 12.7$\pm$1.3 &16.6&45.7&37.7  \\[1pt]
 			\hline
         %    \multicolumn{14}{|c|}{Three-Component fitting} \\
              \hline
		%\hline
	\end{tabular}
    	\begin{tablenotes}
\footnotesize
\item {\it Notes:}\,Subscripts B, N and M denote the {\it broad} and the two {\it narrow} components, respectively. %Lines where the broad component is only tentative show values within brackets.
\item $^a$emission-line intensities are reddening corrected using the extinction coefficients listed in Table~\ref{tab1} and normalised to 100$\times$\hb.
%\item $^a$Fluxes are indicated in units of 10$^{-16}$\,erg\,s$^{-1}$\,cm$^{-2}$.
\item $^b$Fractional emission measure ($F_{\rm i} / F_{\rm tot}$) of each component $i$, in percent (\%).
%\item $^c$Mean errors are $0.01$\AA\ (1.5 km/s).
		\end{tablenotes}
%\end{table*}
\end{table*}

The results from the multi-component line fitting and kinematic properties adopted for $J1429$ are presented in Table~\ref{tab2} and Table~\ref{tab3}.
Table~\ref{tab2} catalogues the %values for the peak position and
total intensity of each line and the intensity contained in each Gaussian component for the measured lines in Figure~\ref{fig4}.
We calculate the total flux ($F_{\rm tot}$) as the sum of the flux within each Gaussian component ($F_{\rm N}$, $F_{\rm M}$ and $F_{\rm B}$, for the narrow, mid-width and broad components respectively).
Uncertainties on the total fluxes are estimated following \citet{Gonzalez-Delgado1994} which includes the standard deviation of the residual of the multi-component Gaussian model with the emission line. Uncertainties on the flux in each component is assumed to contribute equally in quadrature to the standard deviation of the total composite model. The errors computed in this way are typically larger than the ones provided by the $\chi^{2}$ minimisation alone.

Emission-line fluxes were corrected into intrinsic intensities using the extinction coefficient, $c$(\hb), computed from the Balmer decrement \ha/\hb\ and assuming a \citep{Cardelli1989} extinction law. They are presented in  Table~\ref{tab4}. We find  $c$(\hb) to be similar for all three kinematic components and fully consistent within the uncertainties with the value obtained using the total integrated lines, which gives $c$(\hb)$=0.19\pm0.01$. This value suggests a modest dust extinction, which is average for GPs \citep{Amorin2010}.

In Table~\ref{tab2} we present line intensities normalised to 100 times the \hb\ intensity along with the fractional emission measure (EM), defined as the fractional flux contained within each component (EM$_{f,\rm N}$, EM$_{f,\rm M}$ and EM$_{f,\rm B}$ for the narrow, mid-width and broad component, respectively).
It is interesting to note that approximately 80\% of the total flux in each emission line is distributed between the broad and mid-width components, with the broad component contributing $\sim$\,30\% of the \ha\ line. The implications of this result will be discuss later in Section~\ref{sec:discussion}.

\subsection{Emission-line kinematics}

Table~\ref{tab3} details the peak velocity and velocity dispersion for each component in the adopted model for [\oiii] $\lambda$5007 and H$\alpha$, as well as the peak position and full width at zero intensity (FWZI) of the broad component model, defined as the width of the profile where the model falls within 1$\sigma$ of the continuum level \citep[A12b,][]{Bosch2019}.

\begin{table*}
	\centering
	\caption{Multi-Component Fitting: \ha\ and [\oiii] kinematics}
	\label{tab3}
	\begin{tabular}{l c c c c c c c c } % four columns, alignment for each
		\hline
	Em. Line  & $v_{obs}$ & $\Delta_{v, \rm N}^a$ & $\Delta_{v, \rm M}^a$ & $\Delta_{v, \rm B}^a$ & $\sigma_{\rm N}^b$ & $\sigma_{\rm M}^b$ & $\sigma_{\rm B}^b$  & FWZI$^c$  \\[2pt]
    	      &  \kms & \kms & \kms & \kms & \kms & \kms & \kms & \kms (\AA)   \\
    		\hline
        %     \multicolumn{14}{|c|}{Two-Component fitting} \\
              \hline
[O\sc{iii}]\,$\lambda$5007 &  52085.2$\pm$0.9 & 7.0$\pm$0.9 & -53.4$\pm$1.4 & -76.7$\pm$6.2 & 36.4$\pm$1.3 & 108.0$\pm$2.0 & 249.3$\pm$21.0 &  1818$\pm$56 (35.6$\pm$1.1)  \\[1pt]
H$\alpha$                  &  52066.8$\pm$1.0 & 10.4$\pm$1.1 & -48.8$\pm$1.5 & -56.4$\pm$3.1 & 38.3$\pm0.5$ & 102.2$\pm$2.0 & 224.2$\pm$11.8 &  1878$\pm$68 (41.1$\pm$1.5)  \\[1pt]
 			\hline
         %    \multicolumn{14}{|c|}{Three-Component fitting} \\
              \hline
		%\hline
	\end{tabular}
    	\begin{tablenotes}
\footnotesize
\item {\it Note:}\,Subscripts B, N and M denote the {\it broad} and the two {\it narrow} components, respectively.
\item $^a$Velocity shift, $\Delta_{v}=v_{\rm obs}-v_{\rm comp}$, where $v_{\rm obs}$ and $v_{\rm comp}$ are the velocity at the peak of the observed emission line and the corresponding kinematic component, respectively.
\item $^b$Intrinsic velocity dispersion (See text for details).
\item $^c$Full-width at zero intensity (see text for details).
\end{tablenotes}
\end{table*}

The intrinsic velocity dispersion of \ha\ and [\oiii] for the different kinematic components ($\sigma_{\rm N}$, $\sigma_{\rm M}$ and $\sigma_{\rm B}$ respectively) are obtained from the observed velocity dispersion after correcting in quadrature for instrumental ($\sigma _{\rm ins}$) and thermal broadening ($\sigma _{\rm ther}$),

 \begin{equation}
 \sigma = \sqrt{{\sigma_{\rm obs}}^2 - {\sigma _{\rm ins}}^2 - {\sigma _{\rm ther}}^2} \ ,
 \label{eq:1}
 \end{equation}
where $\sigma _{\rm ins}\sim$\,10\,\kms\ is measured from bright sky lines and arc lines, and $\sigma _{\rm ther}$ is derived from

\begin{equation}
\sigma _{\rm ther} = \frac{c \cdot \lambda _{\rm em}}{\lambda _{\rm obs}} \cdot \sqrt{\frac{k_B \cdot T_e}{m_{ion} \cdot c^2}} \ ,
\label{eq:2}
\end{equation}
assuming an electron temperature $T_e = 1.2 \times 10^{4}$\,K typical for GPs \citep{Amorin2010,Amorin2012a} and this galaxy in particular (see Section~4.2).
Again, in both [\oiii] and H$\alpha$, a similar kinematic structure in terms of velocity dispersion can be inferred. This structure in the bright emission lines is comprised of one low dispersion (narrow) component and two high dispersion (mid and broad components), a scheme which also reproduces the fainter emission lines.

Overall, \ourobject\ shows emission-line kinematics comparable to other GPs (A12b).
The three emission-line components found in \ourobject\ show velocity dispersions from $\sim$\,35\,\kms\ for the narrow component to $\sim$250\,\kms\ for the broader component, strongly suggesting the presence of highly turbulent gas. The luminosities found for these components in the brighter emission lines are in the range of $L_{\rm H\alpha}\sim$\,0.5-1.2$\times$10$^{42}$\,\cgs\ and $L_{\rm [OIII]}\sim$\,0.8-2.0$\times$10$^{42}$\,\cgs.
These values are analogous to those found in highly dispersion-dominated low-mass (M$_{*}<$\,10$^{10}$M$_{\odot}$) galaxies at similar redshift in spatially resolved kinematic studies \citep[e.g.][]{Glazebrook2013}. The implications of these results will be discussed further in  Section~\ref{sec:discussion}.

\subsection{Emission-line diagnostics}
\label{sec:diagnostics}

\begin{figure*}
\centering
\includegraphics[scale=0.4, angle=0]{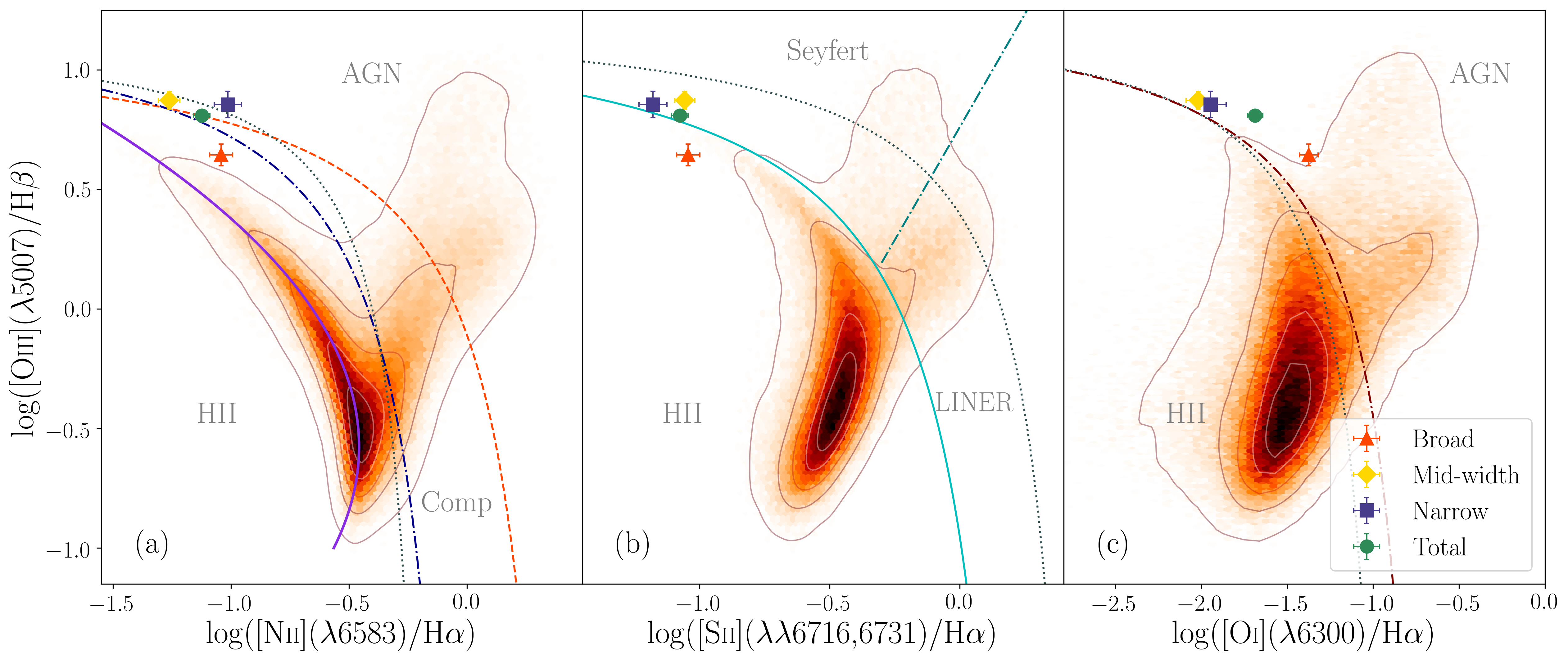}
\caption{Multi-component BPT diagnostics for \ourobject. In each panel, the green marker describes the position of \ourobject\ with all components summed; the narrow, mid-width and broad components are represented by the blue, yellow and red markers respectively. The density plots are obtained from the SDSS-DR7 MPA-JHU catalog, where the contours describe the 25\%, 50\%, 75\%, 90\% and 99\% percentiles.
In panel (a), the ionisation models produced by \citet{Kewley2001}, \citet{Kauffmann2003} and \citet{Brinchmann2004} are plotted in red (dashed), blue (dashed-dotted) and purple (solid) respectively. Panel (b) includes models from \citet[][solid]{Kewley2001} and \citet[][dashed-dotted]{Kewley2006}.
In panel (c) the model from \citet{Kewley2001} is plotted in dark red (dashed-dotted).
Each diagnostic diagram is also plotted with the maximal starburst prediction for an object with a global metallicity of $0.3 \times Z_\odot$ (30\% solar), as determined by \citet{Xiao2018}.
These models are described by the black dotted line in each plot. }
\label{fig6}
\end{figure*}

In Figure~\ref{fig6} we investigate the excitation properties of the three kinematic components using classic BPT diagnostic diagrams, [\oiii]$\lambda$5007/\hb\ vs. [\nii]$\lambda$6583/\ha, [\sii]$\lambda\lambda$6717,6731/\ha\ \citep{Baldwin1981} and [\oi]$\lambda$6300/\ha\ \citep{Veilleux1987}. The emission-line ratios of the three components, together with the total integrated ratio, are compared to those of SDSS star-forming galaxies and different empirical and model-based demarcation lines for stellar and AGN-like photoionization.
We note that caution must be taken in this analysis since the BPT diagrams were originally designed as excitation diagnostics to separate star-forming and AGN dominated galaxies in integrated long-slit spectra.

In the three BPT diagnostic diagrams, we find the narrow and mid-width components (blue and yellow) lying in similar positions, showing very high excitation conditions and tracing the region limited by the classical demarcation lines from \citet{Kewley2001} and \citet{Kauffmann2003}.
These two components also appear representative of the total integrated [\oiii]/\hb\ ratio (green).
Compared to the locus followed by most star-forming galaxies in these diagrams \citep[][purple line]{Brinchmann2004}, the [\oiii]/\hb\ ratios of \ourobject\ are offset upwards by $\gtrsim$\,0.3 dex. This strongly suggests a harder radiation field powering these emission-line components, which dominate the integrated nebular emission of the galaxy.
The excitation conditions shown by the broader component, however, reveal a different behaviour.
Its [\oiii]/\hb\ ratio is lower than the narrower components but still shows high excitation conditions consistent with pure stellar photoionisation.
While the [\nii]/\ha\ and [\sii]/\ha\ of the three components are low and quite typical of low metallicity HII regions, the broader component shows a significantly higher [\oi]/\ha\ ratio compared to the other two components and clearly exceeds the demarcation line by \citet{Kauffmann2003}.
Since [\oi]/\ha\ is highly sensitive to hard radiation fields, particularly shock emission in the neutral ISM \citep{Allen2008}, this may suggest that an additional excitation mechanism other than stellar photoionisation (e.g. shocks) would be required to explain the nature of the broad emission.

\begin{figure}
\centering
\includegraphics[width=0.95\columnwidth, angle=0]{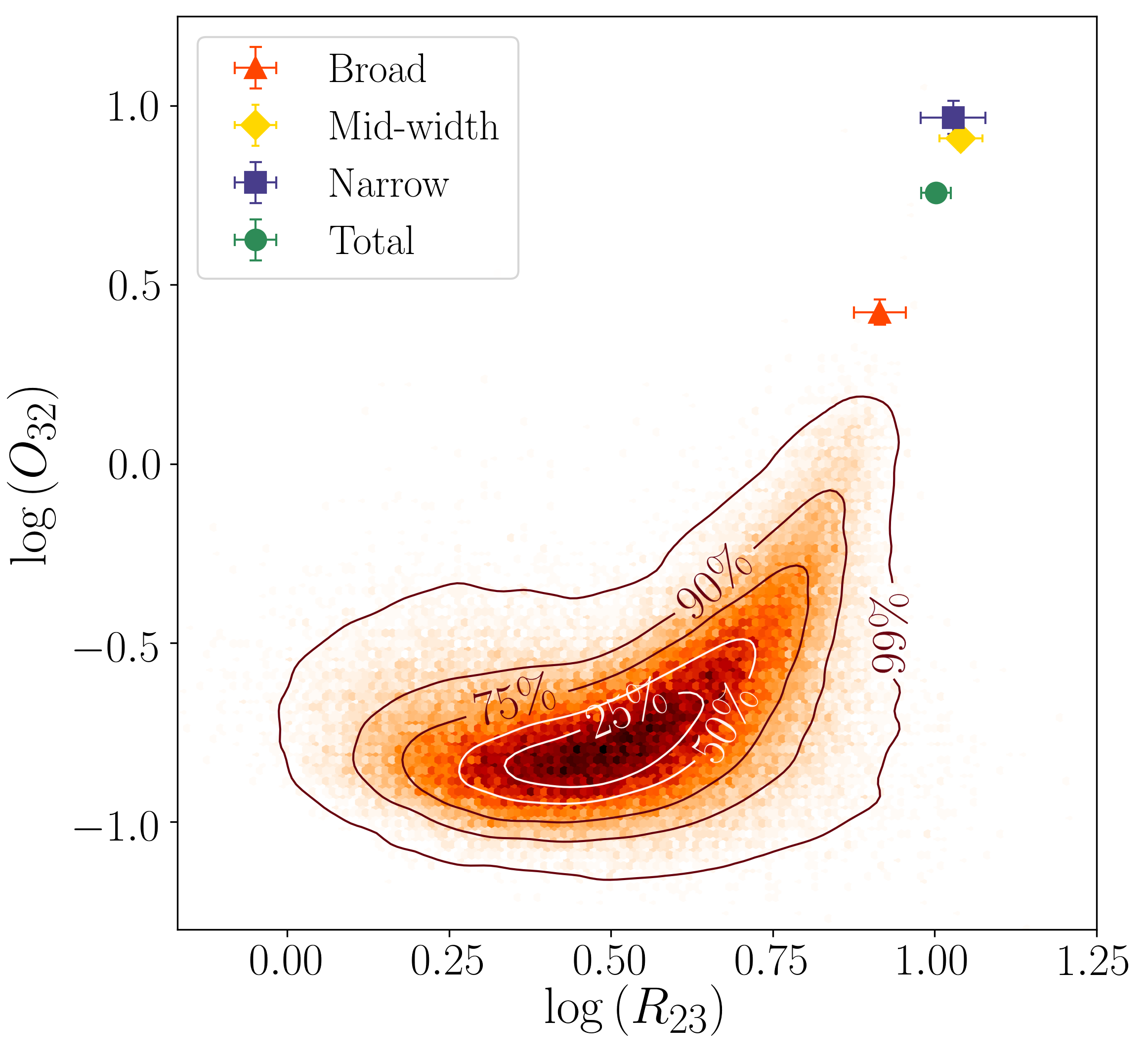}
\caption{Multi-component $O_{32}$ - $R_{23}$ diagnostic diagram for \ourobject. The density plot describes SDSS-DR7 star-forming galaxies with contours at the 25\%, 50\%, 75\%, 90\% and 99\% percentiles. The global profile position for \ourobject\ (i.e. with all components summed) is represented by the green marker. The positions of the narrow, mid-width and broad components are indicated by the blue, yellow and red markers respectively.}
\label{fig7}
\end{figure}

In Figure~\ref{fig7}, we show the three kinematic components compared to SDSS-DR7 star-forming galaxies in the $O_{32}-R_{23}$ plane\footnote{Here, \mbox{$O_{32}=$[\oiii]\,$\lambda\lambda$,4959, 5007\,/\,[\oii]\,$\lambda\lambda$\,3726, 3729}, and
\mbox{$R_{23}=($[\oiii]\,$\lambda\lambda$\,4959, 5007\,$+$\,[\oii]\,$\lambda\lambda$\,3726, 3729$)$/\hb}}.
As in Fig.~\ref{fig6}, the position of \ourobject\ is far shifted towards very high ionisation at low metallicity, as characterised by unusually high $O_{32}$ and $R_{23}$.
Again, the contribution of the narrower components dominates the integrated line ratio and the broad shows $O_{32}$ values a factor of $\sim$\,2-3 lower, indicating less extreme ionisation conditions.
It is noteworthy that extreme ionisation traced by $O_{32}$, like those observed for \ourobject, are exceedingly rare in local universe galaxies \citep[e.g.][]{Jaskot2013} and are exclusively found in extreme emission-line galaxies such as GPs and other strong \lya\ emitters at higher redshifts \citep[e.g.][]{Nakajima2013,Amorin2014,Amorin2015,Erb2016,Calabro2017}.
These extreme conditions can be associated with density-bounding, as we will consider in more depth in Section~\ref{sec:discussion}.

In relation to Figures~\ref{fig6} and \ref{fig7}, we can draw several interesting inferences.
Firstly, we note that the integrated line ratios (green points) are typically dominated by the narrower components, but its value can be shifted towards higher [\oi]/\ha\ values (and lower $O_{32}$, although in much lesser degree) due to the contribution from the broad component.
However, in the case of \ourobject, we do not find evidence of unusually large [\nii]/\ha\ or [\sii]/\ha\ ratios for the broad component. This suggests that the gas responsible for the broad emission, even if it is shocked, is probably of similar ionisation and metallicity conditions to the one producing the narrower components.
The latter components, on the contrary, appear to be nebulae photoionised by harder ionising star clusters than typical HII regions. This is further illustrated in Fig.~\ref{fig6}a and Fig.~\ref{fig6}b by the inclusion of the maximal starburst demarcation presented by \citet{Xiao2018} for pure stellar photoionisation when they include massive star binaries through the BPASS stellar evolution models \citep{Eldridge2017}.
This model provides significantly harder radiation field without the need of invoking non-thermal processes, such as AGNs. In this case, the demarcation line corresponds to a model with a global metallicity of approximately 30\% solar.

Unfortunately, we do not have any independent indication to support or rule out some AGN contribution in \ourobject, but models including binarity in massive star evolution suggest AGN emission is not required to explain the line ratios of \ourobject. The broad component in \ourobject\ comes from a relatively high density gas with line ratios consistent with stellar photoionisation and contribution from shocks as evidenced by its [\oi] emission. This is consistent with the interpretation of the broad component as due to an outflow produced by the collecting effect of SNe remnants and stellar winds of massive stars (A12b). We will elaborate upon this further in Section~\ref{sec:discussion}.

\subsection{Electron densities and temperatures}

Using a multicomponent analysis of all the emission lines of \ourobject, we have studied the main physical properties of \ourobject, namely electron density (from [\oii] and [\sii]), [\oiii] electron temperature ($T_e$), and ionisation parameter ($\log(U)$), which are shown in Table~\ref{tab4}.

Using reddening corrected intensities, $T_e$[\oiii] is calculated for each component from the ratio $R_{O3}=\frac{I(4959) + I(5007)}{I(4363)}$ \citep{OsterbrockFerland2006} and using the equations presented in \citet{PM2017}, which depends on electron density.
We find that the highest $T_e$[\oiii] is in the mid-width component and the lowest in the broad component, with the narrower component having a temperature consistent to the integrated value of 1.15$\times$10$^4$K.

The electron density $n_{\rm e}$[\sii] for each component in the emission spectra and the total integrated value are presented in Table~\ref{tab4}.
They have been derived using the flux ratio $I(6717)/I(6731)$ and the equations provided by \citep{PM2017}.
Since we need both $T_e$[\oii] and $T_e$[\sii] and we do not have access to [\oii] and [\sii] auroral lines, we adopt here the formulation originally proposed by \citet{Hagele2006}, which includes the dependency with $n_{\rm e}$,
and we assume that $T_e$[\oii]$=T_e$[\sii].
We find that $n_{\rm e}$[\sii] is highest in the broad component ($\sim$490 cm$^{-3}$), dramatically lowest in the mid-width component ($\lesssim 155$ cm$^{-3}$) and intermediate in the narrow component ($\sim230$ cm$^{-3}$).
For $n_{\rm e}$([\oii]) we also find very consistent values for each component, but this result is largely driven by the fact that the [\oii] line profile has been performed using as a prior the [\sii] 6717/6731 ratio.

Finally, the ionisation parameter, $\log(U)$ (also presented in Table~\ref{tab4})  is derived from emission-line ratios and photoionisation models following the approach implemented in the code HII-Chi-mistry \citep[HCm,][]{PM2014}.
We find the broader component to show lower ionisation compared to the mid and narrower components, which show values more representative of the total integrated emission.
This naturally agrees with the results for the $O_{32}$ ratio presented in Figure~\ref{fig6} given that $O_{32}$ is a good tracer of $\log(U)$ at low metallicity \citep[e.g.][]{Diaz2000}.

\begin{table}
  \centering
	\caption{Physical properties}
	\label{tab4}
	\begin{tabular}{l c c c c}
		\hline
	Component  &   $n_{\rm e}$[\sii] & $T_{\rm e}$[\oiii] & $T_{\rm e}$[\oii]$^{a}$ & $c(\hb)^{b}$ \\[2pt]
    	  & cm$^{-3}$ & 10$^4$K &  10$^4$K &     \\
    		\hline
              \hline
%	Total (SDSS)& & 219.5 & 1.29$\pm$0.01 & 8.05$\pm$0.02 & $\pm$ & 8.27$\pm$0.02 & -1.04$\pm$0.03 & $\pm$ & -0.86$\pm$0.03 & $\pm$  \\[1pt]
	Total       & 204$\pm$98 & 1.15$\pm$0.05 & 1.03$\pm$0.03 & 0.19$\pm$0.01  \\[1pt]
	Narrow 		& 228$\pm$180& 1.20$\pm$0.12 & 1.04$\pm$0.05 & 0.22$\pm$0.03  \\[1pt]
	Mid-width   &$<$155 & 1.20$\pm$0.07 & 1.16$\pm$0.06 & 0.18$\pm$0.02  \\[1pt]
	Broad	    & 490$\pm$224 & 1.01$\pm$0.09 & 0.86$\pm$0.03 & 0.21$\pm$0.04  \\[1pt]
							\hline
				              \hline
	\end{tabular}
		\begin{tablenotes}
			\footnotesize
			\item $^{a}$ The electron temperature of [OII] is obtained from $T_{\rm e}$[\oiii] and $n_{\rm e}$[\oii]$= n_{\rm e}$[\sii] using the empirical relations in \citet{PM2017}.
      \item $^{b}$ Extinction coefficient derived from the \ha/\hb\ ratio and the \citet{Cardelli1989} extinction curve.
		\end{tablenotes}
\end{table}

\begin{table*}
  \centering
	\caption{Oxygen and nitrogen abundances}
	\label{tab5}
	\begin{tabular}{l c c c c c}
		\hline
	Component  & $12+\log($O/H$)_{\rm Te}$ & $12+\log($O/H$)_{\rm HCm}$  & $\log($N/O$)_{\rm Te}$ & $\log($N/O$)_{\rm HCm}$ & $\log(U)_{\rm HCm}$  \\[2pt]
    	(1)  & (2) & (3) & (4) & (5) & (6)  \\
    		\hline
              \hline
	Total       &  8.29$\pm$0.09 & 8.19$\pm$0.13 & -1.06$\pm$0.08 & -1.07$\pm$0.17 & -2.39$\pm$0.03   \\[1pt]
	Narrow 	    &  8.21$\pm$0.17 & 8.20$\pm$0.17 & -0.80$\pm$0.16 & -0.87$\pm$0.21 & -2.27$\pm$0.08   \\[1pt]
	Mid-width   &  8.22$\pm$0.14 & 8.18$\pm$0.16 & -1.06$\pm$0.11 & -1.18$\pm$0.15 & -2.24$\pm$0.05   \\[1pt]
	Broad		&  8.52$\pm$0.13 & 8.32$\pm$0.19 & -1.33$\pm$0.12 & -1.15$\pm$0.21 & -2.65$\pm$0.07   \\[1pt]
							\hline
				              \hline
	\end{tabular}
		\begin{tablenotes}
			\footnotesize
            \item {\it Colums:} (1) emission-line component; (2) to (5) Metallicities and N/O obtained using the direct $T_{\rm e}$-method \citep{PM2017} and HCm \citep{PM2014}; (6) Ionisation parameter computed from emission-line ratios and photoionisation models using HCm (See text for details).
		\end{tablenotes}
\end{table*}

\subsection{Metallicity and N/O abundance}
\label{sec:metallicity}

\begin{figure}
\centering
\includegraphics[width=\columnwidth, angle=0]{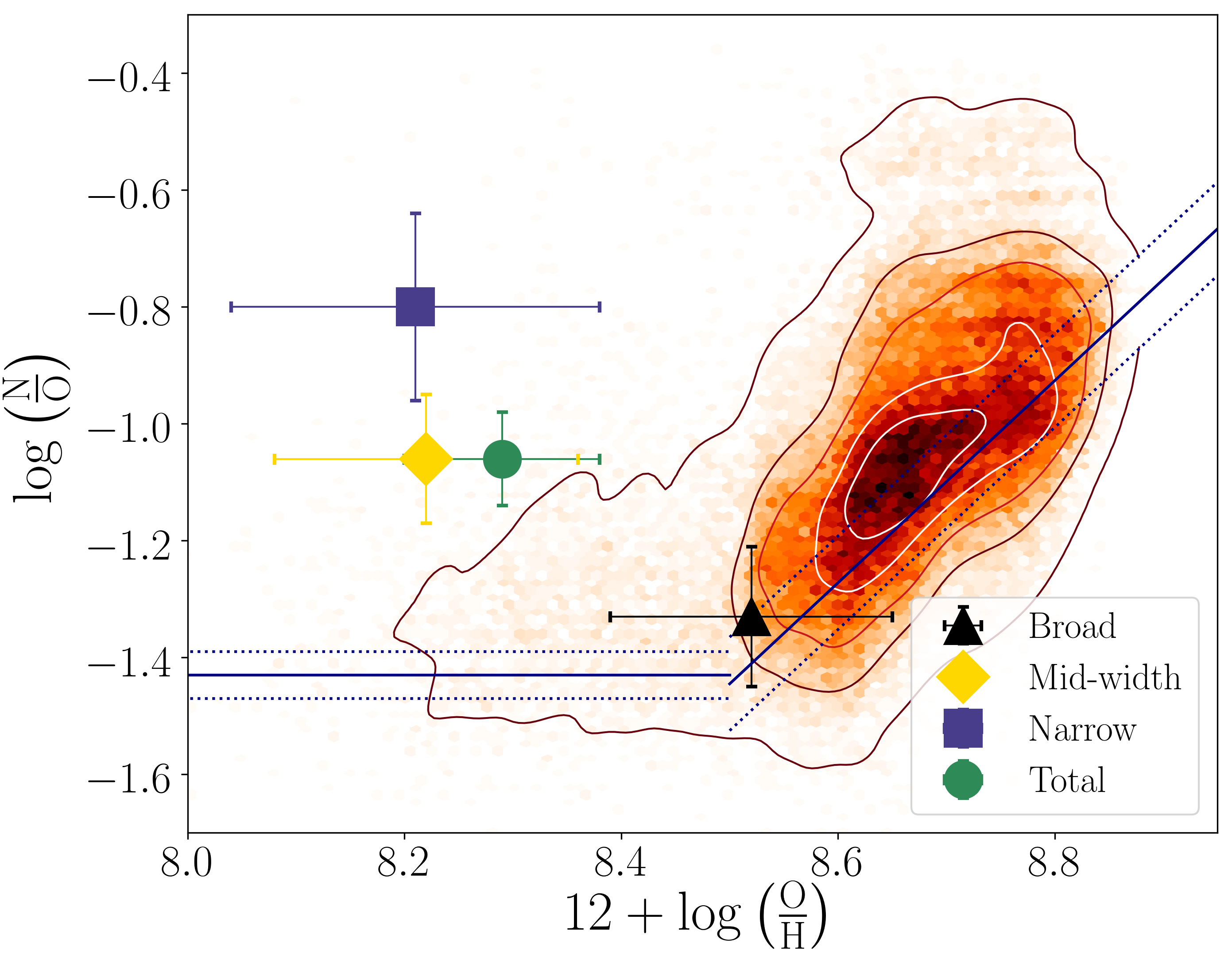}
\caption{$\log$(N/O) versus $12 + \log $(O/H) diagram for multiple spectral components of \ourobject\ as calculated from the direct $T_{\rm e}$-method. The density plot describes SDSS star-forming galaxies with contours at the 25\%, 50\%, 75\%, 90\% and 99\% percentiles.
For them we derived both O/H and N/O simultaneously (also with $\log(U)$) using HCm (see text).
The integrated values for \ourobject\ (i.e. accounting for the flux within the integrated profile) are represented by the green marker.
The positions of the narrow, mid-width and broad components are indicated by the blue, yellow and black markers respectively.
The primary regime, as reported in \citet{Andrews2013}, for N/O is illustrated by a constant $\log $(N/O)$ \ = -1.43 \ \left(\pm 0.04\right)$ for $12 + \log $(O/H) $< 8.5$.
The secondary regime, also reported in \citet{Andrews2013}, is plotted as a linear relation with $\log $(N/O)$ = 1.73 \times (12+\log $(O/H)$) - 16.15 \ \left(\pm 0.08\right)$ for $12 + \log $(O/H) $ > 8.5$.}
\label{fig8}
\end{figure}

In Table~\ref{tab5}, we present the results of metallicity and N/O abundance analyses conducted on \ourobject.
Two different methods are used, the direct $T_{\rm e}$ method and the semi-empirical approach of the code HII-Chi-mistry \citep[HCm, ][]{PM2014}, which uses observed emission lines with predictions of photoionisation models.

For the direct method, we use the intrinsically faint $T_e$-sensitive auroral line [\oiii] $\lambda$4363 to derive O/H and N/O for each emission-line component, following the prescriptions in \citet{PM2017}. The $O^+$, $O^{++}$ and $N^{+}$ ionic abundances are derived based on [\oiii] electron temperature and relevant bright line ratios, assuming empirical relations for $T_{e}$[\oii] and  $T_{e}$[\nii] from $T_{e}$[\oiii], as explained in Sect.~\ref{sec:diagnostics}.
This is due to the absence of [\oii] and [\nii] auroral lines in our observed spectral range.

In order to test possible uncertainties due to the relatively poor S/N ratio of the [\oiii] auroral line, we use the code HCm \citep{PM2014}, version  3.1\footnote{\url{https://www.iaa.csic.es/~epm/HII-CHI-mistry-opt.html}}, to simultaneously derive metallicity, $\log$(N/O) and $\log(U)$.
The code calculates line ratios using the observed lines [\oii]\,3727,29, [\oiii]\,4363, [\oiii]\,5007, [\nii]\,6584 and [\sii]\,6717,31, and internally compares them to a grid of photoionisation models obtained with  Cloudy~v.17\footnote{\url{https://www.nublado.org/}} \citep{Ferland2017}. Errors are computed following a Montecarlo scheme. As shown in \citet{PM2014}, results using all the input lines should give consistent results, within uncertainties, with those from the direct method.

Overall, we find robust results with the above two methods for both metallicity and N/O. Small differences between the two methods for the broad component are attributed to the assumption of $n_e=100$\,cm$^{-3}$ in the HCm calculations, which is more than 4 times lower than its measured value.
When computed for the total integrated profile, the abundances appear largely driven by the narrower components and also suggest that the broad emission corresponds to ionised gas with a slightly lower N/O and higher O/H than the gas responsible for the narrower components (Table~\ref{tab4}).

The results in Table~\ref{tab5} are illustrated in the O/H-N/O plane in Fig.~\ref{fig8} alongside SDSS-DR7 star-forming galaxies. The SDSS sample use O/H and N/O values computed with HCm.
The bimodal fit performed by \citet{Andrews2013} from stacked SDSS spectra using the direct method, is also included. These lines reflects the different regimes where N/O appears dominated by primary and secondary nitrogen production.
From Fig.~\ref{fig8}, \ourobject\ shows N/O abundances typically larger than those of SDSS galaxies with similar metallicity.
This is  consistent with previous results on GPs \citep[cf.][]{Amorin2010,Amorin2012a} and strong [\oiii] {emitting galaxies at $z\sim$\,2-3  \citep{Masters2014,Kojima2017}}.
However, the broad component of \ourobject\ shows a relatively lower N/O value, well within the relation of SDSS galaxies. Thus, we find the largest difference in the abundances is found between the narrower and broader components, which are located well above and within the N/O plateau, respectively. A metallicity difference of about 0.2-0.3 dex is found between these points. This may suggest that the gas residing in the ongoing starburst sites is somehow slightly more metal-poor with a higher N/O than the gas responsible for the broader emission, which is more consistent with the position of normal star-forming galaxies in SDSS. We will discuss this further in Section~\ref{sec:discussion}.

\section{Discussion}
\label{sec:discussion}

The multi-component analysis of the emission-line spectrum of \ourobject\ into three distinctive kinematic components has revealed a nebular structure with a range of velocity dispersion and ISM physical properties. In the following sub-sections, we will discuss the interpretation of these remarkable features and the implications they have for high redshift galaxies in the context of reionisation studies.

\subsection{Narrow and mid-width components: violent starbursts and turbulent ISM}

Within the kinematic components of \ourobject, we find a narrow and a mid-width Gaussian components with diverse kinematics and ISM properties.
The presence of multiple Gaussian components in all of the measured emission lines of \ourobject\ is in line with the multiple narrow kinematic components found by A12b in GP spectra, which can be associated to spatially unresolved star-forming clumps shown in HST UV imaging.
As in A12b, the presence of narrow and mid-width components in \ourobject\ may also be interpreted as the superposition of various spatially unresolved HII regions at different velocities.
This has also been proposed to explain relatively broad and asymmetric bright emission lines in other compact starburst galaxies, such as Haro~11 \citep[$\sigma \sim$\,80-100\,\kms, ][]{Ostlin2015}.
We are unable to identify these spectral components as multiple star-forming regions in the optical SDSS images of \ourobject\ due to insufficient spatial resolution. The narrow and the blue-shifted mid-width components of \ourobject\ are spatially unresolved in our ISIS 2D spectrum, in contrast to the starburst $J1615$ studied by A12b where secondary narrow line components are also spatially detached at kpc scales in the 2D spectra, strongly suggesting the merging of individual clumps.
However, we can test this hypothesis using the HST COS data available for \ourobject\ in the rest-UV, which traces the young stellar population on spatial scales of the order of giant HII complexes, such as 30 Doradus or NGC 2366 \citep[e.g.][]{Micheva2017}.

In contrast to Haro~11 and other GPs in \citet{Cardamone2009} sample, the UV image of \ourobject\ presented in Fig.~\ref{fig1} does not reveal obvious merger features nor any indication of additional star-forming clumps other than the single central region in the HST UV image.
Consequently, \citet{Alexandroff2015} classified \ourobject\ as an extremely compact galaxy with a ``dominant central object'' \citep[DCO,][]{Heckman2011,Overzier2009}
of M$_{*}\sim$\,10$^{9}$M$_{\odot}$ within an aperture of $\sim$\,290\,pc in radius containing 50\% of the UV light and with no other significant substructure in the HST-COS image (Fig.~\ref{fig1}).
DCOs have been associated with highly dense young massive star cluster complexes experiencing violent starburst activity. Moreover, they have also been discussed as possible seeds for central black holes \citep{Overzier2009} and as candidate progenitors of galactic bulges, which might be associated with the coalescence of formerly star-forming knots \citep[e.g.][]{Elmegreen2012}.
If multiple spatially unresolved star-forming clumps are contributing to the mid-width component, they must be located within the central $\sim$\,100-200 pc, while still coexisting as independent giant HII regions moving away from the main narrower component with a relative velocity of  $\sim$\,60\,\kms.
Using the calibration given by \citet{Kennicutt2012} with the IMF of \citet{Kroupa2003}, the extinction-corrected \ha\ SFRs that correspond to the narrow and mid-width components are very high; $1.9 \pm 0.2$\,M$_{\odot}$\,yr$^{-1} $ and $4.0 \pm 0.4$\,M$_{\odot}$\,yr$^{-1}$, respectively.
Their combined contribution after aperture correction\footnote{Note that for this calculation we have corrected upwards the \ha\ luminosities by a factor of $\sim$\,2.5 to account for the different apertures of the SDSS fiber plus an additional factor of 1.7 derived by \citet{Overzier2009} to account for extended flux outside the SDSS fiber and used by \citet{Alexandroff2015} to make their UV and \ha\ SFRs comparable.
We have not applied any additional corrections for possible differences in flux calibration between SDSS and our ISIS spectra.} is about 24\,M$_{\odot}$\,yr$^{-1}$, close to the dust-corrected UV SFR of about 27\,M$_{\odot}$\,yr$^{-1}$ reported by  \citet{Alexandroff2015}.
This implies that most if not all the star formation in \ourobject\ comes from the ongoing starburst, i.e. it has a timescale of $<$\,20 Myr.

The velocity dispersion ($\sim 40$\,\kms) and  \hb \ and [\oiii] luminosities of the narrow component are consistent with the expected position of a compact HII galaxy in the $L-\sigma$ relation \citep{Terlevich2014,Melnick2017}, which traces virial motions through the gravitational potential of a galaxy-wide starburst. The velocity dispersion and SFR of this narrow component are also compatible with observed relations found for \textit{entire galaxies} at similar redshifts  \citep{Green2014, Herenz2016}.

A different situation is noticed for the mid-width component, which is blueshifted by $\sim$60 \kms\ with respect to the global \ha\ profile centroid and the peak of the narrow component. Its larger velocity dispersion ($\sim$\,100 \kms), while broadly consistent with its larger SFR, is higher than that expected from the $L-\sigma$ relation for its \hb\ luminosity even considering possible aperture corrections, therefore suggesting that gravitational virial motions are not sufficient and an additional source of broadening is needed to explain the large velocity dispersion of this component.

Additional sources of turbulence produced by an unusually high star formation activity and feedback, as proposed by some authors \citep[e.g.][]{Green2010,Moiseev2012} or merely the overlap of extremely compact, unresolved gas clumps in the same line of sight interacting with a strong outflow could enhance turbulent motions, as we discuss in Section~5.3.2.
Spatially resolved studies of nearby BCDs have shown that the velocity dispersion of the \ha\ and [\oiii] lines increase by a factor of 2-3 from the SF knots to the intra-knot regions \citep[e.g.][]{Cairos2017, Moiseev2012, Lagos2016,Kumari2017}.
This increase in the line broadening might be related to gas turbulent motions which do not reflect virial motions, but are instead connected with the radiative and mechanical energy injected by stellar feedback into a gaseous disc \citep[e.g.][but see \cite{Krumholz2016}]{Moiseev2015,Dib2006}.
In this context, the broadening of the mid-width component might be related with enhanced turbulence in the lower surface brightness gas, induced by SNe feedback from the DCO \citep[e.g.][]{Avillez2007}. As this ionised gas is not moving at the systemic velocity, the large velocity dispersion of the mid-width component could be representative of turbulent gas from the base of the outflow which is moving away from the DCO and interacts with a clumpy nebular ISM. We will come back to this interpretation in Section~5.3.2.

\subsubsection{Are GPs dispersion dominated?}

Whether  \ourobject\ is dispersion-dominated or a rotating disc goes beyond the possible analysis with our current data. However, we can argue that it is likely that turbulent motions  dominate the global kinematics of \ourobject. Comparison with deep GEMINI-GMOS IFU observations of another GP with similar integrated properties from our WHT-ISIS dataset presented by \citet{Bosch2019} shows a quantitatively similar kinematic structure, i.e. narrow, mid-width, and broad components with similar $\sigma$ values.
While the global profile is certainly dispersion dominated, the narrow component in \citet{Bosch2019} shows a uniform velocity dispersion of $\sigma\sim$\,40\kms\ and line-of-sight velocities consistent with a perturbed disk, showing evidence of rotation. In previous IFU studies \citet{Lofthouse2017} also found evidence of dispersion-dominated disks in a sample of four GPs, two of them showing no evidence of merging and similarly high \textit{integrated} velocity dispersion.
Also \citet{Goncalves2010} and \citet{Herenz2016} presented dispersion-dominated  galaxies in their samples of local Lyman-break analogs (the LBA sample) and  \lya\ emitters (the LARS sample), respectively.
Both samples contain a small number of GPs, such as SDSS\,$J$\,113303 and SDSS\,$J$\,092600 (a.k.a. LARS14), which also show high velocity dispersions and show no clear indication of rotation. We note however that these IFU studies do not include multicomponent analysis of the emission lines, in some cases precluded by a limited velocity resolution, limited S/N, or both.
As shown in \citet{Bosch2019}, the presence of multiple unresolved components may dilute or mask-out a rotation pattern in the disk components.
Therefore, we conclude that systematic high S/N spatially resolved, high-dispersion spectroscopy is essential to elucidate the true dynamical state of GPs. For \ourobject, this will be subject of follow-up observations.

\subsection{Broad emission in GPs as a signature of starburst-driven outflows}

The broad emission component in the wings of the global emission profiles for \ourobject\ is indicative of very high velocity gas.
Previous work has shown that these features in giant extra-galactic HII Regions \citep[e.g.][]{Castaneda1990,Roy1992,Hagele2007,Hagele2009,Hagele2010,Hagele2013, Firpo2010,Firpo2011}
and nearby low-mass compact starburst galaxies  \citep[e.g.][]{Martin1998,Izotov2007b,James2009,James2013a,James2013b,Lagos2014,Ostlin2015,Thone2019}
are typically associated with stellar feedback.
More generally, broad \ha\ emission with FWHM$\sim$\,350\kms\ in star-forming galaxies is also interpreted as signature of outflowing gas  \citep[e.g.][and references therein]{Rodriguez2019}.

Green Pea galaxies are compact, low-metallicity starbursts for which broad emission appears ubiquitous and likely associated with the violent star formation episode and its associated feedback, rather than AGN activity \citep[A12b,][Amor\'in et al. in prep.]{Bosch2019}.
However, as discussed in A12b, a number of non-mutually exclusive mechanisms including strong stellar winds of massive stars, expansion of young SN remnants, SN-driven shells and superbubble blow-ups, as well as turbulent mixing layers, may produce similar broad features.
This, alongside the lack of spatially resolved high-dispersion data for representative samples, makes the analysis of the origin and physical properties of the broad emission in GP galaxies somewhat incomplete.

Even without spatially resolved spectra, our analysis of \ourobject\ can provide new insight on the nature of broad emission in GPs, taking advantage of the derived physical properties from deep high dispersion long-slit spectra.

\subsubsection{Stellar winds and SNe feedback}

The kinematic analysis of \ourobject\ strongly suggest stellar feedback as responsible for the broad components observed in all emission lines, including both permitted and forbidden lines. Furthermore, the broad emission is consistently heavily blueshifted compared to the centroid of the lines, contributing significantly to the asymmetry of the global profiles. Our interpretation is that there is a powerful outflow moving out along the line-of-sight.
This feature has been shown as a signature of strong winds in violent starbursts at low and high redshift  \citep[e.g.][]{LehnertHeckman1996,Martin1998,Genzel2011,Newman2012}
and is expected in starbursts with SFR surface density SFR/Area$>$\,1\,M$_{\odot}$\,yr$^{-1}$\,kpc$^{-2}$ \citep[e.g.][]{Heckman2011,Newman2012}.
The latter condition is largely met by \ourobject, which has an projected SFR surface density\footnote{The projected area has been computed as $A=$2$\pi$\,$r_{\rm 50,UV}^{2}$. We note that $r_{\rm 50,UV}$ is a factor of $\sim$5 smaller than the slit FWHM of our \ha\ observation. } SFR$_{\ha}$/Area$=$\,67.6\,M$_{\odot}$\,yr$^{-1}$\,kpc$^{-2}$.

The broad component in \ourobject\ has a relatively high velocity dispersion of $\sim$\,240\kms\ (FWHM$\sim$560\,\kms) and a FWZI of $\sim$\,1800 \kms\ ($\sim$\,35-40\AA), suggesting very large expansion velocities.
These velocities are close to the terminal velocities of stellar winds from WR, LBV and Of stars \citep{Izotov2007b}, but are smaller than the velocities required for large kpc-scale shells and SNe-driven superbubbles seen in blow-up phase  \citep[i.e. many thousands of \kms, e.g.][]{Castaneda1990,Roy1992}.
Instead, it is likely that we are witnessing a relatively young outflow driven by very recent star formation.

Remarkably, the energetics and kinematics of the broad emission in  \ourobject\ is found to fit well in all bright and faint emission lines probed in this work, in contrast to other other low-z BCDs where the broad component is only seen in  bright lines such as \hb, \ha, and [\oiii] \citep[e.g.][]{Izotov2007b}.
The large contribution of the broad emission to the total flux, ranging from $\sim$\,15\% for [\oiii]\,4363 to $\sim$65\% for [\oi]6300 (Table~\ref{tab2}), is generally slightly lower in higher ionisation lines.
This is in qualitative agreement with the lower ionisation parameter and slightly lower [\oiii] electron temperature found for the broad component.

Conclusions from the three BPT excitation diagnostics explored in this work, particularly from the strong [\oi] $\lambda 6300$ broad emission detected in \ourobject, suggest that fast radiative shocks associated with the broad emission in \ourobject\ may contribute to its photoionisation.
However, these shocks are probably not a dominant source. Fast shocks in a warm ionised ISM can be associated to stellar feedback, such as young SNe remnants. The presence of shocks of a few hundreds \kms\ can be expected from the presence of strong SNe-driven outflows but they typically produce higher [\nii], [\sii] and [\oi] ratios and lower excitation \citep[e.g.][]{Ho2014}.

The broad emission in \ourobject\ has nearly the same \ha/\hb\ ratio as the narrower components, which is close to the recombination value, suggesting relatively low dust extinction.
The electron density computed from the [\sii] lines is $\sim$\,490\,cm$^{-3}$, twice the density of the narrower component, implying a denser ionised medium.
This, as well as its moderate electron temperature ($T_{\rm e}\sim$\,10$^{4}$\,K), indicates that the broad emission is mostly associated with the interstellar medium, as opposed to the dense circumstellar medium of young massive stars, such as Of and LBV, which typically show larger ($n_{\rm e}\gtrsim$\,10$^{3}$\,cm$^{-3}$) densities and lower broad line luminosities \citep[e.g.][]{Izotov2007b}.

The high emission-line EWs of \ourobject\ and the existence of strong nebular HeII emission in the SDSS spectrum with no sign of stellar WR bumps \citep{Shirazi2012} strongly suggests a very young starburst age and the presence of very massive low metallicity stars \citep[e.g.][]{Kehrig2015, Kehrig2018} and/or fast radiative shocks \citep{Izotov2007b}. However, strong star formation over longer timescales is also evident from the high UV luminosity.
The luminosity of the broad emission in \ourobject\ is $L_{\ha}\sim$\,7.6$\times$10$^{41}$ \cgs,
much larger than what would be expected if only stellar winds were producing the broad emission
\citep[10$^{37}$-10$^{39}$ \cgs, e.g.][]{Izotov2007b}.
Non-thermal (synchrotron) radio continuum detection in GPs \citep{Chakraborti2012} with strong magnetic fields ($>$30$\mu G$) provides an additional evidence of SNe feedback and increased turbulence.

We, therefore, conclude that the broad component seen in \ourobject\ is likely dominated by SNe-driven outflows. Its relative high metallicity and low N/O compared to the narrower components can be considered indicative of oxygen enrichment of the outflow material, as we will discuss below.

\subsubsection{Outflow averaged properties}

In this section, we derive the main outflow properties from our \ha\ measurements, namely i) outflow velocity, ii) mass-loading fraction and iii) mass outflow rate, and we also derive the SFR/Area after subtraction of the broad emission.
These quantities are important constraints for models and can be compared with feedback prescriptions used by numerical simulations.

In order to further compare our results with those of star-forming galaxies at $z \sim$2 from long-slit and IFU spectroscopic surveys  \citep[e.g.][]{Freeman2019,Newman2012,Davies2019}, we adopt an outflow model from \citet{Genzel2011}. This model assumes an spherically symmetric outflow with a constant velocity and an outflow rate defined as,

\begin{equation}
    \dot{M}_{\rm out} = \frac{1.36 \ m_H}{\gamma_{H\alpha} \ n_e} \left(\frac{v_{\rm out}}{r_{\rm out}}\right) L_{\rm H\alpha, broad}
    \label{eq:3}
\end{equation}
where, 1.36\,$m_H$ is the atomic mass of hydrogen and 10 per cent helium,
$\gamma_{H\alpha}=$ 3.56$\times$10$^{-25}$\,erg\ s$^{-1}$ cm$^{3}$ is the \ha\ emissivity at
$T_e =$10$^4$K, $n_e$ is the local electron density of the outflowing gas, $v_{\rm out}$ and $r_{\rm out}$ are the maximum velocity and (deprojected) radial extent of the outflow (assumed constant), and $L_{\rm H\alpha, broad}$ is the extinction-corrected \ha\ luminosity of the broad component.

For the electron density we assumed $n_e = n_{\rm e, broad}=$\,490 cm$^{-3}$ (Table~\ref{tab4}).
For the outflow velocity, the \citet{Genzel2011} model assumes $v_{\rm out}=|\Delta v - 2\sigma_{\rm broad}|$, where $\Delta v$ is the velocity offset in the broad component centroid. This yields a $v_{\rm out}=$\,508$\pm$23 km s$^{-1}$ for \ha.
For the radial extent $r_{\rm out}$ we can only constrain its value, because spatially-resolved IFU spectra for \ourobject\ is needed to quantify the true extension of the broad component.
Here we adopt $r_{\rm out}=$\,1.4 kpc, which corresponds to half the slit width and the average seeing FWHM during our observation, and nearly corresponds to the radius of the 2$\sigma_{rms}$ isophote in the near UV image of \ourobject\ ($\sim$4.5 $r_{\rm 50, UV}$) shown in Fig.~\ref{fig1}.
Both, $n_{e}$ and $r_{\rm out}$ are assumed to be independent of the SFR.

Assuming the SFR of the HII regions is the one obtained from $L_{H\alpha}$ of the two combined narrower components, we can obtain the mass-loading factor as,

\begin{equation}
    \eta = \frac{\dot{M}_{\rm out}}{\rm SFR_{\rm narrow}}
    \label{eq:4}
\end{equation}
where SFR$_{\rm narrow}$ is computed from the total $L_{H\alpha}$ after subtraction of the broad emission component (i.e. using the quantity EM$_{\rm broad}$ in Table~\ref{tab2}).

We obtain $\eta =$\,0.24$\pm$0.01. The largest uncertainties, which are not accounted in our result, are the true extent of the broad emission and the uncertain contribution to the SFR of the mid-width component in Eq.~\ref{eq:4}, which scales linearly with $\eta$. The above  $\eta$ value can be therefore considered as a  lower limit: if we consider only the SFR accounted by the \textit{narrower} component we increase $\eta$ by a factor of 2.5 and an additional factor of 4.5
will be added in case we consider the outflow extending out to the \textit{resolved} effective radius of the galaxy in the near UV. This would yield a sort of upper limit for $\eta \sim$\,2.7.

It is interesting to note that the physical properties derived for the outflow in \ourobject\ are quantitatively similar to that obtained by \citet{Davies2019} \citep[see also][]{Newman2012} following the same model prescriptions for star-forming galaxies at $z\sim$\,2 after stacking AO-assisted IFU spectra (their Table~2).
In \citet{Davies2019}, the outflows are resolved at kpc-scales in 11 out of 14 galaxies at $z=2.2$ with stellar masses M$_{*}\lesssim$\,10$^{10}$M$_{\odot}$, which are located above the star formation main sequence and contribute to stacks with higher SFR/Area.
These stacks  are those showing higher broad-to-narrow emission ratio and outflow parameters closer to the values obtained for \ourobject. Therefore, our results may suggest that local analogs such as \ourobject\ show similar outflow properties and follow similar scalings relating galaxy properties such as SFR/Area and outflow properties such as velocities and mass loading factors.

We note, however, that $\eta$ is strongly dependent on the adopted electron density and total flux of the broad component tracing the outflow (therefore on the parametrisation chose to modelling of the emission-line wings).  In the case of \ourobject, the low $\eta$ value is driven by a high $n_{\rm e, broad}$ (Eq.~\ref{eq:3}), at least a factor of $\sim$\,3 higher than that of typical HII regions in normal galaxies.

Assuming the galaxy is face-on (i.e. inclination $i=0^{\circ}$) its circular velocity is $v_{\rm circ}=$\,88\,\kms\ \citep{Heckman2015}\footnote{This value has been derived from the Tully-Fisher relation obtained by \citet{Simons2015} and has an uncertainty ($<$10\%).
A similar value of $\sim$100 \kms\ is obtained from a similar relation found for local BCDs
by \citet{Amorin2009}.} and the ratio $v_{\rm out}/v_{\rm circ}\sim$\,5.5, thus implying an outflow velocity larger than the escape velocity of the galaxy, $v_{\rm esc}\sim$\,3$v_{\rm circ}$ \citep{Chisholm2015}.
The $v_{\rm out}/v_{\rm circ}$ ratio is strongly correlated with the SFR/Area, as shown by \citet{Heckman2015} based on an analytical model were gravity and momentum-driven outflow accelerate gas clouds. A large number of galaxies display fast outflows but only a few of them develop galactic outflows with velocities superseeding their escape velocities \citep{Chisholm2015, Heckman2015} and this regime appears restricted to low-mass galaxies \citep{Rodriguez2019}.

Our results are therefore consistent with the possibility of part of the outflowing material in \ourobject\ is able to escape from the galaxy potential thus contributing to enrich its halo and the IGM with metals.
Caution should be taken when interpreting the above results since they rely on quite simplistic model assumptions \citep{Newman2012} and depend on parameters which cannot be  accurately constrained with our current observations.

\subsubsection{{The implication of the metal-enriched outflow in the mass-metallicity relation of GPs}}

\citet{Amorin2010} studied the position of GPs and LBAs in the mass-metallicity relation (MZR) and found that GPs are typically offset to lower metallicities by $\sim$\,0.3-0.6 dex within their stellar mass range (M$_{*}\sim$\,10$^{8}$-10$^{10}$\,M$_{\odot}$), and showing a shallower MZR slope compared to normal galaxies. Similar properties are
typically found for extreme emission-line galaxies at low and intermediate redshifts suggesting a different evolutionary stage for these galaxies \citep[e.g.][]{Amorin2014,Amorin2017,Ly2016,Calabro2017}
Massive inflows of metal-poor gas and {metal-}enriched outflows were invoked by \cite{Amorin2010} to explain the metallicity offsets but no quantitative analysis was performed.
While the former tend to dilute metallicity with fresh metal-poor gas, an increased metallicity in the outflow may indicate that part of the newly formed metals are being removed from the star-forming regions by the mechanical energy of the outflow.

Here we can partially test this hypothesis using our results for \ourobject\ which have a direct metallicity measured for both the narrow and broad components (Table~\ref{tab4}), i.e. assuming they trace \hii\ regions and outflow, respectively. Compared to the MZR based on the direct $T_{\rm e}$-method by \citet{Andrews2013}, both the integrated and narrow component (\hii\ regions) of \ourobject\ is $\sim$\,0.6 dex offset to lower metallicity.
Considering the metallicity ratio $Z_{\rm broad}$/$Z_{\rm narrow}\equiv Z_{\rm outflow}$/$Z_{\rm ISM}\sim$\,1.4 (Table~\ref{tab4}), we find a metallicity-weighted mass loading factor of
$\eta_{\rm Z}=$\,$\eta$ \ $\frac{Z_{\rm outflow}}{Z_{\rm ISM}}$\,$\sim$\,0.35, smaller than model predictions for galaxies of similar stellar mass \citep{Peeples2011,Chisholm2018}, which require $\eta_{\rm Z}>1$ to match the shape of the MZR.
Only adopting our upper limit in $\eta$ would increase the metal loading factor comparable to values required to match the MZR of normal galaxies \citep{Chisholm2018}. We tentative conclude that the enriched outflowing gas in \ourobject\ is not enough to explain  its low-metallicity offset with respect to the average MZR \citep{Andrews2013}.
Additional dilution, most likely by recent inflows of metal-poor gas fueling the current starburst, appear then as a significant contributor to reduce the global metallicity of our GP.

The above result is also supported considering the  dependence of the MZR scatter on the SFR, known to as the fundamental metallicity relation  \citep[FMR,][]{Mannucci2010} which is typically interpreted as a result of the balance between inflows and outflows in galaxy growth, as predicted by theoretical models and simulations \citep[e.g.][]{Dave2011,Lilly2013}.
However, the existence of the FMR and its interpretation has been debated  in literature and recent work suggests it might be driven by extreme star-forming systems \citep[see e.g.,][and references therein]{Sanchez2019}.
Considering the stellar mass and SFR for \ourobject\ (Table~\ref{tab1}), we find the position of \ourobject\ in {good} agreement within the uncertainties with the FMR of SDSS star-forming galaxies obtained using the direct method by \citet{Andrews2013}.

\subsubsection{Comparison of outflow detection in optical and UV spectra}

Star-formation driven outflows, seen in emission as low surface brightness broad components, are typically associated with a younger and denser phase of the superwind, which can extent out to kpc distances from the launching point \citep[e.g.][]{Genzel2011,Newman2012,Davies2019}.
In contrast, more diffuse outflowing material expelled at usually larger spatial scales and over longer timescales can be detected as blueshifted absorptions in rest-frame UV spectra of star-forming galaxies at low and intermediate redshifts \citep[e.g.][]{Weiner2009,Talia2012, Chisholm2015}.

Previous work exploiting the HST-COS capabilities to examine the rest-UV spectrum of GPs and LBAs has provided insight on the presence of strong outflows in \ourobject.
From the analysis of the blue-shifted absorption in low and mid-ionisation absorption UV lines, \citet{Alexandroff2015} found an outflow velocity $v_{\rm out}=360$\,\kms\ from mid-ionisation Si\,III1206\AA\ absorption lines, while \citet{Chisholm2015} used the same data to obtain an outflow velocity $v_{90}=516$\,\kms\ from the SiII\,1260 line profile.
Later, \citet{Heckman2016} reanalysed the same HST COS spectra and reported a maximum outflow velocity computed from both the SiII\,1260 and CII\,1334 absorption lines of $v_{\rm max}\sim$\,660\,\kms\ for \ourobject.

Within this work and following Section~5.2.2, the FWHM from both \ha\ and [\oiii] blueshifted broad line components of \ourobject\ imply an outflow velocity  $v_{\rm out} \sim$\,508$\pm$23 and 575$\pm$20 \kms, respectively.
In Fig.~\ref{fig10}, we show at the same velocity resolution the \lya\ and UV low- and mid-ionisation interstellar absorption lines detected in HST-COS spectra together with the \ha\ broad and narrow components to compare the velocity structure of these lines.
Considering the relatively large uncertainties and assumptions behind these calculations, this result is well within previous findings based on UV interstellar lines, providing independent evidence of the presence of a strong outflow in \ourobject. We note, however, that the maximum velocity estimated from UV lines is a factor of $\sim$3 lower than the expansion velocity of the ionised outflow in emission, as measured from the FWZI.

\begin{figure}
\centering
\includegraphics[width=\columnwidth, angle=0]{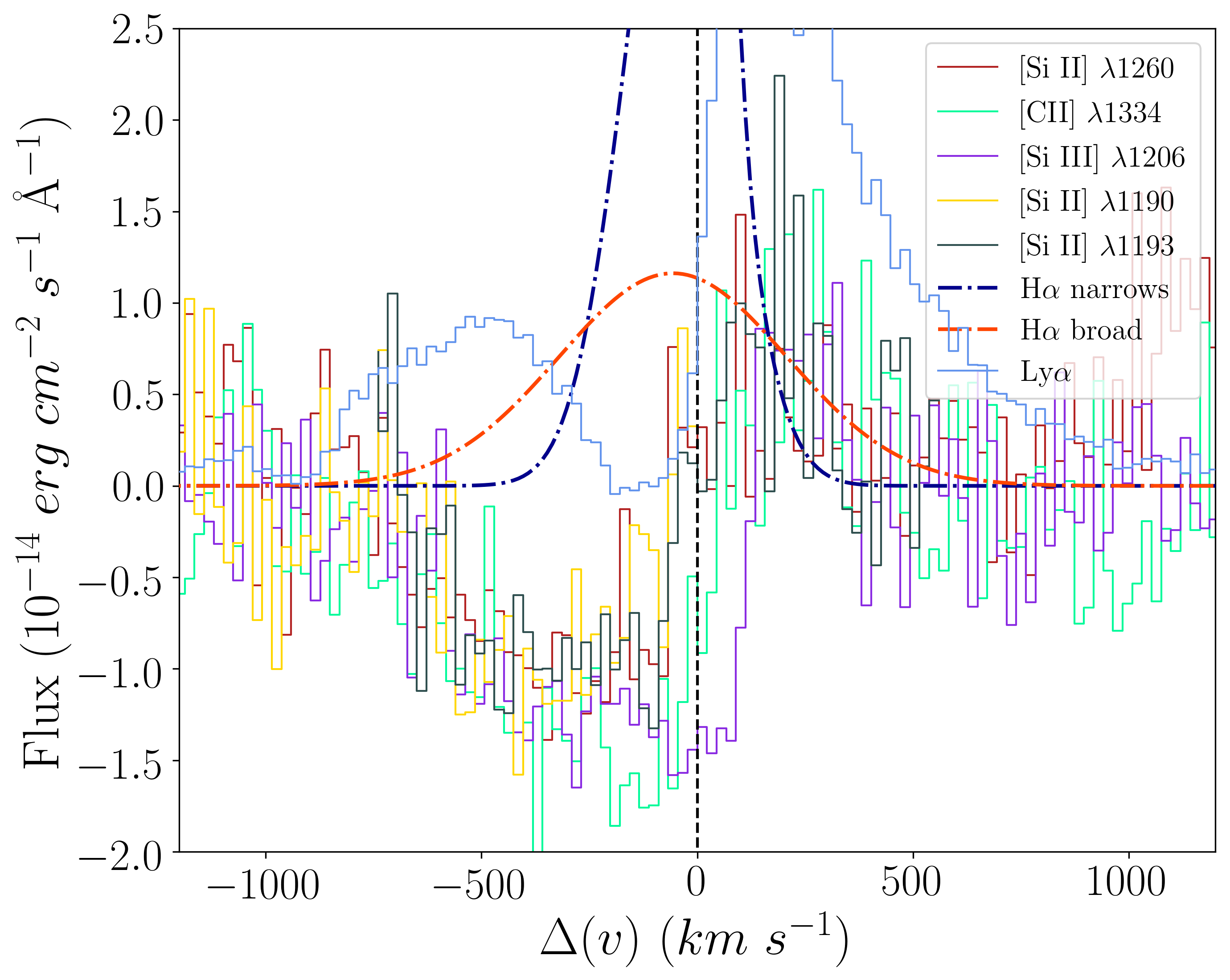}
\caption{\lya\ profile and absorption profiles from HST COS data with H$\alpha$ composite narrow (sum of narrow and mid-width components) and broad components. The Ly$\alpha$ profile is depicted by the solid pale blue line, the composite narrow component by the dashed blue line and the broad component by the dashed orange line. Scaling and corrections are as previously detailed. The absorption profiles for [Si II] $\lambda 1260$, [CII] $\lambda 1334$, [Si III] $\lambda 1206$ and [Si II] $\lambda \lambda 1190, 1193$ from HST COS data are also given in the figure and scaled arbitrarily.}
\label{fig9}
\end{figure}

\subsection{Other possible sources of line broadening}

\subsubsection{AGN activity from the central DCO?}

The presence of AGN activity as a driver for the observed broad emission in \ourobject\ is unlikely according to the emission-line diagnostics.
Broad line emission with powerful \ha\ luminosities ($10^{40} - 10^{42} erg s^{-1}$) and strong ionisation can be expected in galaxies hosting an AGN (A12b), resulting from accretion onto an intermediate-mass black hole.
The existence of relatively faint AGN activity in low metallicity starbursts, such as \ourobject, has also been predicted by models and candidates located in the upper-left part of the BPT diagnostics, which has been discussed in previous studies \citep{IzotovThuan2008}.
However, one of the main differences between \ourobject \ (i.e. low-metallicity AGN candidates) and more massive compact starbursts, likely hosting AGN activity in a central DCO \citep{Overzier2009}, is that for the latter the broad emission is mostly observed in Hydrogen recombination lines and/or strong [\oiii]4957,5007 lines.
In contrast, \ourobject\ possesses broad emission with consistent kinematics in all the observed lines, including low and high ionisation collisionally excited and recombination lines. The contribution of the broad emission to the total emission of different ions is high and similar in each instance, a feature that is not seen in low-metallicity AGN candidates \citep[cf.][]{IzotovThuan2008}.

The emission-line ratios of \ourobject, and particularly those corresponding to the broad emission, can be explained by thermal photoionisation without the need of invoking non-thermal AGN radiation.
Emission-line diagnostics in Fig.~\ref{fig6} and Fig.~\ref{fig7} suggest that harder radiation fields than those in typical star-forming galaxies are necessary to explain the high [\oiii]/\hb\ and high [\oiii]/[\oii] ratios seen in \ourobject.
For example, the maximal starburst predictions from Cloudy photoionisation models considered by \citet{Xiao2018} shifts the classical demarcation lines upwards between the pure star-forming and AGN regions. These predictions account for the harder radiation fields resulting from the inclusion of massive star binaries evolution into the single stellar population models processed by Cloudy.
The displacement observed in the BPT-N2 and BPT-S2 diagrams (Fig.~\ref{fig6})  for the narrow component of \ourobject\ is not large enough to suggest AGN emission. Furthermore, the broad emission appears fully consistent with pure stellar photoionisation, possibly powered by rapdiation from star cluster complexes with very high effective temperatures (above the temperatures of typical HII regions).
Some contribution of fast radiative shocks can be also expected due to the relatively high [\oi] emission, but shocks do not appear to be the dominant mechanism.

Moreover, while \ourobject\ possesses a narrow HeII\,4686\AA\ emission-line of nebular origin in the integrated SDSS spectrum that requires a hard radiation field \citep{Shirazi2012}, the HeII/\hb\ ratio is consistent with thermal photoionionisation, either by very massive stars \citep{Kehrig2015,Kehrig2018}, fast radiative shocks \citep{Izotov2012}, X-ray binaries \citep{Schaerer2019} or a combination of the above.
No further indication of non-thermal emission is found in \ourobject.
We note, however, that without X-ray data we cannot completely disregard the existence of an AGN or probe the contribution of X-ray binaries, which both emit bright X-rays as point sources.

\subsubsection{Turbulent mixing layers}

The consistency of the kinematics, in combination with the fact that the main strong-line ratios are similar for the three components of \ourobject\, adds a stringent constraint for any model attempting to explain the nature of the broader components.

Turbulent mixing layers (TMLs) are a source of emission-line broadening produced at the interface of cold gas in the disk and warmer outflow gas  \citep[e.g.][]{Westmoquette2007,Westmoquette2008, Binette2009}. Using a set of TML models, \citet{Binette2009} studied the remarkably hypersonic (FWHM$\sim$\,2400\kms) broad components present in the \textit{bright} [\oiii]4959,5007, \hb, and \ha\ emission lines of Mrk\,71, a very young
(age $\lesssim$\,4 Myr) giant HII region in the blue dwarf galaxy NGC\,2366
--proposed as the closest "analogue" of the GP's ISM extreme conditions \citep{Micheva2017,Micheva2019}.
While these models successfully reproduce the very high velocity wings of these lines of FWHM$\sim$3500\kms, they do not predict any broad wings underneath the lower excitation lines such as [\nii] or [\sii]. According to \citet{Binette2009}, in models with a high $\log(U)$ the ionisation throughout the turbulent layer is high and, therefore, the gas is dominated by high excitation species, such as O$^{+2}$ or N$^{+2}$. While the TML model of \citet{Binette2009} seems to work well for Mrk71, which has very broad\footnote{FWHM$\sim$1100\kms, a factor of $\gtrsim$\,2 larger than that of \ourobject\
\citep{Micheva2019}} faint \ha\ emission that is not present in low ionisation lines such as [\sii] and [\nii], this is clearly inconsistent with the emission-line structure found in \ourobject.

A different scenario arises when TMLs are invoked to explain broad emission components with FWHM of a few hundred \kms, such as in \ourobject\ and other starburst galaxies
\citep[e.g. M82 or NGC\,1569][]{Westmoquette2007,Westmoquette2008}.
The outflow velocities inferred from these broad components are much lower and, therefore, most likely driven by SNe. In the case of \ourobject, intermediate ionisation lines such as [\nii] and [\sii] may possess broad components if the TML model assumes a sufficiently low ionisation parameter \citep{Binette2009}.
However, the $\log(U)$ parameter for the broad components in \ourobject\ reported in Section~\ref{sec:results} (Table~\ref{tab3}) is modest but not notably low.
As remarked by \citet{Binette2009}, the ionisation conditions of the nebula (i.e. ionisation-bunded or matter(density)-bounded) also affects the quantitative predictions of the models.
Ad-hoc TML models and high-dispersion IFU data would be needed to constrain this hypothesis further.

 Recently, \citet{Bosch2019} presented spatially-resolved, high-dispersion IFU data of the GP $J\,0820$, which shows a similar multi-component structure in all the emission lines probed in their study. The authors discuss the presence of broad and mid-width Gaussian components, with average FWHM$\sim$\,400 and 250\kms respectively.
 As in \ourobject, these components are also found to be shifted in velocity with respect to a narrower component (FWHM$\sim$\,90\kms). Interesting, they find indication of disordered rotation from the narrower component.
 The mid-width component of $J\,0820$ appears instead to follow the kinematics of the broad component, the latter showing indications of a biconical outflow expanding out to a few kpc from the galaxy centre.
 In line with the TML scenario proposed by \citet{Westmoquette2007} for the central starburst in M82, \citet{Bosch2019} suggest the mid-width component might be strongly affected by highly turbulent outer layers of the ionised nebula at the interface between a hot superwind originated in the star-forming clump and colder ISM condensations in a possibly thick disk, resulting from an extraordinarily high SFR surface density.

 In \ourobject, the mid-width component is also found to follow the kinematics (i.e. spatial distribution and radial velocities) of the outflow traced by the broad component. We, therefore, hypothesise that TMLs may arise an  alternative interpretation to that of multiple unresolved HII regions to explain the presence of mid-width components. This type of component would be associated with turbulence in the outflow rather than the bulk flow of the superwind traced by the broader components.

\subsection{Impact on high redshift studies}

\subsubsection{May broad emission affect integrated measurements in compact galaxies at high redshift?}

Galaxies with very similar integrated properties to \ourobject\  appear increasingly common at intermediate and high redshifts $z>1-2$, where present-day spectroscopy of low-mass galaxies are limited due to poor spatial resolution. Recent large surveys have shown that the integrated line ratios of $z\sim$\,2-3 galaxies are offset towards more extreme ionisation properties in emission-line diagnostics (e.g. BPT diagrams) when compared to lower redshift galaxies of similar stellar mass \citep[e.g.][]{Shapley2016,Strom2017}.
The presence of broad emission in low redshift analogs, such as \ourobject, pose the question of whether its contribution may affect the interpretation of integrated spectral measurements in the case of a high-$z$ galaxy observed at low spectral resolution and/or the impossibility to fit a broad component due to poor S/N in the wings of bright emission lines.

Indications of broad emission in $z\sim$\,2-3 galaxies from the MOSDEF survey were found and discussed by \citet{Freeman2019}. They concluded that part of such broad emission show evidence of shock excitation, moving towards the top-right part of the classical BPT diagnostics showing higher N2 and S2 than expected for pure stellar photoionisation.
Results for \ourobject\ does only partially support these conclusions, although differences in the position in the different diagnostics explored here appear not so extreme as in \citet{Freeman2019} and the excitation of the broad component in \ourobject\ can be explained with stellar photoionisation.
The largest difference is seen in the [\oi] BPT and the O32-R23 diagram, where the total integrated line ratios appear offset by $\gtrsim$0.1-0.2~dex compared to the positions of the narrower components which are  associated to giant HII regions. However, we note here that these differences might be larger if high redshift measurements are done using single gaussian fitting and we therefore suggest the use of total integration under the line profile as a most reliable way to obtain total integrated measurements.

\subsubsection{The Role of Star Formation and Feedback on the Escape of Ionising Photons}

In order to keep the Universe reionised at $z>6$, early star-forming galaxies would require a significant escape fraction of Lyman continuum (LyC) ionising photons ($f_{\rm esc}^{\rm LyC}$) of up to 15-20\% \citep{Ouchi2009,Robertson2015,Atek2015}.
At lower redshift, where LyC leakage can be inferred directly or indirectly, normal star-forming galaxies struggle to  achieve these fractions \citep{Grazian2016,Grazian2017,Guaita2016} and only the most extreme star forming low-mass galaxies such as GPs and LBAs, which appear ubiquitous at $z>6$, could make the job \citep[e.g.][and references therein]{Schaerer2016,Izotov2018b}.

The mechanisms and conditions that would drive such extreme LyC leakage are not yet established. Two possible  scenarios, which are likely non-mutually exclusive at galactic scales, are widely discussed in literature. On the one hand, mechanical feedback in the form of stellar winds and SNe-driven outflows from low-mass galaxies with a SFR density above a critical value has been proposed as a mechanism to clear out HI gas and produce a porous ISM, thus enabling LyC photon escape and boosting  escape fractions \citep[e.g.][]{ClarkeOey2002,WiseCen2009,Heckman2011, Alexandroff2015, Sharma2017, Trebitsch2017}.
On the other hand, large escape fractions can be expected from galaxies in which young super star clusters ionise their surrounding neutral ISM, creating density-bounded HII regions where the HI distribution truncates before the LyC photons are absorbed \citep{Jaskot2013,Nakajima2014}.
The latter is typically associated to high excitation (e.g. elevated ionisation [\oiii]/[\oii]) and low HI column densities (e.g. double-peak \lya\ with small peak separation), which appear typical properties of the most extreme GPs and compact HII galaxies.
As discussed in \citet{Jaskot2017}, under the extreme ISM density and pressure conditions of these extremely young ($\lesssim$\,2-3 Myr) super star clusters, the formation of superwinds can be inhibited by strong radiative cooling \citep[e.g.][]{Silich2007,SilichGTT2017} which can also produce faint and broad wings in nebular emission lines \citep{GTT2010}.

Here, we can discuss further these scenarios in the context of the results obtained for \ourobject, classified by \citet{Alexandroff2015} as a LyC leaking candidate.
According to our results, the narrow line components seen in \ourobject\ show ISM conditions consistent with LyC escape, namely very high ionisation ($\log(U)\sim$\,-2.2) with [\oiii]/[\oii] ratios of about 10 (Fig.~\ref{fig7}), low metallicity of about 20\% solar, relatively high electron densities $\gtrsim$200\,cm$^{-3}$ (Table~\ref{tab4}).
The galaxy has a single massive clump of star formation with high SFR density (Section~5.1) and evidence of hard radiation fields from young massive stars  (e.g. the SDSS spectrum shows a strong HeII\,4686 narrow line). All these properties suggest the narrow components are originated in HII regions approaching density-bounded conditions.

\begin{figure}
\centering
\includegraphics[width=\columnwidth, angle=0]{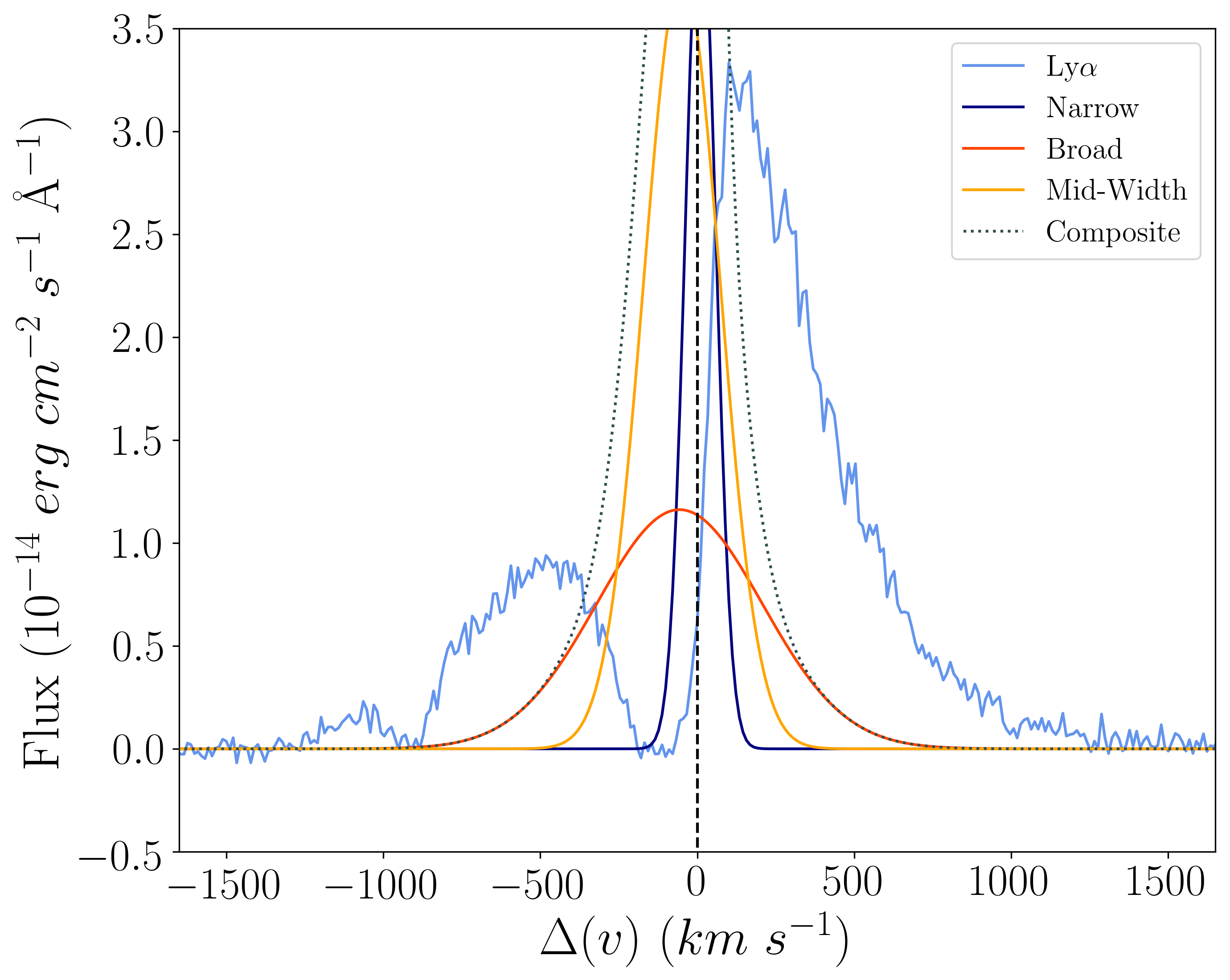}
\caption{\lya\ profile from HST COS data with H$\alpha$ three-component model. The \lya\ profile is depicted by the solid pale blue line and the H$\alpha$ composite model by the dotted grey line. The narrow, mid-width and broad components of the model are given by the solid dark blue, gold and orange lines respectively. The \ha\ model and components have been scaled by a Case B recombination factor of 8.7. The HST COS data has been re-binned to correct for the difference in instrumental resolution with the ISIS data and the \ha\ profile and model scaled to correct for the difference in aperture size.}
\label{fig10}
\end{figure}

Moreover, \ourobject\ show clear evidence of strong outflows driven by star formation from both UV (low- and mid-ionisation) absorption lines and low and high ionisation nebular lines. From hydrogen recombination lines the outflowing ISM is also evident in the shape of both Lyman and Balmer lines (Fig.~\ref{fig9}).
In Figure~\ref{fig10}, we compare at the same spectral resolution the \lya\ and \ha\ profiles of \ourobject\ using for the latter our best multi-component model.
The \ha\ profiles shown in Fig.{\ref{fig10}} have been corrected to match the COS aperture and scaled by the case B recombination factor 8.7 (assuming $T_{\rm e}=$\,10$^4$\,K and $n_{\rm e}=$\,350\,cm$^{-3}$).
From this comparison, we derive a \lya\ escape fraction $f_{\rm esc}^{Ly\alpha}\sim$\,0.1, in good agreement with previous estimates \citep{Alexandroff2015,Yang2017b}.
The \lya\ profile is clearly broader than \ha\ suggesting significant resonant scattering in the neutral ISM.

The central \lya \ absorption of \ourobject\ is nearly coincident with the peak velocity of the broad component in \ha, about 60\kms\ blue-shifted from the systemic velocity measured from the narrower component. It is also interesting to note that a second blue bump is apparent in the wing of \lya, suggesting a triple-peaked profile as seen in other LyC leakers \citep{Izotov2018b,Rivera-Thorsen2017,Vanzella2019}.
This feature is also apparent in the 2D spectrum  \citep[see Fig.~1 in][]{Yang2017a}, suggesting that the intrinsic \lya\ profile would be symmetric with respect to the systemic velocity marked by the narrower component of \ha.

The \lya\ peak separation strongly correlates with HI column density \citep{Dijkstra2006,Verhamme2015}, where peak separations below 300\,\kms\ usually indicate LyC leakage \citep{Verhamme2017,Guaita2017}.
Following the work of \citet{Yang2017b}, the peak separation of \ourobject\ is $\sim$\,490\kms, which according to the outflowing shell model fitted by these authors indicates a relatively high HI column density $\log(n_{\rm HI})=$\,20.4\,cm$^{-2}$ and a very low HI shell velocity of $\sim$\,10\kms.
The HI column density exceed by several orders of magnitude the values allowed for LyC escape, i.e. to produce unit optical depth at the Lyman edge ($\log(n_{\rm HI})=$\,17.2\,cm$^{-2}$), suggesting that if LyC leakage from \ourobject\ will be only possible through ionised holes or channels where the HI column density is significantly lower than its integrated value in the line of sight.
We note, however, that the above results are very dependent on the models adopted to fit the \lya\ emission. As extensively discussed in recent works \citep[see e.g.][]{Gronke2016,Orlitova2018}, modelling with the complex double-peaked \lya\ profiles of starburst galaxies with simple geometries and usual assumptions sometimes makes difficult to reconcile measured properties (e.g. LyC leakage) and model results.

Moreover, \ourobject\ shows a factor of $\sim$2 larger \lya-to-UV continuum size ratio than GP leakers in the sample analysed by \citet{Yang2017b}, with a FWHM(\lya)/FWHM(UV)$\sim$7 that suggests an extended Lya halo.
Interestingly, IFU observations  suggest a connection between extended \lya\ haloes of GPs such as LARS14 \citep{Hayes2013,Herenz2016} with their large \ha\ velocity dispersion.
As all the GPs explored so far with high dispersion spectroscopy, \ourobject\ shows large velocity dispersion suggesting a highly turbulent ISM, i.e., a dispersion dominated kinematics.
As suggested in \citet{Herenz2016}, a highly turbulent ISM may favour the escape of \lya\ photons contributing to increase their resonant scattering and the  formation of an extended halo. In this line, recent numerical simulations \citep{Kimm2019} show the escape fractions of both LyC and \lya\ generally increasing with time if gas clouds are  efficiently dispersed by radiation and supernova feedback.
The presence of a turbulent structure in \ourobject\ can be  interpreted as a driver for the creation of a \lya\ halo and add support to the possible escape of LyC ionising photons through turbulent channels with lower HI column densities carved by the outflowing ionised gas.

Therefore, we hypothesise that if LyC photons escape from \ourobject, they  do it through holes, open cavities, channels or filaments created by the a strong ionised outflow powered by the collective action of stellar winds and SNe feedback in a quasi-static and extended HI halo.
In this scenario, \ourobject\ is experiencing a long ($\gtrsim$\,10 Myr) starburst episode in the central DCO that has started to metal-enrich and clear out part of the surrounding neutral ISM, and a second generation of younger star clusters ($\sim$2-4 Myr) still contain very massive stars illuminating these channels  with the hard radiation fields that create young HII regions approaching density-bounded conditions within the DCO.
The latter appear kinematically represented by highly turbulent narrower emission-line components. The broader, mid-width component is in turn consistent with the turbulent and clumpy ionised medium that is interacting with the strong outflow and could be tracing the channels or filaments from which Lyman photons can escape.
This picture is in line with the conclusions by \citet{Herenz2017} from a detailed MUSE study of the  extremely metal-poor compact starburst SBS\,0335-052E.
Additional support to this picture in GP with similar properties is provided by the spatially resolved kinematics presented by \citet{Bosch2019}.

We note, however, that our current spectroscopic  data cannot provide details on sub-kpc scales and spatially resolved spectroscopy is needed to test the above scenario.
Similar leaking starbursts at much lower distances have been studied using IFU spectroscopy  such as Haro\,11 \citep{Menacho2019}, Mrk\,71 \citep{Micheva2019}, Tol\,1247-232  \citep{Micheva2018} and ESO\,338-IG04 \citep{Bik2018}, providing a closer look into the details of Lyman photon escape.
All these objects show complex kinematics and ionised emission properties similar to GPs. We note however, that all of them show multiple star-forming regions and/or show clear evidence of mergers and/or strong interactions.
Instead, \ourobject\ appear to be a relatively regular face-on stellar host dominated by a single massive central star cluster complex of a few hundred of pc in size. AO-assisted IFU spectroscopy \citep[e.g.][]{Goncalves2010} would be required to provide a more detailed view of the central DCO, while high $S/N$ seeing-limited IFU may be sufficient to map the extended ionised chemodynamical properties allowing spatial analysis of the outflow  and the  identification of the locations from which ionising photons escape.

\section{Summary and Conclusions}

This work presents the first chemodynamical study of a GP galaxy, \ourobject \ at $z\sim$\,0.17, based on deep high-dispersion optical spectra. We fit a detailed  multi-component Gaussian model of resolved emission-line profiles, from [OII]\,3727 to [SII]\,6730, to obtain tight constraints on the ionised gas kinematics, chemical abundances, and ionization properties of \ourobject. The key findings of such analysis are summarised as follows:

\begin{itemize}
\item Our best-fit model consists of a narrow component with intrinsic velocity dispersion $\sigma_{\rm N} \sim 40$\,km\,s$^{-1}$, and a mid-width and broad components with $\sigma_{\rm M} \sim 100$\,km\,s$^{-1}$ and $\sigma_{\rm B} \sim 250$\,km\,s$^{-1}$, respectively.
The mid-width and broad components are blue-shifted by $\gtrsim$\,50\,km\,s$^{-1}$ with respect to the narrow component and their combined contribution to the total \ha\ and [\oiii] luminosity is larger than $\sim$\,50\%.
The broad component shows a full width at zero intensity of about 1500\,km\,s$^{-1}$, suggesting the presence of very high velocity gas. This kinematic structure consistently fits all the relevant optical lines under study in \ourobject.

\item Emission-line ratios are used to compute extinction, electron densities, temperatures, and excitation properties for each kinematic component. While nebular extinction traced by \ha/\hb\ is consistently low for each component, we find that both narrow and mid-width components have higher [\oiii]/\hb, [\oiii]/[\oii], and [\oiii]\,4363/[\oiii]\,5007 ratios, and lower [\nii]/\ha, [\sii]/\ha, and [\oi]/\ha\ ratios compared to the broader component. Using BPT diagnostics, the three kinematic components show line ratios consistent with stellar photoionisation dominating nebular excitation.

\item We observe evidence of higher $n_{\rm e}$ and lower $T_{\rm e}$ for the broader component. Using the direct $T_e$-method, therefore, we measure higher O/H, a higher ionisation parameter and lower N/O for the broad than for the narrow and mid-width components. Furthermore, these results are consistent with values found using a semi-empirical method, combining observed emission-line ratios with predictions, from Cloudy photionisation models.
\end{itemize}

Combining the above results with ancillary HST and SDSS data, we attempt to interpret the observed multiple kinematic components. Moreover, we discuss the role of turbulence and stellar feedback as possible drivers of chemical enrichment and ionising photon escape in low-metallicity starburts (analogues of high-redshift galaxies).
While the narrow component is attributed to the emission of the dominant central star-forming clump of \ourobject, the blue-shifted mid-width component is interpreted as a highly turbulent, possibly clumpy, ionised gas at lower density.
This mid-width component is kinematically associated with the broader component, which traces a denser photoionised galactic outflow showing some evidence of metal-enrichment.

Different broadening effects, such as an AGN, radiative shocks and turbulent mixing layers, are discussed but are largely dismissed as dominant drivers for the broad emission.
Instead, we favour stellar feedback as the origin of broad emission, which is additionally supported by the evidence of outflows from interstellar low- and mid-ionisation UV absorption lines in HST-COS spectra.
Both UV absorptions and optical broad emission possess similar kinematics, with outflow velocities of about 500-600 km s$^{-1}$.
Using a simple model, we find relatively low mass- and metal-loading factors values for the outflow, which are similar to those found in $z\sim$\,2 galaxies of similar stellar mass and SFR density.
While this model implies that the outflow may strip some metals from the central star-forming clump, this would not be sufficient to explain the offset position towards low metallicity of \ourobject\ in the mass-metallicity relation.
Additional effects, such as metallicity dilution by metal-poor inflowing gas, may also need to be considered to account for this result \citep[e.g.][]{Amorin2010}.

Finally, our findings are in agreement with a two-stage starburst picture for \ourobject. Hard radiation fields, from very young massive stars forming in a central starburst of just a few hundred parsecs in size, illuminate a turbulent and clumpy ISM previously eroded by stellar feedback.
Approximately $\sim$\,10\% of the \lya\ photons produced in the star-forming regions escape through a relatively dense, neutral ISM through a prominent, double-peaked \lya\ line.
There is no direct evidence of LyC leakage for \ourobject. However, its UV and optical spectroscopic properties imply that if a fraction of LyC photons are able to leak from \ourobject, they would do so through optically-thin holes or channels of lower \hi\ column density cleared out by the outflow.
We propose that the lower density, highly turbulent mid-width emission component, showing very high [\oiii]/[\oii] and following the outflow kinematics, might be tracing these optically-thin channels.
Similar studies conducted on both confirmed LyC leakers and non-leakers, especially using deep high-dispersion IFU data, will be necessary to further probe this hypothesis.

\section*{Acknowledgements}
We thank the anonymous referee for her/his helpful report. We would like to express our gratitude to David and Bridget Jacob for their support through the Astronomy Summer Internship Programme at University of Cambridge. Their generous funding made this project possible. L.H. thanks the hospitality of the Kavli Institute for Cosmology Cambridge during her Internship and subsequent visits. We would also like to thank Dr. A. Saintonge for her guidance during the compliation of this paper. R.A is also grateful to Roberto Maiolino and Polychronis Papaderos for useful discussions on this project. R.A. thanks the support of DIDULS (project 030503072341) and the ERC Advance Grant "QUENCH". JVM and EPM acknowledge the  support by the Spanish MINECO project  Estallidos 6 AYA2016-79724-C4, and by the Spanish Science Ministry "Centro de Excelencia Severo Ochoa Program under grant SEV-2017-0709.
Supported by the international Gemini Observatory, a program of NSF's OIR Lab, which is managed by the Association of Universities for Research in Astronomy (AURA) under a cooperative agreement with the National Science Foundation, on behalf of the Gemini partnership of Argentina, Brazil, Canada, Chile, the Republic of Korea, and the United States of America.

%\newpage
%\clearpage

%\onecolumn

%%%%%%%%%%%%%%%%%%%%%%%%%%%%%%%%%%%%%%%%%%%%%%%%%%



%\bibliographystyle{mnras}

\begin{thebibliography}{99}
%  \interlinepenalty=10000

\bibitem[\protect\citeauthoryear{Alexandroff et al.}{2015}]{Alexandroff2015} Alexandroff R.~M., Heckman T.~M., Borthakur S., Overzier R., Leitherer C., 2015, ApJ, 810, 104

\bibitem[\protect\citeauthoryear{Amor{\'{\i}}n et al.}{2009}]{Amorin2009} Amor{\'{\i}}n R., Aguerri J.~A.~L., Mu{\~n}oz-Tu{\~n}{\'o}n C., Cair{\'o}s L.~M., 2009, A\&A, 501, 75

\bibitem[\protect\citeauthoryear{Amor{\'{\i}}n, P{\'e}rez-Montero, \& V{\'{\i}}lchez}{Amor{\'{\i}}n et al.}{2010}]{Amorin2010} Amor{\'{\i}}n R.~O., P{\'e}rez-Montero E., V{\'{\i}}lchez J.~M., 2010, ApJ, 715, L128

\bibitem[\protect\citeauthoryear{Amor{\'{\i}}n et al.}{2012a}]{Amorin2012a} Amor{\'{\i}}n R., P{\'e}rez-Montero E., V{\'{\i}}lchez J.~M., Papaderos P., 2012a, ApJ, 749, 185

\bibitem[\protect\citeauthoryear{Amor{\'{\i}}n et al.}{2012b}]{Amorin2012b} Amor{\'{\i}}n R., V{\'{\i}}lchez J.~M., H{\"a}gele G.~F., Firpo V., P{\'e}rez-Montero E., Papaderos P., 2012b, ApJ, 754, L22 (A12b)


\bibitem[\protect\citeauthoryear{Amor{\'{\i}}n et al.}{2014}]{Amorin2014} Amor{\'{\i}}n R., et al., 2014, A\&A, 568, L8

\bibitem[\protect\citeauthoryear{Amor{\'{\i}}n et al.}{2015}]{Amorin2015} Amor{\'{\i}}n R., et al., 2015, A\&A, 578, A105


\bibitem[\protect\citeauthoryear{Amor{\'{\i}}n et al.}{2017}]{Amorin2017} Amor{\'{\i}}n R., et al., 2017, NatAs, 1, 0052

\bibitem[\protect\citeauthoryear{Andrews \& Martini}{2013}]{Andrews2013} Andrews B.~H., Martini P., 2013, ApJ, 765, 140

\bibitem[\protect\citeauthoryear{Allen et al.}{2008}]{Allen2008} Allen M.~G., Groves B.~A., Dopita M.~A., Sutherland R.~S., Kewley L.~J., 2008, ApJS, 178, 20

\bibitem[\protect\citeauthoryear{Arribas et al.}{2014}]{Arribas2014} Arribas S., Colina L., Bellocchi E., Maiolino R., Villar-Mart{\'{\i}}n M., 2014, A\&A, 568, A14

\bibitem[\protect\citeauthoryear{Asplund et al.}{2009}]{Asplund2009} Asplund M., Grevesse N., Sauval A.~J., Scott P., 2009, ARA\&A, 47, 481

\bibitem[\protect\citeauthoryear{Atek et al.}{2014}]{Atek2014} Atek H., et al., 2014, ApJ, 789, 96

\bibitem[\protect\citeauthoryear{Atek et al.}{2015}]{Atek2015} Atek H., et al., 2015, ApJ, 814, 69

%\bibitem[\protect\citeauthoryear{Atek et al.}{2018}]{Atek2018} Atek H., Richard J., Kneib J.-P., Schaerer D., 2018, MNRAS, 479, 5184

\bibitem[\protect\citeauthoryear{Baldwin, Phillips, \& Terlevich}{Baldwin et al.}{1981}]{Baldwin1981} Baldwin J.~A., Phillips M.~M., Terlevich R., 1981, PASP, 93, 5


\bibitem[\protect\citeauthoryear{Bian et al.}{2017}]{Bian2017} Bian F., Fan X., McGreer I., Cai Z., Jiang L., 2017, ApJ, 837, L12

%\bibitem[\protect\citeauthoryear{Bian, Kewley, \& Dopita}{Bian et al.}{2018}]{Bian2018} Bian F., Kewley L.~J., Dopita M.~A., 2018, ApJ, 859, 175

\bibitem[\protect\citeauthoryear{Bik et al.}{2018}]{Bik2018} Bik A., {\"O}stlin G., Menacho V., Adamo A., Hayes M., Herenz E.~C., Melinder J., 2018, A\&A, 619, A131

\bibitem[\protect\citeauthoryear{Binette et al.}{2009}]{Binette2009} Binette L., Flores-Fajardo N., Raga A.~C., Drissen L., Morisset C., 2009, ApJ, 695, 552


%\bibitem[\protect\citeauthoryear{Bloom et al.}{2018}]{2018MNRAS.476.2339B} Bloom J.~V., et al., 2018, MNRAS, 476, 2339


\bibitem[\protect\citeauthoryear{Bordalo \& Telles}{2011}]{2011ApJ...735...52B} Bordalo V., Telles E., 2011, ApJ, 735, 52

\bibitem[\protect\citeauthoryear{Borthakur et al.}{2014}]{Borthakur2014} Borthakur S., Heckman T.~M., Leitherer C., Overzier R.~A., 2014, Sci, 346, 216

\bibitem[\protect\citeauthoryear{Bosch, et al.}{2019}]{Bosch2019} Bosch G., et al., 2019, MNRAS, 489, 1787

\bibitem[\protect\citeauthoryear{Bouwens et al.}{2015}]{Bouwens2015} Bouwens R.~J., Illingworth G.~D., Oesch P.~A., Caruana J., Holwerda B., Smit R., Wilkins S., 2015, ApJ, 811, 140

\bibitem[\protect\citeauthoryear{Bowler et al.}{2017}]{Bowler2017} Bowler R.~A.~A., Dunlop J.~S., McLure R.~J., McLeod D.~J., 2017, MNRAS, 466, 3612

\bibitem[\protect\citeauthoryear{Brinchmann et al.}{2004}]{Brinchmann2004} Brinchmann J., Charlot S., White S.~D.~M., Tremonti C., Kauffmann G., Heckman T., Brinkmann J., 2004, MNRAS, 351, 1151


\bibitem[\protect\citeauthoryear{Cair{\'o}s \& Gonz{\'a}lez-P{\'e}rez}{2017}]{Cairos2017} Cair{\'o}s L.~M., Gonz{\'a}lez-P{\'e}rez J.~N., 2017, A\&A, 600, A125


\bibitem[\protect\citeauthoryear{Calabr{\`o} et al.}{2017}]{Calabro2017} Calabr{\`o} A., et al., 2017, A\&A, 601, A95


\bibitem[\protect\citeauthoryear{Cardamone et al.}{2009}]{Cardamone2009} Cardamone C., et al., 2009, MNRAS, 399, 1191

\bibitem[\protect\citeauthoryear{Cardelli et al.}{1989}]{Cardelli1989} Cardelli J., et al., 1989, ApJ, vol. 345, Oct. 1, 1989, p. 245-256

\bibitem[\protect\citeauthoryear{Casta\~neda, Vilchez, \& Copetti}{Casta\~neda et al.}{1990}]{Castaneda1990} Castaneda H.~O., Vilchez J.~M., Copetti M.~V.~F., 1990, ApJ, 365, 164

\bibitem[\protect\citeauthoryear{Castellano et al.}{2016}]{Castellano2016} Castellano M., et al., 2016, ApJ, 823, L40

\bibitem[\protect\citeauthoryear{Castellano et al.}{2017}]{Castellano2017} Castellano M., et al., 2017, ApJ, 839, 73
%\bibitem[\protect\citeauthoryear{Chen et al.}{2015}]{Chen2015} Chen T.-W., et al., 2015, MNRAS, 452, 1567

\bibitem[\protect\citeauthoryear{Chakraborti et al.}{2012}]{Chakraborti2012} Chakraborti S., Yadav N., Cardamone C., Ray A., 2012, ApJ, 746, L6

\bibitem[\protect\citeauthoryear{Ch{\'a}vez et al.}{2014}]{Chavez2014} Ch{\'a}vez R., Terlevich R., Terlevich E., Bresolin F., Melnick J., Plionis M., Basilakos S., 2014, MNRAS, 442, 3565

%\bibitem[\protect\citeauthoryear{Ch{\'a}vez et al.}{2012}]{Chavez2012} Ch{\'a}vez R., Terlevich E., Terlevich R., Plionis M., Bresolin F., Basilakos S., Melnick J., 2012, MNRAS, 425, L56

%\bibitem[\protect\citeauthoryear{Chevalier \& Clegg}{1985}]{Chevalier1985} Chevalier R.~A., Clegg A.~W., 1985, Natur, 317, 44

\bibitem[\protect\citeauthoryear{Chisholm et al.}{2015}]{Chisholm2015} Chisholm J., Tremonti C.~A., Leitherer C., Chen Y., Wofford A., Lundgren B., 2015, ApJ, 811, 149

%\bibitem[\protect\citeauthoryear{Chisholm et al.}{2016}]{Chisholm2016} Chisholm J., Tremonti C.~A., Leitherer C., Chen Y., Wofford A., 2016, MNRAS, 457, 3133

\bibitem[\protect\citeauthoryear{Chisholm et al.}{2017}]{Chisholm2017} Chisholm J., Orlitov{\'a} I., Schaerer D., Verhamme A., Worseck G., Izotov Y.~I., Thuan T.~X., Guseva N.~G., 2017, A\&A, 605, A67

\bibitem[\protect\citeauthoryear{Chisholm et al.}{2018}]{Chisholm2018} Chisholm J., et al., 2018, A\&A, 616, A30

\bibitem[\protect\citeauthoryear{Clarke \& Oey}{2002}]{ClarkeOey2002} Clarke C., Oey M.~S., 2002, MNRAS, 337, 1299

%\bibitem[\protect\citeauthoryear{Contini et al.}{2016}]{Contini2016} Contini T., et al., 2016, A\&A, 591, A49

%\bibitem[\protect\citeauthoryear{Curti et al.}{2017}]{Curti2017} Curti M., Cresci G., Mannucci F., Marconi A., Maiolino R., Esposito S., 2017, MNRAS, 465, 1384
\bibitem[\protect\citeauthoryear{Dav{\'e}, Finlator, \& Oppenheimer}{2011}]{Dave2011} Dav{\'e} R., Finlator K., Oppenheimer B.~D., 2011, MNRAS, 416, 1354

\bibitem[\protect\citeauthoryear{Davies et al.}{2019}]{Davies2019} Davies R.~L., et al., 2019, ApJ, 873, 122

\bibitem[\protect\citeauthoryear{de Avillez \& Breitschwerdt}{2007}]{Avillez2007} de Avillez M.~A., Breitschwerdt D., 2007, ApJL, 665, L35

\bibitem[\protect\citeauthoryear{de Barros et al.}{2016}]{deBarros2016} de Barros S., et al., 2016, A\&A, 585, A51

\bibitem[\protect\citeauthoryear{Dib, Bell, \& Burkert}{2006}]{Dib2006} Dib S., Bell E., Burkert A., 2006, ApJ, 638, 797

\bibitem[\protect\citeauthoryear{Dijkstra, Haiman, \& Spaans}{Dijkstra et al.}{2006}]{Dijkstra2006} Dijkstra M., Haiman Z., Spaans M., 2006, ApJ, 649, 14

\bibitem[\protect\citeauthoryear{D{\'{\i}}az et al.}{2000}]{Diaz2000} D{\'{\i}}az A.~I., Castellanos M., Terlevich E., Luisa Garc{\'{\i}}a-Vargas M., 2000, MNRAS, 318, 462

\bibitem[\protect\citeauthoryear{Dors et al.}{2018}]{Dors2018} Dors O.~L., Agarwal B., H{\"a}gele G.~F., Cardaci M.~V., Rydberg C.-E., Riffel R.~A., Oliveira A.~S., Krabbe A.~C., 2018, MNRAS, 479, 2294

\bibitem[\protect\citeauthoryear{Dottori \& Bica}{1981}]{Dottori1981} Dottori H.~A., Bica E.~L.~D., 1981, A\&A, 102, 245

\bibitem[\protect\citeauthoryear{Eldridge et al.}{2017}]{Eldridge2017} Eldridge J.~J., Stanway E.~R., Xiao L., McClelland L.~A.~S., Taylor G., Ng M., Greis S.~M.~L., Bray J.~C., 2017, PASA, 34, e058


%\bibitem[\protect\citeauthoryear{Elmegreen \& Burkert}{2010}]{2010ApJ...712..294E} Elmegreen B.~G., Burkert A., 2010, ApJ, 712, 294


\bibitem[\protect\citeauthoryear{Elmegreen, Zhang, \& Hunter}{Elmegreen et al.}{2012}]{Elmegreen2012} Elmegreen B.~G., Zhang H.-X., Hunter D.~A., 2012, ApJ, 747, 105

\bibitem[\protect\citeauthoryear{Erb}{2015}]{Erb2015} Erb D.~K., 2015, Natur, 523, 169

%\bibitem[\protect\citeauthoryear{Erb, Steidel, \& Chen}{2018}]{Erb2018} Erb D.~K., Steidel C.~C., Chen Y., 2018, arXiv, arXiv:1807.00065


\bibitem[\protect\citeauthoryear{Erb et al.}{2016}]{Erb2016} Erb D.~K., Pettini M., Steidel C.~C., Strom A.~L., Rudie G.~C., Trainor R.~F., Shapley A.~E., Reddy N.~A., 2016, ApJ, 830, 52

\bibitem[\protect\citeauthoryear{Esteban \& Vilchez}{1992}]{EstebanVilchez1992} Esteban C., Vilchez J.~M., 1992, ApJ, 390, 536
\bibitem[\protect\citeauthoryear{Ferland et al.}{2017}]{Ferland2017} Ferland G.~J., et al., 2017, RMxAA, 53, 385

\bibitem[\protect\citeauthoryear{Firpo et al.}{2010}]{Firpo2010} Firpo V., Bosch G., H{\"a}gele G.~F., Morrell N., 2010, MNRAS, 406, 1094


\bibitem[\protect\citeauthoryear{Firpo, Bosch, \& Morrell}{Firpo et al.}{2005}]{Firpo2005} Firpo V., Bosch G., Morrell N., 2005, MNRAS, 356, 1357


\bibitem[\protect\citeauthoryear{Freeman et al.}{2019}]{Freeman2019} Freeman W.~R., et al., 2019, ApJ, 873, 102

\bibitem[\protect\citeauthoryear{Firpo et al.}{2011}]{Firpo2011} Firpo V., Bosch G., H{\"a}gele G.~F., D{\'{\i}}az {\'A}.~I., Morrell N., 2011, MNRAS, 414, 3288

%\bibitem[\protect\citeauthoryear{Gazagnes et al.}{2018}]{Gazagnes2018} Gazagnes S., Chisholm J., Schaerer D., Verhamme A., Rigby J.~R., Bayliss M., 2018, A\&A, 616, A29

\bibitem[\protect\citeauthoryear{Genzel et al.}{2011}]{Genzel2011} Genzel R., et al., 2011, ApJ, 733, 101


\bibitem[\protect\citeauthoryear{Glazebrook}{2013}]{Glazebrook2013} Glazebrook K., 2013, PASA, 30, e056


\bibitem[\protect\citeauthoryear{Gon{\c c}alves et al.}{2010}]{Goncalves2010} Gon{\c c}alves T.~S., et al., 2010, ApJ, 724, 1373

\bibitem[\protect\citeauthoryear{Gonzalez-Delgado, et al.}{1994}]{Gonzalez-Delgado1994} Gonzalez-Delgado R.~M., et al., 1994, ApJ, 437, 239

%\bibitem[\protect\citeauthoryear{Gordon et al.}{2003}]{Gordon2003} Gordon K.~D., Clayton G.~C., Misselt K.~A., Landolt A.~U., Wolff M.~J., 2003, ApJ, 594, 279

\bibitem[\protect\citeauthoryear{Grazian et al.}{2017}]{Grazian2017} Grazian A., et al., 2017, A\&A, 602, A18

\bibitem[\protect\citeauthoryear{Grazian et al.}{2016}]{Grazian2016} Grazian A., et al., 2016, A\&A, 585, A48


\bibitem[\protect\citeauthoryear{Green et al.}{2014}]{Green2014} Green A.~W., et al., 2014, MNRAS, 437, 1070

\bibitem[\protect\citeauthoryear{Green et al.}{2010}]{Green2010} Green A.~W., et al., 2010, Natur, 467, 684

\bibitem[\protect\citeauthoryear{Gronke et al.}{2016}]{Gronke2016} Gronke M., Dijkstra M., McCourt M., Oh S.~P., 2016, ApJ, 833, L26

\bibitem[\protect\citeauthoryear{Guaita et al.}{2017}]{Guaita2017} Guaita L., et al., 2017, A\&A, 606, A19

\bibitem[\protect\citeauthoryear{Guaita et al.}{2016}]{Guaita2016} Guaita L., et al., 2016, A\&A, 587, A133

%\bibitem[\protect\citeauthoryear{Guo et al.}{2016}]{Guo2016} Guo Y., et al., 2016, ApJ, 833, 37

%\bibitem[\protect\citeauthoryear{Guzman et al.}{1996}]{Guzman1996} Guzman R., Koo D.~C., Faber S.~M., Illingworth G.~D., Takamiya M., Kron R.~G., Bershady M.~A., 1996, ApJ, 460, L5

\bibitem[\protect\citeauthoryear{H{\"a}gele, P{\'e}rez-Montero, D{\'\i}az, Terlevich \& Terlevich}{2006}]{Hagele2006} H{\"a}gele G.~F., P{\'e}rez-Montero E., D{\'\i}az {\'A}. I., Terlevich E., Terlevich R., 2006, MNRAS, 372, 293

\bibitem[\protect\citeauthoryear{H{\"a}gele et al.}{2007}]{Hagele2007} H{\"a}gele G.~F., D{\'{\i}}az {\'A}.~I., Cardaci M.~V., Terlevich E., Terlevich R., 2007, MNRAS, 378, 163

\bibitem[\protect\citeauthoryear{H{\"a}gele et al.}{2009}]{Hagele2009} H{\"a}gele G.~F., D{\'{\i}}az {\'A}.~I., Cardaci M.~V., Terlevich E., Terlevich R., 2009, MNRAS, 396, 2295

\bibitem[\protect\citeauthoryear{H{\"a}gele et al.}{2010}]{Hagele2010} H{\"a}gele G.~F., D{\'{\i}}az {\'A}.~I., Cardaci M.~V., Terlevich E., Terlevich R., 2010, MNRAS, 402, 1005

\bibitem[\protect\citeauthoryear{H{\"a}gele et al.}{2012}]{Hagele2012} H{\"a}gele G.~F., Firpo V., Bosch G., D{\'{\i}}az {\'A}.~I., Morrell N., 2012, MNRAS, 422, 3475

\bibitem[\protect\citeauthoryear{H{\"a}gele et al.}{2013}]{Hagele2013} H{\"a}gele G.~F., D{\'{\i}}az {\'A}.~I., Terlevich R., Terlevich E., Bosch G.~L., Cardaci M.~V., 2013, MNRAS, 432, 810

%\bibitem[\protect\citeauthoryear{H{\"a}gele et al.}{2020}]{Hagele2020} H{\"a}gele G., et al., in preparation

\bibitem[\protect\citeauthoryear{Hayes et al.}{2013}]{Hayes2013} Hayes M., et al., 2013, ApJ, 765, L27

\bibitem[\protect\citeauthoryear{Heckman et al.}{2015}]{Heckman2015} Heckman T.~M., Alexandroff R.~M., Borthakur S., Overzier R., Leitherer C., 2015, ApJ, 809, 147

\bibitem[\protect\citeauthoryear{Heckman et al.}{2011}]{Heckman2011} Heckman T.~M., et al., 2011, ApJ, 730, 5

\bibitem[\protect\citeauthoryear{Heckman et al.}{2005}]{Heckman2005} Heckman T.~M., et al., 2005, ApJ, 619, L35


\bibitem[\protect\citeauthoryear{Heckman et al.}{2001}]{Heckman2001} Heckman T.~M., Sembach K.~R., Meurer G.~R., Leitherer C., Calzetti D., Martin C.~L., 2001, ApJ, 558, 56

\bibitem[\protect\citeauthoryear{Heckman \& Borthakur}{2016}]{Heckman2016} Heckman T.~M., Borthakur S., 2016, ApJ, 822, 9


\bibitem[\protect\citeauthoryear{Henry et al.}{2015}]{Henry2015} Henry A., Scarlata C., Martin C.~L., Erb D., 2015, ApJ, 809, 19

\bibitem[\protect\citeauthoryear{Herenz et al.}{2016}]{Herenz2016} Herenz E.~C., et al., 2016, A\&A, 587, A78

\bibitem[\protect\citeauthoryear{Herenz et al.}{2017}]{Herenz2017} Herenz E.~C., Hayes M., Papaderos P., Cannon J.~M., Bik A., Melinder J., {\"O}stlin G., 2017, A\&A, 606, L11

\bibitem[\protect\citeauthoryear{Ho et al.}{2014}]{Ho2014} Ho I.-T., et al., 2014, MNRAS, 444, 3894

\bibitem[\protect\citeauthoryear{Huang et al.}{2016}]{Huang2016} Huang K.-H., et al., 2016, ApJ, 817, 11

%\bibitem[\protect\citeauthoryear{Hunt et al.}{2012}]{Hunt2012} Hunt L., et al., 2012, MNRAS, 427, 906


\bibitem[\protect\citeauthoryear{Izotov et al.}{2018b}]{Izotov2018b} Izotov Y.~I., Worseck G., Schaerer D., Guseva N.~G., Thuan T.~X., Fricke V., A., Orlitov{\'a} I., 2018b, MNRAS, 478, 4851

\bibitem[\protect\citeauthoryear{Izotov et al.}{2018a}]{Izotov2018a} Izotov Y.~I., Schaerer D., Worseck G., Guseva N.~G., Thuan T.~X., Verhamme A., Orlitov{\'a} I., Fricke K.~J., 2018a, MNRAS, 474, 4514

\bibitem[\protect\citeauthoryear{Izotov et al.}{2016b}]{Izotov2016b} Izotov Y.~I., Schaerer D., Thuan T.~X., Worseck G., Guseva N.~G., Orlitov{\'a} I., Verhamme A., 2016b, MNRAS, 461, 3683

\bibitem[\protect\citeauthoryear{Izotov et al.}{2016a}]{Izotov2016a} Izotov Y.~I., Orlitov{\'a} I., Schaerer D., Thuan T.~X., Verhamme A., Guseva N.~G., Worseck G., 2016a, Natur, 529, 178

\bibitem[\protect\citeauthoryear{Izotov, Thuan, \& Privon}{Izotov et al.}{2012}]{Izotov2012} Izotov Y.~I., Thuan T.~X., Privon G., 2012, MNRAS, 427, 1229

\bibitem[\protect\citeauthoryear{Izotov, Guseva, \& Thuan}{Izotov et al.}{2011}]{Izotov2011} Izotov Y.~I., Guseva N.~G., Thuan T.~X., 2011, ApJ, 728, 161

\bibitem[\protect\citeauthoryear{Izotov \& Thuan}{2008}]{IzotovThuan2008} Izotov Y.~I., Thuan T.~X., 2008, ApJ, 687, 133

%\bibitem[\protect\citeauthoryear{Izotov, Thuan \& Stasi{\'n}ska}{Izotov et al.}{2007a}]{Izotov2007a} Izotov Y.~I., Thuan T.~X., Stasi{\'n}ska G., 2007, ApJ, 662, 15

\bibitem[\protect\citeauthoryear{Izotov, Thuan, \& Guseva}{Izotov et al.}{2007}]{Izotov2007b} Izotov Y.~I., Thuan T.~X., Guseva N.~G., 2007, ApJ, 671, 1297

\bibitem[\protect\citeauthoryear{James et al.}{2009}]{James2009} James B.~L., Tsamis Y.~G., Barlow M.~J., Westmoquette M.~S., Walsh J.~R., Cuisinier F., Exter K.~M., 2009, MNRAS, 398, 2

\bibitem[\protect\citeauthoryear{James et al.}{2013a}]{James2013a} James B.~L., Tsamis Y.~G., Walsh J.~R., Barlow M.~J., Westmoquette M.~S., 2013a, MNRAS, 430, 2097

\bibitem[\protect\citeauthoryear{James et al.}{2013b}]{James2013b} James B.~L., Tsamis Y.~G., Barlow M.~J., Walsh J.~R., Westmoquette M.~S., 2013b, MNRAS, 428, 86

\bibitem[\protect\citeauthoryear{James et al.}{2016}]{James2016} James B.~L., Auger M., Aloisi A., Calzetti D., Kewley L., 2016, ApJ, 816, 40

\bibitem[\protect\citeauthoryear{Jaskot \& Oey}{2013}]{Jaskot2013} Jaskot A.~E., Oey M.~S., 2013, ApJ, 766, 91

\bibitem[\protect\citeauthoryear{Jaskot \& Oey}{2014}]{Jaskot2014} Jaskot A.~E., Oey M.~S., 2014, ApJ, 791, L19

\bibitem[\protect\citeauthoryear{Jaskot et al.}{2017}]{Jaskot2017} Jaskot A.~E., Oey M.~S., Scarlata C., Dowd T., 2017, ApJ, 851, L9

\bibitem[\protect\citeauthoryear{Ji, et al.}{2019}]{Ji2019} Ji Z., et al., 2019, arXiv, arXiv:1908.00556

\bibitem[\protect\citeauthoryear{Jones et al.}{2001}]{Jones2001} Jones~E., Oliphant~E., Peterson~P., et al., ``SciPy: Open Source Scientific Tools for Python''

\bibitem[\protect\citeauthoryear{Kauffmann et al.}{2003}]{Kauffmann2003} Kauffmann G., et al., 2003, MNRAS, 346, 1055

\bibitem[\protect\citeauthoryear{Kassin et al.}{2012}]{Kassin2012} Kassin S.~A., et al., 2012, ApJ, 758, 106

\bibitem[\protect\citeauthoryear{Kennicutt \& Evans}{2012}]{Kennicutt2012} Kennicutt R.~C., Evans N.~J., 2012, ARA\&A, 50, 531

\bibitem[\protect\citeauthoryear{Kehrig et al.}{2018}]{Kehrig2018} Kehrig C., V{\'{\i}}lchez J.~M., Guerrero M.~A., Iglesias-P{\'a}ramo J., Hunt L.~K., Duarte-Puertas S., Ramos-Larios G., 2018, MNRAS, 480, 1081

\bibitem[\protect\citeauthoryear{Kehrig et al.}{2015}]{Kehrig2015} Kehrig C., V{\'{\i}}lchez J.~M., P{\'e}rez-Montero E., Iglesias-P{\'a}ramo J., Brinchmann J., Kunth D., Durret F., Bayo F.~M., 2015, ApJ, 801, L28

\bibitem[\protect\citeauthoryear{Kewley et al.}{2001}]{Kewley2001} Kewley L.~J., Dopita M.~A., Sutherland R.~S., Heisler C.~A., Trevena J., 2001, ApJ, 556, 121

\bibitem[\protect\citeauthoryear{Kewley et al.}{2006}]{Kewley2006} Kewley L.~J., Groves B., Kauffmann G., Heckman T., 2006, MNRAS, 372, 961

%\bibitem[\protect\citeauthoryear{Koo et al.}{1995}]{Koo1995} Koo D.~C., Guzman R., Faber S.~M., Illingworth G.~D., Bershady M.~A., Kron R.~G., Takamiya M., 1995, ApJ, 440, L49

\bibitem[\protect\citeauthoryear{Khostovan et al.}{2016}]{Khostovan2016} Khostovan A.~A., Sobral D., Mobasher B., Smail I., Darvish B., Nayyeri H., Hemmati S., Stott J.~P., 2016, MNRAS, 463, 2363

\bibitem[\protect\citeauthoryear{Kimm et al.}{2019}]{Kimm2019} Kimm T., Blaizot J., Garel T., Michel-Dansac L., Katz H., Rosdahl J., Verhamme A., Haehnelt M., 2019, MNRAS, 486, 2215

\bibitem[\protect\citeauthoryear{Kojima, et al.}{2017}]{Kojima2017} Kojima T., Ouchi M., Nakajima K., Shibuya T., Harikane Y., Ono Y., 2017, PASJ, 69, 44

\bibitem[\protect\citeauthoryear{Kroupa \& Weidner}{2003}]{Kroupa2003} Kroupa P., Weidner C., 2003, ApJ, 598, 1076

\bibitem[\protect\citeauthoryear{Krumholz \& Burkhart}{2016}]{Krumholz2016} Krumholz M.~R., Burkhart B., 2016, MNRAS, 458, 1671

\bibitem[\protect\citeauthoryear{Kumari et al.}{2018}]{Kumari2018} Kumari N., James B.~L., Irwin M.~J., Amor{\'{\i}}n R., P{\'e}rez-Montero E., 2018, MNRAS, 476, 3793

\bibitem[\protect\citeauthoryear{Kumari, James, \& Irwin}{Kumari et al.}{2017}]{Kumari2017} Kumari N., James B.~L., Irwin M.~J., 2017, MNRAS, 470, 4618

\bibitem[\protect\citeauthoryear{Kunth \& {\"O}stlin}{2000}]{Kunth2000} Kunth D., {\"O}stlin G., 2000, A\&ARv, 10, 1

\bibitem[\protect\citeauthoryear{Olmo-Garc{\'{\i}}a et al.}{2017}]{OlmoGarcia2017} Olmo-Garc{\'{\i}}a A., S{\'a}nchez Almeida J., Mu{\~n}oz-Tu{\~n}{\'o}n C., Filho M.~E., Elmegreen B.~G., Elmegreen D.~M., P{\'e}rez-Montero E., M{\'e}ndez-Abreu J., 2017, ApJ, 834, 181

\bibitem[\protect\citeauthoryear{Lagos et al.}{2018}]{Lagos2018} Lagos P., Scott T.~C., Nigoche-Netro A., Demarco R., Humphrey A., Papaderos P., 2018, MNRAS, 477, 392

\bibitem[\protect\citeauthoryear{Lagos et al.}{2016}]{Lagos2016} Lagos P., Demarco R., Papaderos P., Telles E., Nigoche-Netro A., Humphrey A., Roche N., Gomes J.~M., 2016, MNRAS, 456, 1549


\bibitem[\protect\citeauthoryear{Lagos et al.}{2014}]{Lagos2014} Lagos P., Papaderos P., Gomes J.~M., Smith Castelli A.~V., Vega L.~R., 2014, A\&A, 569, A110

\bibitem[\protect\citeauthoryear{Law et al.}{2009}]{Law2009} Law D.~R., Steidel C.~C., Erb D.~K., Larkin J.~E., Pettini M., Shapley A.~E., Wright S.~A., 2009, ApJ, 697, 2057

\bibitem[\protect\citeauthoryear{Leitherer et al.}{1999}]{Leitherer1999} Leitherer C., et al., 1999, ApJS, 123, 3

\bibitem[\protect\citeauthoryear{Lehnert \& Heckman}{1996}]{LehnertHeckman1996} Lehnert M.~D., Heckman T.~M., 1996, ApJ, 462, 651

%\bibitem[\protect\citeauthoryear{Lehnert et al.}{2009}]{Lehnert2009} Lehnert M.~D., Nesvadba N.~P.~H., Le Tiran L., Di Matteo P., van Driel W., Douglas L.~S., Chemin L., Bournaud F., 2009, ApJ, 699, 1660

%\bibitem[\protect\citeauthoryear{Leloudas et al.}{2015}]{Leloudas2015} Leloudas G., et al., 2015, MNRAS, 449, 917

\bibitem[\protect\citeauthoryear{Lilly et al.}{2013}]{Lilly2013} Lilly S.~J., Carollo C.~M., Pipino A., Renzini A., Peng Y., 2013, ApJ, 772, 119

\bibitem[\protect\citeauthoryear{Lofthouse, Houghton, \& Kaviraj}{Lofthouse et al.}{2017}]{Lofthouse2017} Lofthouse E.~K., Houghton R.~C.~W., Kaviraj S., 2017, MNRAS, 471, 2311

\bibitem[\protect\citeauthoryear{Ly et al.}{2016}]{Ly2016} Ly C., Malkan M.~A., Rigby J.~R., Nagao T., 2016, ApJ, 828, 67

\bibitem[\protect\citeauthoryear{Martin}{1998}]{Martin1998} Martin C.~L., 1998, ApJ, 506, 222

\bibitem[\protect\citeauthoryear{Maiolino \& Mannucci}{2019}]{Maiolino2018} Maiolino R., Mannucci F., 2019, A\&ARv, 27, 3

\bibitem[\protect\citeauthoryear{Mannucci et al.}{2010}]{Mannucci2010} Mannucci F., Cresci G., Maiolino R., Marconi A., Gnerucci A., 2010, MNRAS, 408, 2115

\bibitem[\protect\citeauthoryear{Marino et al.}{2013}]{Marino2013} Marino R.~A., et al., 2013, A\&A, 559, A114

\bibitem[\protect\citeauthoryear{Maseda et al.}{2014}]{Maseda2014} Maseda M.~V., et al., 2014, ApJ, 791, 17

%\bibitem[\protect\citeauthoryear{Maseda et al.}{2018}]{Maseda2018} Maseda, M.~V., van der Wel, A., Rix, H.-W., et al.\ 2018, ApJ, 854, 29

\bibitem[\protect\citeauthoryear{Mason et al.}{2017}]{Mason2017} Mason C.~A., et al., 2017, ApJ, 838, 14

\bibitem[\protect\citeauthoryear{Masters et al.}{2014}]{Masters2014} Masters D., et al., 2014, ApJ, 785, 153

%\bibitem[\protect\citeauthoryear{McKinney et al.}{2019}]{McKinney2019} McKinney J.~H., Jaskot A.~E., Oey M.~S., Yun M.~S., Dowd T., Lowenthal J.~D., 2019, ApJ, 874, 52

\bibitem[\protect\citeauthoryear{Melnick et al.}{2017}]{Melnick2017} Melnick J., et al., 2017, A\&A, 599, A76

\bibitem[\protect\citeauthoryear{Menacho et al.}{2019}]{Menacho2019} Menacho V., et al., 2019, arXiv, arXiv:1903.11662


\bibitem[\protect\citeauthoryear{Micheva et al.}{2017}]{Micheva2017} Micheva G., Oey M.~S., Jaskot A.~E., James B.~L., 2017, ApJ, 845, 165

\bibitem[\protect\citeauthoryear{Micheva et al.}{2018}]{Micheva2018} Micheva G., Oey M.~S., Keenan R.~P., Jaskot A.~E., James B.~L., 2018, ApJ, 867, 2

\bibitem[\protect\citeauthoryear{Micheva et al.}{2019}]{Micheva2019} Micheva G., Christian Herenz E., Roth M.~M., {\"O}stlin G., Girichidis P., 2019, A\&A, 623, A145
\bibitem[\protect\citeauthoryear{Moiseev \& Lozinskaya}{2012}]{Moiseev2012} Moiseev A.~V., Lozinskaya T.~A., 2012, MNRAS, 423, 1831

\bibitem[\protect\citeauthoryear{Moiseev, Tikhonov, \& Klypin}{Moiseev et al.}{2015}]{Moiseev2015} Moiseev A.~V., Tikhonov A.~V., Klypin A., 2015, MNRAS, 449, 3568

\bibitem[\protect\citeauthoryear{Nakajima \& Ouchi}{2014}]{Nakajima2014} Nakajima K., Ouchi M., 2014, MNRAS, 442, 900


\bibitem[\protect\citeauthoryear{Nakajima et al.}{2013}]{Nakajima2013} Nakajima K., Ouchi M., Shimasaku K., Hashimoto T., Ono Y., Lee J.~C., 2013, ApJ, 769, 3

%\bibitem[\protect\citeauthoryear{Nakajima et al.}{2016}]{Nakajima2016} Nakajima K., Ellis R.~S., Iwata I., Inoue A.~K., Kusakabe H., Ouchi M., Robertson B.~E., 2016, ApJ, 831, L9

\bibitem[\protect\citeauthoryear{Nakajima et al.}{2018}]{Nakajima2018} Nakajima K., Fletcher T., Ellis R.~S., Robertson B.~E., Iwata I., 2018, MNRAS, 477, 2098

\bibitem[\protect\citeauthoryear{Newman et al.}{2012}]{Newman2012} Newman S.~F., et al., 2012, ApJ, 761, 43

%\bibitem[\protect\citeauthoryear{Noeske et al.}{2007}]{Noeske2007} Noeske K.~G., et al., 2007, ApJ, 660, L47

\bibitem[\protect\citeauthoryear{Olmo-Garc{\'{\i}}a et al.}{2017}]{Olmo-Garcia2017} Olmo-Garc{\'{\i}}a A., S{\'a}nchez Almeida J., Mu{\~n}oz-Tu{\~n}{\'o}n C., Filho M.~E., Elmegreen B.~G., Elmegreen D.~M., P{\'e}rez-Montero E., M{\'e}ndez-Abreu J., 2017, ApJ, 834, 181

\bibitem[\protect\citeauthoryear{Orlitov{\'a} et al.}{2018}]{Orlitova2018} Orlitov{\'a} I., Verhamme A., Henry A., Scarlata C., Jaskot A., Oey M.~S., Schaerer D., 2018, A\&A, 616, A60


\bibitem[\protect\citeauthoryear{Osterbrock \& Ferland}{2006}]{OsterbrockFerland2006} Osterbrock D.~E., Ferland G.~J., 2006, agna.book,

%\bibitem[\protect\citeauthoryear{{\"O}stlin et al.}{2001}]{Ostlin2001} {\"O}stlin G., Amram P., Bergvall N., Masegosa J., Boulesteix J., M{\'a}rquez I., 2001, A\&A, 374, 800

\bibitem[\protect\citeauthoryear{{\"O}stlin et al.}{2015}]{Ostlin2015} {\"O}stlin G., Marquart T., Cumming R.~J., Fathi K., Bergvall N., Adamo A., Amram P., Hayes M., 2015, A\&A, 583, A55

\bibitem[\protect\citeauthoryear{Overzier et al.}{2009}]{Overzier2009} Overzier R.~A., et al., 2009, ApJ, 706, 203

\bibitem[\protect\citeauthoryear{Overzier et al.}{2008}]{Overzier2008} Overzier R.~A., et al., 2008, ApJ, 677, 37-62

\bibitem[\protect\citeauthoryear{Ouchi et al.}{2009}]{Ouchi2009} Ouchi M., et al., 2009, ApJ, 706, 1136

\bibitem[\protect\citeauthoryear{Paulino-Afonso et al.}{2018}]{Paulino-Afonso2018} Paulino-Afonso A., et al., 2018, MNRAS, 476, 5479

\bibitem[\protect\citeauthoryear{Peeples \& Shankar}{2011}]{Peeples2011} Peeples M.~S., Shankar F., 2011, MNRAS, 417, 2962

\bibitem[\protect\citeauthoryear{P{\'e}rez-Montero}{2017}]{PM2017} P{\'e}rez-Montero E., 2017, PASP, 129, 043001

\bibitem[\protect\citeauthoryear{P{\'e}rez-Montero}{2014}]{PM2014} P{\'e}rez-Montero E., 2014, MNRAS, 441, 2663


\bibitem[\protect\citeauthoryear{P{\'e}rez-Montero \& Contini}{2009}]{PMC2009} P{\'e}rez-Montero E., Contini T., 2009, MNRAS, 398, 949

\bibitem[\protect\citeauthoryear{Rivera-Thorsen et al.}{2017}]{Rivera-Thorsen2017} Rivera-Thorsen T.~E., et al., 2017, A\&A, 608, L4

\bibitem[\protect\citeauthoryear{Roberts-Borsani et al.}{2016}]{Roberts-Borsani2016} Roberts-Borsani G.~W., et al., 2016, ApJ, 823, 143
\bibitem[\protect\citeauthoryear{Robertson et al.}{2015}]{Robertson2015} Robertson B.~E., Ellis R.~S., Furlanetto S.~R., Dunlop J.~S., 2015, ApJ, 802, L19

\bibitem[\protect\citeauthoryear{Rodr{\'{\i}}guez del Pino et al.}{2019}]{Rodriguez2019} Rodr{\'{\i}}guez del Pino B., Arribas S., Piqueras L{\'o}pez J., Villar-Mart{\'{\i}}n M., Colina L., 2019, MNRAS, 486, 344

\bibitem[\protect\citeauthoryear{Roy et al.}{1992}]{Roy1992} Roy J.-R., Aube M., McCall M.~L., Dufour R.~J., 1992, ApJ, 386, 498

\bibitem[\protect\citeauthoryear{S{\'a}nchez et al.}{2019}]{Sanchez2019} S{\'a}nchez S.~F., et al., 2019, MNRAS, 484, 3042

%\bibitem[\protect\citeauthoryear{S{\'a}nchez Almeida et al.}{2014}]{SanchezAlmeida2014} S{\'a}nchez Almeida J., Elmegreen B.~G., Mu{\~n}oz-Tu{\~n}{\'o}n C., Elmegreen D.~M., 2014, A\&ARv, 22, 71

\bibitem[\protect\citeauthoryear{S{\'a}nchez Almeida et al.}{2015}]{SanchezAlmeida2015} S{\'a}nchez Almeida J., et al., 2015, ApJ, 810, L15

\bibitem[\protect\citeauthoryear{Santini et al.}{2017}]{Santini2017} Santini P., et al., 2017, ApJ, 847, 76

\bibitem[\protect\citeauthoryear{Schaerer et al.}{2016}]{Schaerer2016} Schaerer D., Izotov Y.~I., Verhamme A., Orlitov{\'a} I., Thuan T.~X., Worseck G., Guseva N.~G., 2016, A\&A, 591, L8

\bibitem[\protect\citeauthoryear{Schaerer, Fragos, \& Izotov}{Schaerer et al.}{2019}]{Schaerer2019} Schaerer D., Fragos T., Izotov Y.~I., 2019, A\&A, 622, L10

%\bibitem[\protect\citeauthoryear{Schlegel, Finkbeiner, \& Davis}{Schlegel et al.}{1998}]{Schlegel1998} Schlegel D.~J., Finkbeiner D.~P., Davis M., 1998, ApJ, 500, 525

%\bibitem[\protect\citeauthoryear{Schulze et al.}{2018}]{Schulze2018} Schulze S., et al., 2018, MNRAS, 473, 1258

\bibitem[\protect\citeauthoryear{Shapley et al.}{2016}]{Shapley2016} Shapley A.~E., Steidel C.~C., Strom A.~L., Bogosavljevi{\'c} M., Reddy N.~A., Siana B., Mostardi R.~E., Rudie G.~C., 2016, ApJ, 826, L24

\bibitem[\protect\citeauthoryear{Sharma et al.}{2017}]{Sharma2017} Sharma M., Theuns T., Frenk C., Bower R.~G., Crain R.~A., Schaller M., Schaye J., 2017, MNRAS, 468, 2176

\bibitem[\protect\citeauthoryear{Shibuya, Ouchi, \& Harikane}{Shibuya et al.}{2015}]{Shibuya2015} Shibuya T., Ouchi M., Harikane Y., 2015, ApJS, 219, 15

\bibitem[\protect\citeauthoryear{Shim \& Chary}{2013}]{Shim2013} Shim H., Chary R.-R., 2013, ApJ, 765, 26


\bibitem[\protect\citeauthoryear{Shirazi \& Brinchmann}{2012}]{Shirazi2012} Shirazi M., Brinchmann J., 2012, MNRAS, 421, 1043

\bibitem[\protect\citeauthoryear{Silich \& Tenorio-Tagle}{2017}]{SilichGTT2017} Silich S., Tenorio-Tagle G., 2017, MNRAS, 465, 1375

\bibitem[\protect\citeauthoryear{Silich, Tenorio-Tagle, \& Mu{\~n}oz-Tu{\~n}{\'o}n}{Silich et al.}{2007}]{Silich2007} Silich S., Tenorio-Tagle G., Mu{\~n}oz-Tu{\~n}{\'o}n C., 2007, ApJ, 669, 952

\bibitem[\protect\citeauthoryear{Smit et al.}{2014}]{Smit2014} Smit R., et al., 2014, ApJ, 784, 58

\bibitem[\protect\citeauthoryear{Smit et al.}{2015}]{Smit2015} Smit R., et al., 2015, ApJ, 801, 122

\bibitem[\protect\citeauthoryear{Simons et al.}{2015}]{Simons2015} Simons R.~C., Kassin S.~A., Weiner B.~J., Heckman T.~M., Lee J.~C., Lotz J.~M., Peth M., Tchernyshyov K., 2015, MNRAS, 452, 986

\bibitem[\protect\citeauthoryear{Sobral et al.}{2018}]{Sobral2018} Sobral D., et al., 2018, MNRAS, 477, 2817

\bibitem[\protect\citeauthoryear{Spergel et al.}{2007}]{Spergel2007} Spergel D.~N., et al., 2007, ApJS, 170, 377

%\bibitem[\protect\citeauthoryear{Stanway et al.}{2014}]{Stanway2014} Stanway E.~R., Eldridge J.~J., Greis S.~M.~L., Davies L.~J.~M., Wilkins S.~M., Bremer M.~N., 2014, MNRAS, 444, 3466

\bibitem[Stark(2016)]{Stark2016} Stark, D.~P.\ 2016, \araa, 54, 761

\bibitem[\protect\citeauthoryear{Stark et al.}{2017}]{Stark2017} Stark D.~P., et al., 2017, MNRAS, 464, 469

%\bibitem[\protect\citeauthoryear{Stasi{\'n}ska}{1990}]{Stasinska1990} Stasi{\'n}ska G., 1990, A\&AS, 83, 501

\bibitem[\protect\citeauthoryear{Stasi{\'n}ska et al.}{2015}]{Stasinska2015} Stasi{\'n}ska G., Izotov Y., Morisset C., Guseva N., 2015, A\&A, 576, A83

%\bibitem[\protect\citeauthoryear{Steidel et al.}{2014}]{Steidel2014} Steidel C.~C., et al., 2014, ApJ, 795, 165

\bibitem[\protect\citeauthoryear{Strom et al.}{2017}]{Strom2017} Strom A.~L., Steidel C.~C., Rudie G.~C., Trainor R.~F., Pettini M., Reddy N.~A., 2017, ApJ, 836, 164

\bibitem[\protect\citeauthoryear{Talia et al.}{2012}]{Talia2012} Talia M., et al., 2012, A\&A, 539, A61

\bibitem[\protect\citeauthoryear{Tasca et al.}{2015}]{Tasca2015} Tasca L.~A.~M., et al., 2015, A\&A, 581, A54

%\bibitem[\protect\citeauthoryear{Th{\"o}ne et al.}{2015}]{Thone2015} Th{\"o}ne C.~C., de Ugarte Postigo A., Garc{\'{\i}}a-Benito R., Leloudas G., Schulze S., Amor{\'{\i}}n R., 2015, MNRAS, 451, L65

\bibitem[\protect\citeauthoryear{Th{\"o}ne et al.}{2019}]{Thone2019} Th{\"o}ne C.~C., et al., 2019, arXiv, arXiv:1904.05935

%\bibitem[\protect\citeauthoryear{Tenorio-Tagle et al.}{1997}]{GTT1997} Tenorio-Tagle G., Mu{\~n}oz-Tu{\~n}{\'o}n C., P{\'e}rez E., Melnick J., 1997, ApJ, 490, L179

\bibitem[\protect\citeauthoryear{Tenorio-Tagle et al.}{2010}]{GTT2010} Tenorio-Tagle G., W{\"u}nsch R., Silich S., Mu{\~n}oz-Tu{\~n}{\'o}n C., Palou{\v s} J., 2010, ApJ, 708, 1621


%\bibitem[\protect\citeauthoryear{Tenorio-Tagle, Silich, \& Mu{\~n}oz-Tu{\~n}{\'o}n}{2003}]{GTT2003} Tenorio-Tagle G., Silich S., Mu{\~n}oz-Tu{\~n}{\'o}n C., 2003, ApJ, 597, 279


%\bibitem[\protect\citeauthoryear{Tenorio-Tagle et al.}{2005}]{GTT2005} Tenorio-Tagle G., Silich S., Rodr{\'{\i}}guez-Gonz{\'a}lez A., Mu{\~n}oz-Tu{\~n}{\'o}n C., 2005, ApJ, 628, L13

\bibitem[\protect\citeauthoryear{Terlevich et al.}{2004}]{Terlevich2004} Terlevich R., Silich S., Rosa-Gonz{\'a}lez D., Terlevich E., 2004, MNRAS, 348, 1191

\bibitem[\protect\citeauthoryear{Terlevich \& Melnick}{1981}]{Terlevich1981} Terlevich R., Melnick J., 1981, MNRAS, 195, 839

\bibitem[\protect\citeauthoryear{Terlevich et al.}{2014}]{Terlevich2014} Terlevich R., Terlevich E., Bosch G., D{\'{\i}}az {\'A}., H{\"a}gele G., Cardaci M., Firpo V., 2014, MNRAS, 445, 1449

\bibitem[\protect\citeauthoryear{Terlevich et al.}{2015}]{Terlevich2015} Terlevich R., Terlevich E., Melnick J., Ch{\'a}vez R., Plionis M., Bresolin F., Basilakos S., 2015, MNRAS, 451, 3001

\bibitem[\protect\citeauthoryear{Trebitsch et al.}{2017}]{Trebitsch2017} Trebitsch M., Blaizot J., Rosdahl J., Devriendt J., Slyz A., 2017, MNRAS, 470, 224

\bibitem[\protect\citeauthoryear{Turner et al.}{2017}]{Turner2017} Turner O.~J., et al., 2017, MNRAS, 471, 1280

\bibitem[\protect\citeauthoryear{Vanzella et al.}{2016}]{Vanzella2016} Vanzella E., et al., 2016, ApJ, 825, 41

\bibitem[\protect\citeauthoryear{Vanzella et al.}{2017}]{Vanzella2017} Vanzella E., et al., 2017, ApJ, 842, 47

\bibitem[\protect\citeauthoryear{Vanzella et al.}{2018}]{Vanzella2018} Vanzella E., et al., 2018, MNRAS, 476, L15

\bibitem[\protect\citeauthoryear{Vanzella et al.}{2019}]{Vanzella2019} Vanzella E., et al., 2019, arXiv, arXiv:1904.07941

\bibitem[\protect\citeauthoryear{Verhamme et al.}{2017}]{Verhamme2017} Verhamme A., Orlitov{\'a} I., Schaerer D., Izotov Y., Worseck G., Thuan T.~X., Guseva N., 2017, A\&A, 597, A13

\bibitem[\protect\citeauthoryear{Verhamme et al.}{2015}]{Verhamme2015} Verhamme A., Orlitov{\'a} I., Schaerer D., Hayes M., 2015, A\&A, 578, A7

\bibitem[\protect\citeauthoryear{Veilleux \& Osterbrock}{1987}]{Veilleux1987} Veilleux S., Osterbrock D.~E., 1987, ApJS, 63, 295

%\bibitem[\protect\citeauthoryear{Veilleux, Cecil, \& Bland-Hawthorn}{2005}]{Veilleux2005ARA&A} Veilleux S., Cecil G., Bland-Hawthorn J., 2005, ARA\&A, 43, 769

\bibitem[\protect\citeauthoryear{Weiner et al.}{2006}]{Weiner2006} Weiner B.~J., et al., 2006, ApJ, 653, 1027

\bibitem[\protect\citeauthoryear{Weiner et al.}{2009}]{Weiner2009} Weiner B.~J., et al., 2009, ApJ, 692, 187

\bibitem[\protect\citeauthoryear{Westmoquette et al.}{2007}]{Westmoquette2007} Westmoquette M.~S., Exter K.~M., Smith L.~J., Gallagher J.~S., 2007, MNRAS, 381, 894

\bibitem[\protect\citeauthoryear{Westmoquette, Smith, \& Gallagher}{Westmoquette et al.}{2008}]{Westmoquette2008} Westmoquette M.~S., Smith L.~J., Gallagher J.~S., 2008, MNRAS, 383, 864

%\bibitem[\protect\citeauthoryear{Wise et al.}{2014}]{2014MNRAS.442.2560W} Wise J.~H., Demchenko V.~G., Halicek M.~T., Norman M.~L., Turk M.~J., Abel T., Smith B.~D., 2014, MNRAS, 442, 2560

\bibitem[\protect\citeauthoryear{Wise \& Cen}{2009}]{WiseCen2009} Wise J.~H., Cen R., 2009, ApJ, 693, 984

%\bibitem[\protect\citeauthoryear{Wisnioski et al.}{2015}]{Wisnioski2015} Wisnioski E., et al., 2015, ApJ, 799, 209

\bibitem[\protect\citeauthoryear{Yang et al.}{2017c}]{Yang2017a} Yang H., Malhotra S., Rhoads J.~E., Leitherer C., Wofford A., Jiang T., Wang J., 2017, ApJ, 838, 4

\bibitem[\protect\citeauthoryear{Yang et al.}{2017b}]{Yang2017b} Yang H., et al., 2017, ApJ, 844, 171

\bibitem[\protect\citeauthoryear{Yang et al.}{2017a}]{Yang2017c} Yang H., Malhotra S., Rhoads J.~E., Wang J., 2017, ApJ, 847, 38


\bibitem[\protect\citeauthoryear{Xiao et al.}{2018}]{Xiao2018} Xiao L., Stanway E.~R., Eldridge J.~J., 2018, MNRAS, 477, 904-934

\end{thebibliography}


%%%%%%%%%%%%%%%%%%%%%%%%%%%%%%%%%%%%%%%%%%%%%%%%%%

%%%%%%%%%%%%%%%%% APPENDICES %%%%%%%%%%%%%%%%%%%%%
%\clearpage
%\newpage

%\appendix
%\section{Online material}
%\subsection{Additional figures on multicomponent fitting}

%If you want to present additional material which would interrupt the flow of the main paper, it can be placed in an Appendix which appears after the list of references.

%%%%%%%%%%%%%%%%%%%%%%%%%%%%%%%%%%%%%%%%%%%%%%%%%%

% Don't change these lines
\bsp	% typesetting comment
\label{lastpage}
\end{document}